\documentclass[journal=psy]{CUP-JNL-DTM}%

\usepackage{graphicx}
\usepackage{multicol,multirow}
\usepackage{amsmath,amssymb,amsfonts}
\usepackage{amsthm}
\usepackage{rotating}
\usepackage{appendix}
\usepackage{ifpdf}
\usepackage[T1]{fontenc}
\usepackage{newtxtext}
\usepackage{newtxmath}
\usepackage{textcomp}
\usepackage{lipsum}
\usepackage[colorlinks,allcolors=blue]{hyperref}

\usepackage{enumitem}
\usepackage[table]{xcolor}

\newcommand{\mirm}[1]{MiRM$(#1)$}

\newtheorem{theorem}{Theorem}[section]

\newtheorem{corollary}[theorem]{Corollary}
\theoremstyle{definition}
\newtheorem{remark}[theorem]{Remark}

\numberwithin{equation}{section}

\jname{Preprint}
\jyear{16 November 2026}

\begin{document}

\begin{Frontmatter}

\title[Article Title]{Microstate Response Model: A Unified IRT Framework from Latent Nominal States to a Theory of Tests}

\author[1]{Seongsu Lee}
\author[1]{Junghyo Jo}

\authormark{Lee \& Jo}
 
\address[1]{
  \orgdiv{Department of Physics Education, College of Education},
  \orgname{Seoul National University},
  \orgaddress{\city{Seoul}, \postcode{08826}, \country{Republic of Korea}}}



\abstract{Several features outside of the conventional assumptions in traditional item response theory (IRT) models have been revealed to be related, yet no unifying framework explains the features simultaneously. They can be unified by treating each cause as a distinct reason, and then regarding every reason as a choice in a latent multiple-choice item. Here, we propose the Microstate Response Model (MiRM), which decomposes each response category (macrostate) into multiple latent reasons (microstates) using Boltzmann weights. The consequences of this simple framework are substantial; for example, it reads the lower and upper asymptotes of item characteristic curves (ICCs) as additional states, and ICC asymmetry and item complexity as multiple states. The model unifies many classical models from the dichotomous logistic family to mixture models. Moreover, the algebra of the model provides the intuition that items and tests are the same object. Two consequences naturally come out: partial credits and test scores are unified, and a new test-level monotonicity condition is proposed. We examine item characteristics including nonmonotonicity on PISA 2015, and show that MiRM provides a deeper understanding of items while achieving the strongest overall performance across fit, parsimony, and prediction.}


\end{Frontmatter}

\section{Introduction}\label{sec:intro}
Item response theory (IRT) aims to estimate a person's latent trait or ability from item response data, and further understand each item's characteristics that enable the measurement. Since the true latent ability cannot be observed, the theory typically assumes certain properties of the items \citep{rasch1960studies} and the trait \citep{holland1986conditional,sijtsma1998methodology}, thereby making persons and items identifiable and interpretable on a common scale. For example, the characteristics of an item should not depend on which group of respondents takes it, the same ability should be obtained regardless of which items are used, no variable other than the ability should explain the responses, and a higher ability should lead to a higher probability of a correct response. In fact, not every assumption holds perfectly in real-world data. For example, respondents sometimes answer correctly by chance or wrongly by mistake, and items may work differently for different groups of respondents. We call these exceptions \emph{departures} from the conventional IRT assumptions, and many of them appear in the shapes of item characteristic curves (ICCs), which show how the probability of a correct response changes with ability.

One approach to resolving the tension between theory and practice is to model the mechanism that is believed to generate each departure, by adding a parameter for it. The 3-parameter logistic (3PL) model adds a guessing parameter $p_c$, as low-ability respondents sometimes `guess' the answer, thereby keeping the lower asymptote of its ICC --- the probability of a correct response at the lowest abilities --- above zero \citep{birnbaum1968some}. The 4-parameter logistic (4PL) model \citep{barton1981upper} likewise adds a parameter for high-ability respondents who `slip' through careless mistakes, thereby keeping the upper asymptote below one. Furthermore, some empirical ICCs are asymmetric; they rise steeply on one side and slowly on the other. The Logistic Positive Exponent (LPE; \cite{samejima2000logistic}) model and the Residual Heteroscedasticity (RH; \cite{molenaar2015heteroscedastic}) model tried to capture this asymmetry by adding interpretable parameters: an exponent parameter in the LPE model, and a population-level heteroscedasticity parameter in the RH model, which lets the randomness of the responses differ across ability levels. Mixture IRT models \citep{rost1990rasch,mislevy1990modeling} assume population-level latent classes that make people respond differently. 

As can be seen, each departure has usually been studied individually, and the shapes of ICCs are categorized differently across fields. However, several recent studies have revealed that such departures are closely related to each other, and also to the complexity of an item, such as the number of steps needed to solve it \citep{lee2018asymmetric,bolt2018application,liao2021item,bolt2022item}. For example, \citet{lee2018alternative} reported that the guessing parameter $p_c$ of the 3PL model can help model the asymmetry of an ICC, and vice versa. In addition, \citet{culpepper2016revisiting} and \citet{meng2020marginalized} showed that the 4PL model can be understood as a mixture model with careless and guessing groups. These relationships also mean that a parameter does not always describe what its name suggests; the guessing parameter, for example, may in fact be describing asymmetry, and what we learn about an item then depends on which model we choose. Although these relationships have been formalized through systematic definitions \citep{feuerstahler2026defining,shim2026asymmetry}, a unified model conceptually combining the departures remains absent.

If we take a step back, the various mechanisms above can be reduced to a single cause. A correct answer from a lucky guess is recorded simply as correct, and a wrong answer from a careless mistake simply as wrong. A difficult and complex item makes people fail for different reasons, while an item open to many shortcuts lets people succeed for different reasons. Seen within a single item, the fact that people respond differently according to their ability levels or dispositions also means that the same response can be chosen for different reasons. Therefore, every response suffers an information loss, as it does not record the `reason' for the choice.

Item formats that record such reasons already exist. Interviews and essay questions are typical examples, and \emph{polytomous items}, which record which option was chosen rather than only whether it was correct, are a more standardized form. Many models have brought such detailed information into IRT; let us consider the most basic one, the Nominal Response Model (NRM; \cite{bock1972estimating}). Noting that not all distractors, the wrong options of an item, are chosen in the same pattern, Bock gave each distractor its own exponential term so that they could be analyzed individually. Applying this idea to any dichotomous item, one could suspect that several latent distractors lie behind its single wrong response. If so, what prevents us from asking whether information about the correct response has likewise been lost? Any response, correct or wrong, can have more than one reason, as the lucky guess above already shows. Both can therefore be decomposed into latent `reasons for the response' from the perspective of polytomous IRT, even though the reasons are not observed.

This approach is not new: \citet{maris1995psychometric} introduced latent responses and their condensation into an observed response as general modeling principles, and related ideas appear in the Nested Logit Model \citep[{NLM;}][]{suh2010nested} and in IRTree models, which describe a response as the end of a sequence of latent decisions \citep{de2012irtrees,bockenholt2012modeling}. Even within ratios-of-sums-of-exponentials, there have been `don't know' models \citep{samejima1979new,thissen1984response} and their subsequent extensions \citep{thissen1989multiple,shigemasu1993new,jin2022exploring} that assume a latent `guessed response' category. However, our model differs from these models because we assume that the number of reasons and their exact cognitive counterparts are also unknown. Hence, we allow an arbitrary number of terms, which comes at a price: each term can be identified only partially from the responses, and the number of terms has to be chosen for each item. Yet being unable to identify a reason exactly does not mean that we should leave it unused. We therefore make room for latent reasons, and let data decide the configuration. 

Under this simple assumption, applying the NRM to the latent `reasons' yields a model that we call the \textbf{Microstate Response Model} (MiRM). The terminology comes from statistical mechanics \citep{landau2013statistical}, where \emph{microstates} are the distinct states of a system, some of which cannot be distinguished by macroscopic observation, and a \emph{macrostate} is a coarse-grained state, that is, a group of microstates that show the same observation. Thus, a macrostate corresponds to an observed response as an aggregation of its latent reasons, and microstates correspond to the reasons. This correspondence is more than an analogy, as it brings mathematical tools and interpretations. \citet{smith2022motivations} captured the similarity between the NRM and the canonical ensemble, the Boltzmann distribution over the states of a system, by relating the energy of each state to the discrimination of each response option. They further examined whether the concept of coarse-graining can be applied. However, they did not define IRT counterparts to `microstates'; the absence is where our approach begins.

The microstates put the `reasons' that have been modeled separately on a single layer. If we regard guessing as a microstate, it keeps a share of correct responses even among the least able respondents, and MiRM reduces to the 3PL model. More generally, MiRM reads departures such as the asymptotes, asymmetry, item complexity, and nonmonotonicity directly from its item parameters, instead of adding a new parameter for each of them. The item discrimination $a$ and difficulty $b$ of the 2-parameter logistic (2PL; \cite{birnbaum1968some}) model, which tell how sharply and at which ability an item distinguishes respondents, have been diluted by such additional parameters. MiRM restores them as functions of ability, $a(\theta)$ and $b(\theta)$, that can be identified from the ICC. Furthermore, the framework covers nonparametric IRT models through the Universal Approximation Theorem (Theorem~\ref{thm:uat}), providing them with an interpretable algebraic lens.

This approach does not stop at combining existing item models; it naturally covers an object beyond the item level as well. Just as an item discards the reasons behind a response and compresses them into a single observed response, psychometrics has one more response system that discards the hidden information behind several responses and compresses them into a single result: a \emph{test}. Although a test has been regarded as a set of items, it is an item itself. Following this intuition, we prove algebraically that an item and a test are the same object under MiRM, and from this we unify partial credits and total scores, which have been regarded as different objects. Therefore, the monotonicity condition that has been imposed on each item can be moved to the test, the unit that actually measures ability. We can thereby model items that show nonmonotonic ICCs in real data while keeping the ability scale meaningful.

The core contributions of this paper are itemwise model unification, item--test equivalence and a new test-level condition, and validation through large-scale datasets. Overall, by proposing MiRM, we aim to observe how the theoretical microstates represent practical structures in empirical data, without introducing departure-specific item parameters to resolve the tension between theory and practice. In doing so, we recover part of the information lost in the responses, thereby identifying individual ability precisely and providing a better understanding of items.

The remaining sections are organized as follows. Sections~\ref{sec:item} and \ref{sec:test} define MiRM at the item and test levels, respectively, thereby establishing its generative structure. Section~\ref{sec:gen} introduces how MiRM unifies existing models, and describes how the nonstandard item characteristics are read from microstates throughout the unification process. Section~\ref{sec:est} shows that the estimation of MiRM follows the standard protocol of existing IRT models, and illustrates how the test-level condition is satisfied in implementation. The resulting model is assessed with respect to its fit performance, prediction, ability recovery, identification, and interpretation of item characteristics on both synthetic (Section~\ref{sec:sim}) and empirical (Section~\ref{sec:emp}) data. Further implications and limitations are discussed in Section~\ref{sec:disc}. Unless stated otherwise, proofs of all formal results are collected in Appendix~\ref{appendix:proof}.

\section{The Item Model}\label{sec:item}
Let $\theta_i$ denote the ability of a person $i$. For an item $j$ with $K_j$ response categories, call each category $k\in\{1,\ldots,K_j\}$ a \emph{macrostate}, and suppose that it contains $M_{jk}$ response pathways that are not directly observed. These pathways are called \emph{microstates} and are indexed by $m\in\{1,\ldots,M_{jk}\}$; equivalently, let $\mathcal{S}_{jk}=\{(j,k,m):m=1,\ldots,M_{jk}\}$. Microstate $(j,k,m)$ has slope $\alpha_{j,k,m}$ and offset $\delta_{j,k,m}$, and its weight at ability $\theta$ is $e^{-\alpha_{j,k,m}\theta+\delta_{j,k,m}}$. The probability of selecting category $k$ is the sum of the weights of its microstates divided by the sum of the weights over the entire item.
\begin{align}\label{eq:def_complex}
  P(X_j=k|\theta=\theta_i)=\frac{\sum_{m=1}^{M_{jk}}e^{-\alpha_{j,k,m}\theta_i+\delta_{j,k,m}}}{\sum_{k'=1}^{K_j}\sum_{m=1}^{M_{jk'}}e^{-\alpha_{j,k',m}\theta_i+\delta_{j,k',m}}}.
\end{align}
Suppressing the person and item indices $i$ and $j$, and reindexing $(k,m)$ into $\mathcal{S}=\bigsqcup_k\mathcal{S}_k$ by $s\in\{1,\ldots,S\}$ with $\alpha_s$ monotonically nondecreasing over $s$ and $\alpha_1=\delta_1=0$ fixed as a reference state (see Appendix~\ref{app:sub:micro}), and writing $\epsilon_s(\theta)=\alpha_s\theta-\delta_s$ so that the weight of microstate $s$ is $e^{-\epsilon_s}$, the model is rewritten compactly as
\begin{align}\label{eq:def_simple}
  P_k(\theta)=\frac{Z_k(\theta)}{Z(\theta)}=\frac{\sum_{s\in\mathcal{S}_k}e^{-\epsilon_s}}{\sum_{s\in\mathcal{S}}e^{-\epsilon_s}}.
\end{align}
We call this model the Microstate Response Model (MiRM). 

Each variable has a counterpart in statistical mechanics: $\theta$ is inverse temperature, $\alpha_s$ is energy level, and $\delta_s$ is log-degeneracy. The correct macrostate usually contains the ground state, so rising ability is cooling into the correct response; the negative sign in the exponent follows this energy convention. The framework imposes no physical assumptions on respondents; statistical mechanics supplies the algebra. In particular, differentiating the logarithm of a partition function $Z_k$ gives averages and variances, as is common in physics: $Z_k$ corresponds to the moment-generating function of the microstate slopes, and $\log Z_k$ to their cumulant-generating function. We define the following quantities in this way, where primes denote derivatives with respect to $\theta$.
\begin{align}\label{eq:notation}
  \Psi_k=\log Z_k,\qquad\bar{\alpha}_k=-\Psi_k',\qquad\sigma_{\alpha,k}^2=\Psi_k'',\qquad\bar{\beta}_k=\bar{\alpha}_k\theta+\Psi_k.
\end{align}
As the notation suggests, $\bar{\alpha}_k$ and $\sigma_{\alpha,k}^2$ are the mean and variance of the microstate slopes $\alpha_s$ within category $k$ (see Eq.~\ref{eq:cgf_alpha}). The calculation and the meaning of $\bar{\beta}_k$ are shown in Section~\ref{ssub:char}, and these quantities are used throughout the paper. When the subscript $k$ is omitted, the same definition is applied to the entire item by replacing $Z_k$ with $Z$. Additional notation is collected in Appendix~\ref{appendix:notation}.

\subsection{A Dichotomous Model}\label{sub:bin}
A dichotomous item with one correct and one distractor response is denoted \mirm{c,d}, meaning that it has $c$ correct microstates, $C_1,\ldots,C_c$, and $d$ distractor microstates, $D_1,\ldots,D_d$. The response probability of the correct answer is
\begin{align}
  P_C(\theta)=\frac{e^{-\epsilon_{C_1}}+\cdots+e^{-\epsilon_{C_c}}}{(e^{-\epsilon_{C_1}}+\cdots+e^{-\epsilon_{C_c}})+(e^{-\epsilon_{D_1}}+\cdots+e^{-\epsilon_{D_d}})}=\frac{Z_C}{Z_C+Z_D}.
\end{align}

\subsubsection{Item Characteristics and Information}\label{ssub:char}
\begin{figure}[t]%
  \FIG{\includegraphics[width=\textwidth]{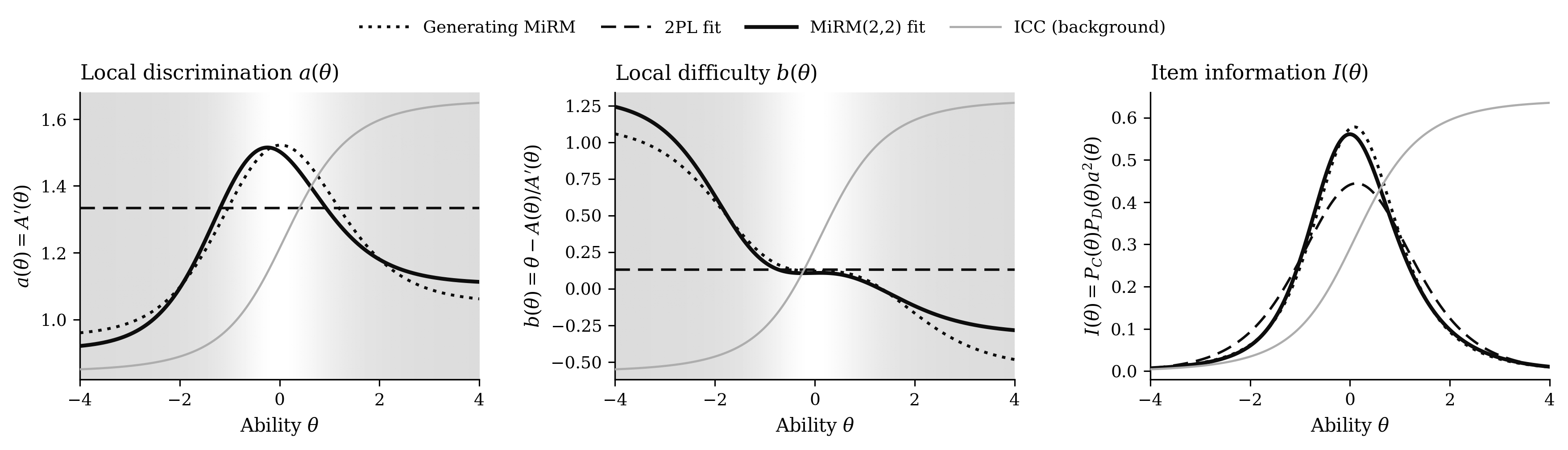}}
  {\caption{Local discrimination, difficulty, and information in MiRM on a simulated item. Dotted lines: generating (true) MiRM; dashed lines: 2PL fit; solid black lines: \mirm{2,2} fit; thin gray curves: underlying ICC. Gray backgrounds on the first two panels denote low information}\label{fig:local_2pl}}
\end{figure}

Using $\Psi_k$, the correct response probability can be written in a direct logistic form, subsuming the model into the Generalized Logit Model \citep{zhang2022bayesian}:
\begin{align}
  P_C(\theta)=\frac{Z_C}{Z_C+Z_D}=\frac{e^{\Psi_C}}{e^{\Psi_C}+e^{\Psi_D}}=\frac{1}{1+e^{-(\Psi_C-\Psi_D)}}.
\end{align}
To interpret the equation further, let $A=\Psi_C-\Psi_D$ be the logit of a dichotomous item. Its derivative is $a=A'$, which we call \emph{local discrimination}. Defining \emph{local difficulty} by $b=\theta-A/a$ where $a\neq0$ gives the following identity.
\begin{align}\label{eq:ab}
  P_C(\theta)=\frac{1}{1+e^{-A(\theta)}}=\frac{1}{1+e^{-a(\theta)[\theta-b(\theta)]}}.
\end{align}
The value $b(\theta)$ is the ability at which the tangent to the logit at $\theta$ crosses zero. Unlike the exact value and count of the microstates, the suggested item characteristics are uniquely identified where $a\neq0$ if the item's ICC is given under a prescribed latent ability scale, since they are functionals of the logit $A(\theta)$. Although the equation follows directly from the definition, its meaning is substantial: every MiRM item is locally a 2PL model at every ability where $a\neq0$, with its two parameters moving along the ability scale. Figure~\ref{fig:local_2pl} illustrates how the functions $a(\theta)$ and $b(\theta)$ are read in practice.

The relation above is more than a relabeling in MiRM, as the local characteristics are obtained directly from the microstate statistics. According to the definitions in Eq.~\ref{eq:notation},
\begin{align}
  a(\theta)=\bar{\alpha}_D(\theta)-\bar{\alpha}_C(\theta),\qquad b(\theta)=\frac{\bar{\beta}_D(\theta)-\bar{\beta}_C(\theta)}{\bar{\alpha}_D(\theta)-\bar{\alpha}_C(\theta)}.
\end{align}
Indeed, $\bar{\alpha}_k$ and $\bar{\beta}_k$ for $k\in\{C,D\}$ take the meaning of their notation as averages over the microstates within category $k$. Differentiating $\Psi_k$ yields
\begin{align}\label{eq:cgf_alpha}
  \bar{\alpha}_k(\theta)=-\Psi_k'=-\frac{d}{d\theta}\left(\log\sum_{s\in\mathcal{S}_k}e^{-\alpha_s\theta+\delta_s}\right)=\sum_{s\in\mathcal{S}_k}\frac{e^{-\epsilon_s}}{Z_k}\alpha_s,
\end{align}
showing that $\bar{\alpha}_k(\theta)$ is the probability-weighted average of each microstate's slope $\alpha_s$ within a single response category $k$. Similarly, $\bar{\beta}_k(\theta)$ is the corresponding effective intercept:
\begin{align}
  \bar{\beta}_k(\theta)&=\bar{\alpha}_k\theta+\Psi_k\\&=\sum_{s\in\mathcal{S}_k}\frac{e^{-\alpha_s\theta+\delta_s}}{Z_k}\left(\alpha_s\theta+\log Z_k\right)\\&=\sum_{s\in\mathcal{S}_k}\left(\frac{e^{-\epsilon_s}}{Z_k}\delta_s\right)-\sum_{s\in\mathcal{S}_k}\left(\frac{e^{-\epsilon_s}}{Z_k}\log\frac{e^{-\epsilon_s}}{Z_k}\right),
\end{align}
representing the probability-weighted average of the offsets $\delta_s$ incorporating the Shannon entropy of the weights. Thus, $a(\theta)$ and $b(\theta)$ vary with ability simply because the dominant pathways for answering correctly or incorrectly shift across ability levels, rather than because the individual microstate slopes or offsets change. This is how MiRM differs from the Generalized Logit Model.

Fisher information measures how precisely an item estimates ability at $\theta$. It also shows where $a(\theta)$ and $b(\theta)$ are important, since they are meaningful only where the item gives enough information. In MiRM, Fisher information is defined at two levels. If the microstates were observed, the score function would be $\bar{\alpha}-\alpha_s$ and the information would be $I_{\rm mic}=\sigma_\alpha^2$. The information in the actually observed response is the variance of the score function $\bar{\alpha}-\bar{\alpha}_k$, namely $I_{\rm mac}=\sum_k P_k(\bar{\alpha}_k-\bar{\alpha})^2$. Therefore,
\begin{align}\label{eq:info}
  I_{\rm mic}=I_{\rm mac}+\sum_k P_k\sigma_{\alpha,k}^2,
\end{align}
where the remaining term $\sum_kP_k\sigma_{\alpha,k}^2$ represents the information lost when microstates are coarse-grained into macrostates. For a binary item, $I_{\rm mac}=P_CP_Da^2$.

\subsubsection{Conditions for Nonmonotonicity}\label{ssub:nonmono}
The model's algebra yields an exact condition for local nonmonotonicity of an ICC.

\begin{theorem}[local condition for nonmonotonicity]\label{thm:nonmono}
  At ability $\theta$, $P_C'(\theta)<0$ iff $\bar{\alpha}_D(\theta)<\bar{\alpha}_C(\theta)$.
\end{theorem}

\begin{proof}
  From the definition of the item logit $A(\theta)=\log[P_C(\theta)/P_D(\theta)]$, an item characteristic curve $P_C$ has derivative:
  \begin{align}
      P_C'(\theta)&=P_C(\theta)P_D(\theta)a(\theta)\\&=P_C(\theta)P_D(\theta)(\bar{\alpha}_D(\theta)-\bar{\alpha}_C(\theta)).
  \end{align}
  Hence, the exact condition of $P_C'(\theta)<0$ at $\theta$ is $\bar{\alpha}_D(\theta)<\bar{\alpha}_C(\theta)$.
\end{proof}

Two consequences follow immediately; the corollaries below translate the theorem into the language of counting the number of microstates.

\begin{corollary}[necessary condition for a peak]\label{cor:peak}
  For an ICC to have a strict local maximum, there must be at least two distractor microstates ($d\geq2$).
\end{corollary}

\begin{corollary}[necessary condition for a dip]\label{cor:dip}
  For an ICC to have a strict local minimum, there must be at least two correct microstates ($c\geq2$).
\end{corollary}

\begin{proof}[Proof of Corollaries~\ref{cor:peak}, \ref{cor:dip}]
  Because $\bar{\alpha}_k'=-\Psi_k''=-\sigma_{\alpha,k}^2\leq0$, $\bar{\alpha}_k$ is nonincreasing. If $d=1$, then $\bar{\alpha}_D=\alpha_D$ is constant, so $a'=\sigma_{\alpha,C}^2\geq0$ and $a$ is nondecreasing. A strict local maximum would require $P_C'>0$ to its left and $P_C'<0$ to its right, meaning that $a$ changes from positive to negative, which a nondecreasing function cannot do. Symmetrically, if $c=1$, then $a'=-\sigma_{\alpha,D}^2\leq0$ and $a$ is nonincreasing; it cannot change from negative to positive and hence there is no strict local minimum.
\end{proof}

The readings of the corollaries are simple. If there is only one possible reason for choosing a wrong answer ($d=1$), there are no peaks in its ICC. Similarly, if there is only one way to answer an item correctly ($c=1$), there are no dips in its ICC.

There are two types of reversal in ICC geometry. \emph{Extremal reversal} denotes reversed asymptotes such as $\lim\limits_{\theta\to\infty}P_C(\theta)=0$ or $\lim\limits_{\theta\to-\infty}P_C(\theta)=1$, and is caused by the highest correct slope or the lowest distractor slope. Existing models corresponding to this type include unfolding or ideal-point models that exhibit a local peak. These models typically interpret $\theta$ as a trait rather than as a monotonically increasing ability. By contrast, \emph{interior reversal}, which denotes an ICC decreasing in a finite interval, can arise when the slope of a microstate lies between the slopes of two microstates of the other macrostate (e.g., $\cdots<\alpha_{C_1}<\alpha_D<\alpha_{C_2}<\cdots$). Assuming that there is no extremal reversal, an interior reversal implies at least one local peak or one local dip. In particular, if the correct-response probability is highest as $\theta\to\infty$, the curve must rise again after any interior drop, producing at least one local dip. Such a curve appears to violate the ordering premise of unidimensional IRT; Section~\ref{sub:prm} relocates this requirement to the test level.

\subsection{A Polytomous Model}\label{sub:mul}
MiRM can model polytomous items, and is denoted as follows. An item with one correct option and several distractors is \mirm{c,d_1,\ldots,d_{K-1}}. A partial-credit item assigning credit $w_k$ to each category is \mirm{\mathbf{M};\mathbf{w}}, where $\mathbf{M}=(M_1,\ldots,M_K)$ and $\mathbf{w}=(w_1,\ldots,w_K)$. Here $\mathbf{M}$ plays the same role as $(c,d)$, giving the number of microstates in each category, and $\mathbf{w}$ gives the credit of each category. For example, the dichotomous item \mirm{c,d} can be rewritten as \mirm{c,d;1,0}. Likewise, scoring a multiple-choice item dichotomously, with unit credit $w_1=1$ for the correct response and zero credit $w_{k\neq1}=0$ for every distractor, yields \mirm{c,d_1,\ldots;1,0,\ldots}. 

For a partial-credit item, we examine the expected-credit curve $\bar{w}(\theta)=\sum_kw_kP_k(\theta)$ instead of an ICC. This curve equals the ICC for a polytomous item scored dichotomously.

\section{The Test Model}\label{sec:test}
Let the set of items $\mathcal{J}=\{1,\ldots,J\}$ denote the \emph{test}, and the probability of a response pattern $\mathbf{X}=(X_1,\ldots,X_J)$ at ability $\theta$ is given by $P_{\mathcal{J}}(\mathbf{X}|\theta)$. Under local independence, the probability of response pattern $\mathbf{k}=(k_1,\ldots,k_J)$ is
\begin{align}
  P_{\mathcal{J}}(\mathbf{X}=\mathbf{k}|\theta)=\prod_{j=1}^JP(X_j=k_j|\theta),
\end{align}
where $P(X_j=k_j|\theta)$ is the selection probability in Eq.~\ref{eq:def_complex}. Here, we point out that the only difference between items and tests is whether the response is scalar or vector. They both collect responses from multiple respondents, and model the response function based on some assumptions in order to understand their ability and traits. Furthermore, the only difference between them can even be resolved if we recognize that both responses are, in fact, neither scalar nor vector, but the elements of a finite set. To clarify this argument, let us consider a test $\mathcal{J}$ with two items ($J=2$). In this example, the two items are both dichotomous; i.e., $K_j=2$ for $j=1,2$. The possible patterns in this test are $(k_1,k_2)\in\{(C,C),(C,D),(D,C),(D,D)\}$. Assigning labels $k^\star_1$, $k^\star_2$, $k^\star_3$, and $k^\star_4$ to $(k_1,k_2)$ in sequence yields an exact polytomous item.

The intuition, which was previously proposed in some models \citep{rosenbaum1988items,wilson1995rasch}, is developed into a formalized theorem under the algebra of MiRM. This is because each item's probability $P_j$ is a ratio of exponential sums, and a product of exponential sums is again an exponential sum, yielding another, larger MiRM. The pattern distribution itself is therefore a single MiRM on the pattern space $\mathcal{K}_\mathcal{J}$. In the two-item case, where $P_{1C}(\theta)=Z_{1C}/(Z_{1C}+Z_{1D})$ and $P_{2C}(\theta)=Z_{2C}/(Z_{2C}+Z_{2D})$, the response probability of this `polytomous item' is given by
\begin{align}
  P_{k^\star_1}(\theta)&=P_{(C,C)}(\theta)\\&=P_{1C}(\theta)P_{2C}(\theta)\\&=\frac{Z_{1C}}{Z_{1C}+Z_{1D}}\cdot\frac{Z_{2C}}{Z_{2C}+Z_{2D}}\\&=\frac{Z_{1C}Z_{2C}}{Z_{1C}Z_{2C}+Z_{1C}Z_{2D}+Z_{1D}Z_{2C}+Z_{1D}Z_{2D}},
\end{align}
and similarly for $k^\star_2$ to $k^\star_4$. The theorem below generalizes the case to every composition of MiRM items.

\begin{theorem}[giant MiRM]\label{thm:giant}
  For a test of locally independent finite-state MiRM items, the MiRM $\mathcal{P}^\star$ on $\mathcal{K}_\mathcal{J}$ defined by $Z_\mathbf{k}^\star=\prod_j Z_{j,k_j}$ and $Z^\star=\prod_j Z_j$ satisfies $P_\mathcal{J}(\mathbf{X}=\mathbf{k}|\theta)=Z_\mathbf{k}^\star/Z^\star$. Moreover, for any $\theta$-independent credit rule $w_\mathcal{J}:\mathcal{K}_\mathcal{J}\rightarrow\mathcal{R}$, the coarsened response $w_\mathcal{J}(\mathbf{X})$ also follows a MiRM.
\end{theorem}

\subsection{Scoring}\label{sub:score} 
One can argue that there is another difference between items and tests: in the two-item example, the single-correct response patterns are distinguished in the test composition, yet remain unscored in the giant item. MiRM readily resolves this: combining two items with scores $\omega_1$, $\omega_2$ yields a giant MiRM with partial credits $\omega_1+\omega_2$, $\omega_1$, $\omega_2$, and 0. On top of that, if $\omega_1=\omega_2$, the two patterns $k^\star_2$ and $k^\star_3$ are \emph{coarsened} into a single score category. Even when each item follows a classical model such as the 2PL (a single-state NRM), coarsening same-score patterns adds their exponential terms together; the NRM is closed under products of items, but not under coarsening. Thus, anyone who has assigned a total score to a test while treating each item as a classical model has already been using a MiRM at the test level.

Furthermore, this extends our view of items and tests: partial credits and test scores can be modeled under a single framework, generalizing the equivalence previously shown for the Rasch family \citep{huynh1994equivalence,huynh1996decomposition}. The formula we suggest below is one natural generalization of existing scoring rules and partial credits.

For a test made of dichotomous items with unit credit $w_{jC}=1$ and $w_{jD}=0$, the credit assigned to a student's pattern $\mathbf{k}=(k_1,\ldots,k_J)$ by a test $\mathcal{J}$ is defined using two anchors --- all-correct response $\mathbf{C}=(C,C,\ldots,C)$, and all-distractor response $\mathbf{D}=(D,D,\ldots,D)$ --- as algebraic reference points, to avoid any prescribed weights before assigning the credits. Writing the giant macrostate mean slope as $\bar{\alpha}_\mathbf{k}^\star(\theta)=-(\log Z_\mathbf{k}^\star)'=\sum_j\bar{\alpha}_{j,k_j}(\theta)$, on the domain $\Theta_T=\{\theta\in\Theta:a_\mathcal{J}(\theta)>0\}$ where $a_\mathcal{J}(\theta)=\bar{\alpha}_\mathbf{D}^\star(\theta)-\bar{\alpha}_\mathbf{C}^\star(\theta)=\sum_j a_j(\theta)$ is positive, we define the \emph{generalized credit} as follows.
\begin{align}\label{eq:credit}
  T(\mathbf{k};\theta)=J\frac{\bar{\alpha}_\mathbf{D}^\star(\theta)-\bar{\alpha}_\mathbf{k}^\star(\theta)}{\bar{\alpha}_\mathbf{D}^\star(\theta)-\bar{\alpha}_\mathbf{C}^\star(\theta)}=\sum_j\omega_j(\theta)w_{j,k_j},\qquad\omega_j(\theta)=J\frac{a_j(\theta)}{a_\mathcal{J}(\theta)}.
\end{align}
The all-distractor pattern receives 0, the all-correct pattern receives $J$, and the weights sum to $J$. The \emph{generalized total score} is 
\begin{align}
  \hat{T}(\mathbf{k})=T(\mathbf{k};\hat{\theta}(\mathbf{k})),
\end{align}
the generalized credit evaluated at the ability estimate. This definition requires $\hat{\theta}(\mathbf{k})\in\Theta_T$ for every pattern to be scored, and patterns outside this domain are reported as having an uncertified score direction. $\hat{T}$ reduces to familiar scores: the sum score under the Rasch model, $J\sum_ja_jw_{j,k_j}/\sum_ja_j$ under the 2PL, the sum under the partial credit model (PCM; \cite{masters1982rasch}), a weighted sum under the generalized partial credit model (GPCM; \cite{muraki1992generalized}), and a score using Birnbaum's weights \citep{birnbaum1968some} under the 3PL (see Appendix~\ref{app:sub:gpcm}).

\subsection{Perfect-Response Monotonicity}\label{sub:prm}
One reason for developing a single framework for items and tests is to allow individual items to have locally nonmonotonic ICCs without losing the meaning of ability. If $\theta$ served only as a coordinate that improves model fit, it would not constitute a measurement scale. IRT therefore requires a condition under which greater ability provides stronger evidence of successful performance. Itemwise monotonicity offers a direct way to impose this requirement and has consequently been treated as a central assumption of unidimensional IRT. It may be sufficient, however, rather than necessary: previous studies have suggested that respondents may remain ordered even when some individual ICCs are nonmonotonic \citep{strout1990new,junker1993conditional,feuerstahler2024monotonicity}.

We likewise seek to identify individual ability by imposing a monotonicity constraint at the test level, as we measure ability using tests, but usually not a single item. However, the major challenge of defining monotonicity at the test level is that there are a lot of aggregate statistics for multiple items, but none of them are fully credible. We do not know which items weigh more, how to average them, and even whether their importance varies over ability, before we assess the data.

Hence, we rely on two criteria: the all-correct pattern $\mathbf{C}$ and the all-distractor pattern $\mathbf{D}$, which were used above to define the generalized credit. When neither the ordering of items, the number of similar or duplicated items, nor any arbitrary scores or weights assigned to them are taken as trustworthy, these are the only two states that remain invariant under any monotone rescoring. Therefore, for $\mathbf{C}$ and $\mathbf{D}$, we expect a test to satisfy
\begin{align}
    \frac{d}{d\theta}P_\mathcal{J}(\mathbf{C}|\theta)\geq0,\qquad\frac{d}{d\theta}P_\mathcal{J}(\mathbf{D}|\theta)\leq0\qquad\forall\theta\in\Theta
\end{align}
meaning that higher ability always increases the probability of a full score and decreases the probability of a zero score. Note that although $\mathbf{C}$ and $\mathbf{D}$ are extreme response patterns, these inequalities must hold at every ability $\theta\in\Theta$, not only at extreme ability levels.

Since logarithmic differentiation does not change the signs of functions,
\begin{align}
    a_{\mathcal{J}C}(\theta)\equiv\frac{d}{d\theta}\log P_\mathcal{J}(\mathbf{C}|\theta)\geq0,\qquad a_{\mathcal{J}D}(\theta)\equiv-\frac{d}{d\theta}\log P_\mathcal{J}(\mathbf{D}|\theta)\geq0.
\end{align}
Hence, we state that a test $\mathcal{J}$ satisfies \textbf{Perfect-Response Monotonicity} (PRM) if $a_{\mathcal{J}C}\geq0$ and $a_{\mathcal{J}D}\geq0$ for every $\theta\in\Theta$. PRM can be applied to any test composition of unidimensional IRT models; in a giant MiRM, however, the two conditions are obtained algebraically since $a_{\mathcal{J}C}=\sum_jP_{jD}a_j$ and $a_{\mathcal{J}D}=\sum_jP_{jC}a_j$, which enables applying the conditions as computable restrictions during estimation (see Section~\ref{sub:prml}).

\emph{Stochastic ordering of the manifest variable} (SOM; \cite{hemker1997stochastic}) requires that the probability of reaching any given sum score or higher is nondecreasing in ability. SOM implies PRM (see Appendix~\ref{app:sub:som}). Therefore,
\begin{align}
  \operatorname{PRM}\Longleftarrow\operatorname{SOM}.
\end{align}
PRM is intentionally weaker than SOM, to allow us to capture individual items' local nonmonotonicity. Although PRM may appear to weaken the overall monotonicity condition, it still gives four guarantees together with the generalized credit (Theorem~\ref{thm:instrument}). First, different abilities give different response distributions, so ability is identifiable. Second, the ability scale keeps a single direction, even when the true trait is multidimensional. Third, a pattern with a higher generalized credit gives stronger evidence of higher ability. Fourth, the expected generalized score $\bar{T}(\theta)=\mathbb{E}\left[T(\mathbf{X};\theta)\,|\,\theta\right]$ stays between 0 and $J$. Conversely, without PRM, models that allow nonmonotonic ICCs, including MiRM, are prone to recover a wrong ability direction. The proofs are given in Appendix~\ref{app:sub:instrument}, and the guarantees are illustrated with experimental results in Sections~\ref{sub:nonmono_sim} and \ref{sub:ndim}.

In practice, one can also distinguish a \emph{bank} from a \emph{form} as two types of test: a bank is the entire pool of items $\mathcal{J}$, whereas a form is an operational subset $\mathcal{J}'\subseteq\mathcal{J}$ administered to an individual respondent. Although we impose PRM on the full item set $\mathcal{J}$ throughout our analyses, the same conditions $a_{\mathcal{J}'C}(\theta)\geq0$ and $a_{\mathcal{J}'D}(\theta)\geq0$ can be applied or diagnosed on any operational form $\mathcal{J}'\subseteq\mathcal{J}$ when needed. In the extreme case where a test or form collapses to a single item ($|\mathcal{J}'|=1$), PRM reduces to $a_{1C}=P_{1D}a_1\geq0$ and $a_{1D}=P_{1C}a_1\geq0$, which is simply $a_1(\theta)\geq0$ for all $\theta\in\Theta$. That is, for a single item to measure ability on its own, its ICC must be monotonically nondecreasing everywhere; this might be the reason that itemwise monotonicity has traditionally been treated as a core assumption of unidimensional IRT.

\begin{remark}[PRM on a single item]\label{rem:single}
    A single item functioning as a test must have a monotonic ICC.
\end{remark}

\section{Relations to Existing Models}\label{sec:gen}

\begingroup
\renewcommand\tablebodyfont{\fontsize{9}{10}\selectfont}
\renewcommand\tablecolheadfont{\fontsize{9}{10}\selectfont}
\begin{table}[t]
  \tabcolsep=0pt%
  \TBL{\caption{Summary of relations between MiRM and existing IRT models\label{tab:rel}}}
  {\begin{fntable}
  \begin{tabular*}{\textwidth}{@{\extracolsep{\fill}}lcll@{}}\toprule%
  \TCH{Model} & \TCH{MiRM form} & \TCH{Relation (condition)} & \TCH{Structural reading}\\\midrule
  \TCH{NRM}& (1,$\ldots$,1) & Exact & Single microstate for one response \\
  \TCH{Rasch}& (1,1) & Nested ($a_j(\theta)=1$) & Single global discrimination \\
  \TCH{2PL}& (1,1) & Exact & Itemwise discrimination/difficulty \\
  \TCH{3PL}& (2,1) & Nested ($\alpha_{C_1}<\alpha_{C_2}=\alpha_{D_1}$)& Guessing as a microstate \\
  \TCH{4PL}& (2,2) & Nested (+ $\alpha_{C_1}=\alpha_{D_2}$)& + Slipping as a microstate \\
  \TCH{LPE}& (1,$\xi$) & Nested for $\xi\in\mathbb{N}$ & Item complexity $\xi$ as latent stages \\
  \multirow[c]{2}{*}{\TCH{RH}}& ($c$,1) & Local correspondence ($h_1<0$) & Heteroscedasticity as $\alpha_C$ heterogeneity \\ & (1,$d$) & Local correspondence ($h_1>0$) & Heteroscedasticity as $\alpha_D$ heterogeneity \\
  \TCH{Continuous ICC\footnotemark[1]}& ($c$,$d$) & Uniform approximation & Arbitrary ICC atomization \\
  \TCH{MNRM}& ($c$,$d$) & Exact piecewise linear readout & Ordering criteria for microstates \\
  \multirow[c]{2}{*}{\TCH{3PLQ}}& ($\infty$,1)\footnotemark[2] & Nested for $q_2>0$ & $\{\alpha_C\}$ on Gaussian spectra \\ & (1,$\infty$)\footnotemark[2] & Nested for $q_2<0$ & $\{\alpha_D\}$ on Gaussian spectra \\
  \TCH{GGUM}& (2,$\ldots$,2) & Nested & Unfolding as paired microstates \\
  \TCH{GPCM}& (1,$\ldots$,1;$\mathbf{w}$) & Nested ($\mathbf{w}=(0,1,2,\ldots)$) & Test-level MiRM within an item \\
  \TCH{Mixture 2PL}& ($2^G-1$,$2^G-1$) & Nested for $G$ groups & Group-aware microstates \\
  \botrule
  \end{tabular*}%
  \footnotetext[]{{Note:} NRM: Nominal Response Model; Rasch: the Rasch model; 2PL: 2-parameter logistic model; 3PL: 3-parameter logistic model; 4PL: 4-parameter logistic model; LPE: Samejima's Logistic Positive Exponent; RH: Molenaar's 3-parameter residual heteroscedasticity model; MNRM: Multidimensional NRM; 3PLQ: 3-parameter quadratic-logit model; GGUM: Generalized Graded Unfolding Model; GPCM: Generalized Partial Credit Model. $c,d\in\mathbb{N}$ are allowed to vary.}
  \footnotetext[1]{This includes any unidimensional continuous binary item response function on a
  compact ability domain.}
  \footnotetext[2]{The sums are replaced by integrals with respect to positive microstate spectral measures.}
  \end{fntable}}
\end{table}
\endgroup

Many existing IRT models are exact equivalents, submodels, or approximations of MiRM (Table~\ref{tab:rel}). Some are described in this section to illustrate how the microstate-reading works, while others are provided in Appendices~\ref{app:sub:gpcm}, \ref{app:sub:instrument}, and \ref{appendix:relations}.

\subsection{Binary Logistic Models}
We start from the 4PL model \citep{barton1981upper}, which follows the formula:
\begin{align}
  P_C(\theta)&=p_c+(1-p_c-p_d)\frac{1}{1+e^{-a(\theta-b)}}\\&=\frac{(1-p_d)+p_ce^{-a(\theta-b)}}{(1-p_d)+p_d+p_ce^{-a(\theta-b)}+(1-p_c)e^{-a(\theta-b)}}.
\end{align}
Thus the 4PL is an exact restriction of \mirm{2,2}. Removing the second distractor microstate $D_2$ of weight $p_d$ ($p_d=0$) gives the 3PL model \citep{birnbaum1968some}, and further removing the second correct microstate $C_2$ of weight $p_ce^{-a(\theta-b)}$ ($p_c=0$) gives the 2PL \citep{birnbaum1968some} with two microstates $C_1$ and $D_1$. From the MiRM perspective, the guessing pathway can be represented as a correct microstate with the distractor's energy $\alpha_{C_2}=a$, and the slipping pathway can be understood as a distractor microstate with the correct choice's energy $\alpha_{D_2}=0$.

In particular, in the 2PL model ($p_c=p_d=0$), the local discrimination $a(\theta)$ and local difficulty $b(\theta)$ collapse into constants $a$ and $b$, indicating that the 2PL assumes itemwise global discrimination and difficulty over the ability space. Extending the assumption to every item derives the Rasch model \citep{rasch1960studies} with $a=1$.

\subsection{Logistic Positive Exponent with Integer Exponent}
The typical expression of the LPE model \citep{samejima2000logistic} is
\begin{align}
  P_C(\theta)=\left[\frac{1}{1+e^{-a(\theta-b)}}\right]^\xi.
\end{align}
Here, we assume an integer $\xi\in\mathbb{N}$ for an exact derivation. Applying the binomial expansion to the denominator yields:
\begin{align}
  P_C(\theta)&=\frac{1}{\sum_{m=0}^\xi\binom{\xi}{m}e^{-ma(\theta-b)}}\\&=\frac{1}{\sum_{m=0}^\xi e^{-ma\theta+\left(mab+\ln\binom{\xi}{m}\right)}},
\end{align}
which equivalently corresponds to \mirm{1,\xi}, with $\alpha_s=ma$ and $\delta_s=mab+\ln\binom{\xi}{m}$ for $s=m+1$. Since the numerator equals 1, the correct microstate corresponds to the case of $m=0$, with $\alpha_1=0$ and $\delta_1=0$. Since $ma>0$ for $m\geq1$, the correct microstate has the lowest $\alpha$, consistent with the intuition that it dominates at high ability.

Moreover, the expansion gives the LPE parameters a direct interpretation. The exponent $\xi$, which may represent item complexity or the number of problem-solving stages \citep{lee2018asymmetric}, equals the number of distractor microstates. Microstate $m$ represents failure at $m$ stages: $ma$ gives its cumulative dependence on ability, $\binom{\xi}{m}$ counts the equivalent failure patterns, and $mab$ locates their transition at difficulty $b$. This interpretation follows because finite sums of exponentials are closed under addition and multiplication. Products represent the AND of required successes, while sums represent the OR of alternative pathways. The LPE model is therefore a perfect-response giant MiRM (Theorem~\ref{thm:giant}): only the all-success pattern is correct, and every other pattern is a distractor response.

\subsection{Residual Heteroscedasticity Model}\label{sub:rh}
\citet{molenaar2015heteroscedastic} proposed the RH model, which has received considerable attention in recent studies. Its dichotomous form is
\begin{align}
  P_C(\theta)=\Phi\left[\frac{a_h(\theta-b_h)}{h(\theta)}\right],
\end{align}
where $h(\theta)=[2h_0/(1+e^{-h_1\theta})]^{1/2}$, and $\Phi$ denotes the standard normal distribution function. Here, $h_0$ determines the baseline residual variance and $h_1$ determines how the residual variance changes with ability. \citet{molenaar2015heteroscedastic} captures varying population-level heteroscedasticity over ability through the $h_1$ parameter, with $h_0=1$ fixed for identification.
 
The structural relation between the RH model and MiRM is restricted locally around $\theta_0=b_h$, where $P_C(\theta_0)=1/2$, rather than exact; however, the matching provides clear intuitions for reading each model in terms of the other. Expanding the RH logit $A_h(\theta)=\log[P_C/(1-P_C)]$ and the MiRM logit $A(\theta)=a(\theta)[\theta-b(\theta)]$ about $\theta=\theta_0$ through second order gives
\begin{align}
  A_h(\theta)&\approx\sqrt{\frac{8}{\pi}}\frac{a_h}{h(\theta_0)}(\theta-\theta_0)-\frac{1}{2}\sqrt{\frac{8}{\pi}}\frac{a_h}{h(\theta_0)}\frac{h_1}{1+e^{h_1\theta_0}}(\theta-\theta_0)^2,\\
  A(\theta)&\approx a(\theta_0)(\theta-\theta_0)+\frac{1}{2}a'(\theta_0)(\theta-\theta_0)^2,
\end{align}
since $A_h(\theta_0)=A(\theta_0)=0$ for $b(\theta_0)=b_h$. Matching the first- and second-order coefficients gives $a(\theta_0)=\sqrt{8/\pi}\,a_h/h(\theta_0)$ and $a'(\theta_0)/a(\theta_0)=-h_1/(1+e^{h_1\theta_0})$, which reduces to
\begin{align}
  \left.\partial_\theta\log a(\theta)\right|_{\theta=\theta_0}=\left.-\partial_\theta\log h^2(\theta)\right|_{\theta=\theta_0}.
\end{align}
This shows that the local discrimination $a(\theta)$ moves opposite to the residual variance $h^2(\theta)$, meaning that higher discrimination at $\theta$ corresponds to lower response uncertainty at that ability.

Because $a'(\theta)=\Psi_C''(\theta)-\Psi_D''(\theta)=\sigma_{\alpha,C}^2(\theta)-\sigma_{\alpha,D}^2(\theta)$, the change of local discrimination is simply the difference in slope variance between the correct and distractor macrostates. Consider \mirm{1,d}, where $\sigma_{\alpha,C}^2=0$ since a single correct state has zero variance; then $a'(\theta_0)=-\sigma_{\alpha,D}^2(\theta_0)$, implying $h_1>0$ when $a(\theta_0)>0$. More intuitively, at the crossing point $\theta=b_h$, the RH model with $h_1>0$ is matched through second order by a symmetric \mirm{1,2} form:
\begin{align}
  P_C(\theta)=\Phi\left[\frac{a_h(\theta-b_h)}{h(\theta)}\right]\simeq\frac{e^{a_0(\theta-b_h)}}{e^{a_0(\theta-b_h)}+\frac{1}{2}e^{\sigma_D(\theta-b_h)}+\frac{1}{2}e^{-\sigma_D(\theta-b_h)}},
\end{align}
where $a_0=a(b_h)$ and $\sigma_D^2=\sigma_{\alpha,D}^2(b_h)=a_0h_1/(1+e^{h_1b_h})$. This reads the single distractor term of the 2PL as splitting into two distractor microstates with slopes $\pm\sigma_D$ around $\theta = b_h$, where each weight $1/2$ can be absorbed into the offset as $\delta=-\log2$. In the RH model, $h_1>0$ means that residual variance $h^2(\theta)$, and thus response uncertainty, increases at higher ability; in \mirm{1,d}, this occurs because having only one way to succeed alongside multiple ways to fail ($\sigma_{\alpha,D}^2>0$) leaves high-ability respondents with several distinct failure modes, slowing the rise of the ICC ($a'<0$). The RH model with $h_1>0$ thereby bridges to the LPE model through \mirm{1,\xi}. Hence, MiRM links item complexity to ICC asymmetry: asymmetry may occur when different problem-solving pathways interact either in addition or multiplication, and the multiple pathways may be driven by the population variance.

Symmetrically, the RH model with $h_1<0$ splits the numerator instead of the denominator and corresponds locally to \mirm{c,1}, or more generally $(c,d)$, which are the subjects of Corollary~\ref{cor:dip}. \citet{feuerstahler2024monotonicity} reported that most of the classical parametric models enforce monotonicity, while the RH model does not. In MiRM terms, this is the point at which the macrostate mean-slope contrast $a(\theta)$ changes sign. The RH model's possible nonmonotonicity and its asymmetric information profile therefore arise from the same underlying mechanism: an ability-dependent imbalance in macrostate slope heterogeneity.

The same direction of asymmetry holds across the ability scale. For \mirm{1,d}, concavity of the logit and $A(\theta_0)=0$ imply $A(\theta_0+\Delta\theta)<-A(\theta_0-\Delta\theta)$ for $\Delta\theta>0$. For RH with $h_1>0$, increasing $h(\theta)$ gives the same ordering of the probit arguments. By the symmetry of the logistic and normal links, both models therefore satisfy
\begin{align}
  P_C(\theta_0+\Delta\theta)+P_C(\theta_0-\Delta\theta)<1,\qquad\Delta\theta>0.
\end{align}
Thus, the average probability at two abilities equally far from the crossing point lies below $1/2$. The inequality reverses for \mirm{c,1} and RH with $h_1<0$.

\subsection{The Universal Approximation Theorem}
Theoretically, MiRM with sufficiently many finite microstates can approximate any continuous item response model, including nonparametric models.

\begin{theorem}[The Universal Approximation Theorem of MiRM]\label{thm:uat}
  For every continuous function $f:[\theta_\mathbf{D},\theta_\mathbf{C}]\to(0,1)$ on the compact interval $[\theta_\mathbf{D},\theta_\mathbf{C}]$ and every $\epsilon_0>0$, there exists a finite-state binary MiRM whose ICC is uniformly within $\epsilon_0$ of $f$.
\end{theorem}

The theorem has three consequences. First, the linear exponent of each microstate does not restrict the shape of an ICC; the flexibility comes from the number of microstates. Second, under local independence, a unidimensional test model is determined by its ICCs, so MiRM can approximate not only a single item but the whole test model. Third, a many-state MiRM fitted to a nonparametric model can serve as an atomization filter: the local characteristics $a(\theta)$ and $b(\theta)$ and the generalized credit can be computed for it, which gives the nonparametric model the same algebraic reading. The theorem only shows existence, and does not tell how many microstates are needed; we examine this empirically in Section~\ref{sub:pfm}.

\section{Estimation}\label{sec:est}
For response data given as $\mathcal{D}=\{(i,j,k)\}$, the true microstate indices $m$ are generally not observed, as the true abilities $\theta_i$ are not. The distribution $\theta\sim\mathcal{N}(0,1)$ for any data $\mathcal{D}$ is the working prior adopted in this paper, but not a structural assumption. It makes the model's results comparable to traditional models; any other unidimensional distribution can be used with the model.

All models under the data $\mathcal{D}$ are estimated by marginal maximum likelihood \citep{bock1981marginal}, the standard method for item parameters in IRT. Since ability is unidimensional and follows a normal prior, it is integrated out by Gauss--Hermite quadrature with $Q=41$ nodes. The marginal likelihood is maximized directly by a quasi-Newton method \citep[{L-BFGS with a strong Wolfe line search;}][]{liu1989limited} rather than by the EM algorithm, since the PRM constraint can simply be added to its loss (Section~\ref{sub:prml}). Every model is fitted with the same implementation, quadrature, and optimizer, so that differences in fit come from the models and not from the estimation. MiRM-specific estimation procedures are described below, with full details in Appendix~\ref{appendix:estimation}.

\subsection{A Normalized PRM Loss}\label{sub:prml}
By default, we impose PRM during estimation as a constraint on the $C$- and $D$-side margins over a grid $\Theta_{\rm P}$ ($Q_{\rm P}=2Q-1$). Because $a_{\mathcal{J}C}$ and $a_{\mathcal{J}D}$ are proportional to the scale of the slopes and the number of items, the sum of itemwise contributions is normalized by their absolute sum. When the denominator is zero at a certain $\theta$, the ability point is excluded from the loss estimation to prevent $0/0$.
\begin{align}
  \tilde{a}_{\mathcal{J}C}(\theta)=\frac{\sum_jP_{jD}a_j}{\sum_j|P_{jD}a_j|},\qquad\tilde{a}_{\mathcal{J}D}(\theta)=\frac{\sum_jP_{jC}a_j}{\sum_j|P_{jC}a_j|}.
\end{align}
For the constraints $g_q^C=-\tilde{a}_{\mathcal{J}C}(\theta_q)\leq0$ and $g_q^D=-\tilde{a}_{\mathcal{J}D}(\theta_q)\leq0$, we use the Powell--Hestenes--Rockafellar augmented Lagrangian \citep{powell1969method,hestenes1969multiplier,rockafellar1974augmented} penalty, which is added to the negative log-likelihood to form the total loss. For more details, see Appendix~\ref{app:sub:prml}.

\subsection{Configuration Selection}\label{sub:cfg} 
The number of microstates can differ across items. The model having a different microstate configuration $(c,d)$ per item is called the \emph{per-item MiRM}, or \mirm{c_j,d_j}. To search for the best configuration for each item, four adjacent models \mirm{c_j,d_j}, \mirm{c_j,d_j+1}, \mirm{c_j+1,d_j}, and \mirm{c_j+1,d_j+1} are trained simultaneously for every item $j\in\{1,\ldots,J\}$, with all four models sharing $c_j$ correct microstates and $d_j$ distractor microstates. Without sharing these base microstates, fitting the four models separately always favors the largest configuration, whereas sharing them forces the base states to remain meaningful across all four models and prevents unconstrained growth of the state counts.

After the training is done, $(c_j,d_j)$ is shifted to a preferred direction under a shared ability posterior $\hat{\theta}$, and then the training resumes. The preferred direction is determined by four conditions: for example, to shift toward a larger $d_j$ for each item $j$, $(c_j,d_j+1)$ must improve on $(c_j,d_j)$, $(c_j+1,d_j+1)$ must improve on $(c_j+1,d_j)$, and both improvements must still hold after subtracting their mean across all items. The loss of each model is calculated under the geometric mean of the four models' ability posteriors. The cycle terminates when no $(c_j,d_j)$ is shifted after training. To select the smallest number of microstates, we start from $(c_j,d_j)=(1,1)$ for every item at the first cycle. A full illustration of the estimation procedure is provided in Appendix~\ref{app:sub:peritem}.

However, since estimating per-item MiRM is computationally heavy, we also use \mirm{c,d} with constant $c$, $d$. In empirical use, \mirm{2,2} is usually sufficient for both obtaining enough performance (see Figure~\ref{fig:grid}) and analyzing macrostate-level characteristics. We additionally use \mirm{5,5} as an analog to flexible nonparametric models.

\section{Simulation}\label{sec:sim}
Some important properties of MiRM cannot be checked on empirical data, because the true ICCs, abilities, microstate configurations, and local dips are not known. In this section, we therefore generate data with known truth and evaluate the accuracy of ICC and ability estimation when the generating model is not known (Section~\ref{sub:prec}), the recovery of itemwise microstate configurations and parameters (Section~\ref{sub:param}), and the detection of local dips with and without PRM (Section~\ref{sub:nonmono_sim}).

\subsection{Design}\label{sub:sim_design}
The main experiments used 4,096 students with abilities drawn from a standard normal distribution and 128 items over 30 replications, with item parameters taken from models fitted to the PISA data (Section~\ref{sec:emp}). For Section~\ref{sub:prec}, responses were generated from the 2PL, 3PL, LPE, I-spline \citep{ramsay1989binomial}, and MiRM models. The MiRM generator used monotone items of the per-item MiRM, so that ($c$,$d$) configurations differ across items. We fitted the 2PL, 3PL, LPE, and \mirm{2,2} to each dataset. Under the MiRM generator, we also varied the number of students from 128 to 2,048 and the number of items from 8 to 64. For Section~\ref{sub:param}, each bank had all monotone (1,1) and (2,1) items of the per-item MiRM fits (13 and 44 items) and monotone (1,2) and (2,2) items in proportion to their numbers (32 and 39 items), which were resampled in each replication; the per-item MiRM was fitted in addition. The monotonicity study differs in its items, respondents, and the number of replications; its design is given in Section~\ref{sub:nonmono_sim}.

\subsection{Measurement Precision}\label{sub:prec}

\begingroup
\renewcommand\tablebodyfont{\fontsize{8}{9}\selectfont}
\renewcommand\tablecolheadfont{\fontsize{8}{9}\selectfont}
\renewcommand\arraystretch{1.5}%
\begin{table}[t]
  \tabcolsep=3pt%
  \TBL{\caption{RMSE between estimated and true ICCs and abilities under each generator\label{tab:sim_precision}}}
  {\begin{fntable}
  \begin{tabular*}{\textwidth}{@{}p{5em}*{10}{>{\centering\arraybackslash}p{\dimexpr(\textwidth-5em-20\tabcolsep)/10\relax}}@{}}\toprule%
  \multirow[c]{3}{*}[-0.8ex]{\TCH{Model}} & \multicolumn{10}{c}{\TCH{Generator}} \\[-0.4ex]
  \cmidrule{2-11}\noalign{\vskip -0.4ex}
  & \multicolumn{2}{c}{\TCH{2PL}} & \multicolumn{2}{c}{\TCH{3PL}} & \multicolumn{2}{c}{\TCH{LPE}}
  & \multicolumn{2}{c}{\TCH{MiRM}} & \multicolumn{2}{c}{\TCH{Spline}} \\[-0.4ex]
  \cmidrule(l{3pt}r{3pt}){2-3}\cmidrule(l{3pt}r{3pt}){4-5}\cmidrule(l{3pt}r{3pt}){6-7}\cmidrule(l{3pt}r{3pt}){8-9}\cmidrule(l{3pt}r{3pt}){10-11}\noalign{\vskip -0.4ex}
  & \TCH{ICC} & \TCH{$\theta$} & \TCH{ICC} & \TCH{$\theta$} & \TCH{ICC} & \TCH{$\theta$}
  & \TCH{ICC} & \TCH{$\theta$} & \TCH{ICC} & \TCH{$\theta$} \\\midrule
  \TCH{2PL} & \cellcolor{gray!20}\textbf{0.012}\textsubscript{$\pm$1} & \cellcolor{gray!20}\textbf{0.192}\textsubscript{$\pm$3}
            & 0.067\textsubscript{$\pm$3} & 0.229\textsubscript{$\pm$6} 
            & 0.041\textsubscript{$\pm$2} & 0.197\textsubscript{$\pm$4}
            & 0.066\textsubscript{$\pm$3} & 0.208\textsubscript{$\pm$5} 
            & 0.078\textsubscript{$\pm$2} & 0.215\textsubscript{$\pm$4} \\
  \TCH{3PL} & 0.022\textsubscript{$\pm$3} & 0.195\textsubscript{$\pm$4} 
            & \cellcolor{gray!20}\textbf{0.021}\textsubscript{$\pm$2} & \cellcolor{gray!20}\textbf{0.200}\textsubscript{$\pm$5} 
            & 0.045\textsubscript{$\pm$3} & 0.202\textsubscript{$\pm$6}
            & 0.056\textsubscript{$\pm$3} & 0.234\textsubscript{$\pm$6}
            & 0.078\textsubscript{$\pm$3} & 0.234\textsubscript{$\pm$6} \\
  \TCH{LPE} & \underline{0.020}\textsubscript{$\pm$1} & \underline{0.193}\textsubscript{$\pm$3}
            & 0.049\textsubscript{$\pm$3} & 0.211\textsubscript{$\pm$4}
            & \cellcolor{gray!20}\textbf{0.020}\textsubscript{$\pm$2} & \cellcolor{gray!20}\textbf{0.189}\textsubscript{$\pm$4}
            & \underline{0.049}\textsubscript{$\pm$2} & \underline{0.202}\textsubscript{$\pm$5}
            & \underline{0.067}\textsubscript{$\pm$3} & \underline{0.206}\textsubscript{$\pm$3} \\
  \cmidrule{2-11}\noalign{\vskip -0.4ex}
  \TCH{\mirm{2,2}}  
            & 0.034\textsubscript{$\pm$4} & 0.198\textsubscript{$\pm$4}
            & \underline{0.043}\textsubscript{$\pm$4} & \underline{0.207}\textsubscript{$\pm$5}
            & \underline{0.035}\textsubscript{$\pm$4} & \underline{0.195}\textsubscript{$\pm$4}
            & \cellcolor{gray!20}\textbf{0.037}\textsubscript{$\pm$4} & \cellcolor{gray!20}\textbf{0.196}\textsubscript{$\pm$5}
            & \textbf{0.060}\textsubscript{$\pm$4} & \textbf{0.201}\textsubscript{$\pm$6} \\
  \botrule
  \end{tabular*}%
  \footnotetext[]{{Note:} ICC: RMSE between the estimated and true ICCs over $\theta \in [-4, 4]$; $\theta$: RMSE between the
  ability estimates and the true abilities. Means over 30 replicates with 4,096 students and 128 items. Gray cells denote
  the generator's own model family (the MiRM generator has diverse configurations; the fitted MiRM is (2,2)). Bold and
  underline denote the best and second-best values in each column. Standard deviations are in units of $10^{-3}$.}
  \end{fntable}}
\end{table}
\endgroup

Table~\ref{tab:sim_precision} compares the root mean squared errors (RMSEs) of the estimated ICCs and abilities across fitted models within each generator. As expected, for data generated by the 2PL, 3PL, or LPE, fitting the corresponding model gave the lowest error on both measures. However, the generator is not known in practice, so comparing the models other than the generator's own model is also important. In most of these comparisons, \mirm{2,2} recovered the ICCs and abilities more accurately than the other models. Under the 3PL generator, the ICC and ability RMSEs of \mirm{2,2} were 0.043 and 0.207, whereas the RMSEs of the 3PL were 0.021 and 0.200. Under the LPE generator, the corresponding values were 0.035 and 0.195 for \mirm{2,2}, and 0.020 and 0.189 for the LPE. Even when the data were generated by MiRM with different itemwise configurations, a common \mirm{2,2} had the lowest RMSEs, 0.037 and 0.196. It also performed best under the spline generator, with RMSEs of 0.060 and 0.201. Thus, when the generating model is unknown, a small fixed MiRM covers a range of ICC shapes.

\begin{figure}[t]%
  \FIG{\includegraphics[width=\textwidth]{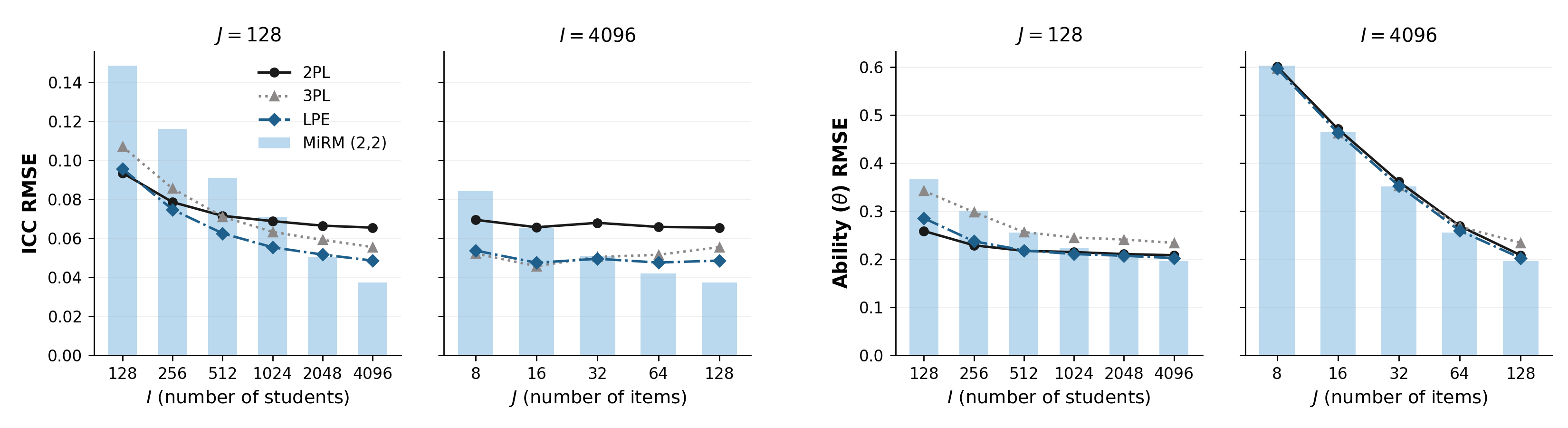}}
  {\caption{ICC RMSE (left two panels) and ability RMSE (right two panels) under MiRM-generated data across 30 replications as the number of students $I$ or items $J$ varies. In each panel, the other sample size is held fixed as indicated above the plot. Bars show \mirm{2,2}; lines with markers show 2PL, 3PL, and LPE. Standard error bars are omitted as they are negligible across replications}\label{fig:sens}} 
\end{figure}

The 2PL generator was an exception: the 2PL had the lowest ICC and ability RMSEs, 0.012 and 0.192, whereas \mirm{2,2} had the highest errors among the fitted models. Estimating additional microstates comes at a cost in finite samples when the true ICC is simple. Figure~\ref{fig:sens} makes this cost, and the eventual gain, clearer as the amount of data varies. With fewer students or items, MiRM had higher errors than the competing models. Its errors then fell more rapidly as the data increased, becoming the lowest with larger samples and longer tests. MiRM thus benefits most from large data; in the next section, we show that the per-item MiRM can mitigate this cost by adjusting its number of microstates.

\subsection{Parameter Recovery}\label{sub:param}
Restricting or increasing the model complexity properly is important for accurate estimation, as shown in Table~\ref{tab:sim_precision}, and item parameters can be compared only when the number of microstates matches. Therefore, we examined whether the $(c,d)$ configuration selection algorithm in Section~\ref{sub:cfg} and Appendix~\ref{app:sub:peritem} recovers true item characteristics and parameters. We measured both ICC and parameter recovery for the per-item MiRM, since microstate parameters intrinsically have identification limits (Appendix~\ref{app:sub:micro}).

\begin{figure}[t]%
  \FIG{\includegraphics[width=\textwidth]{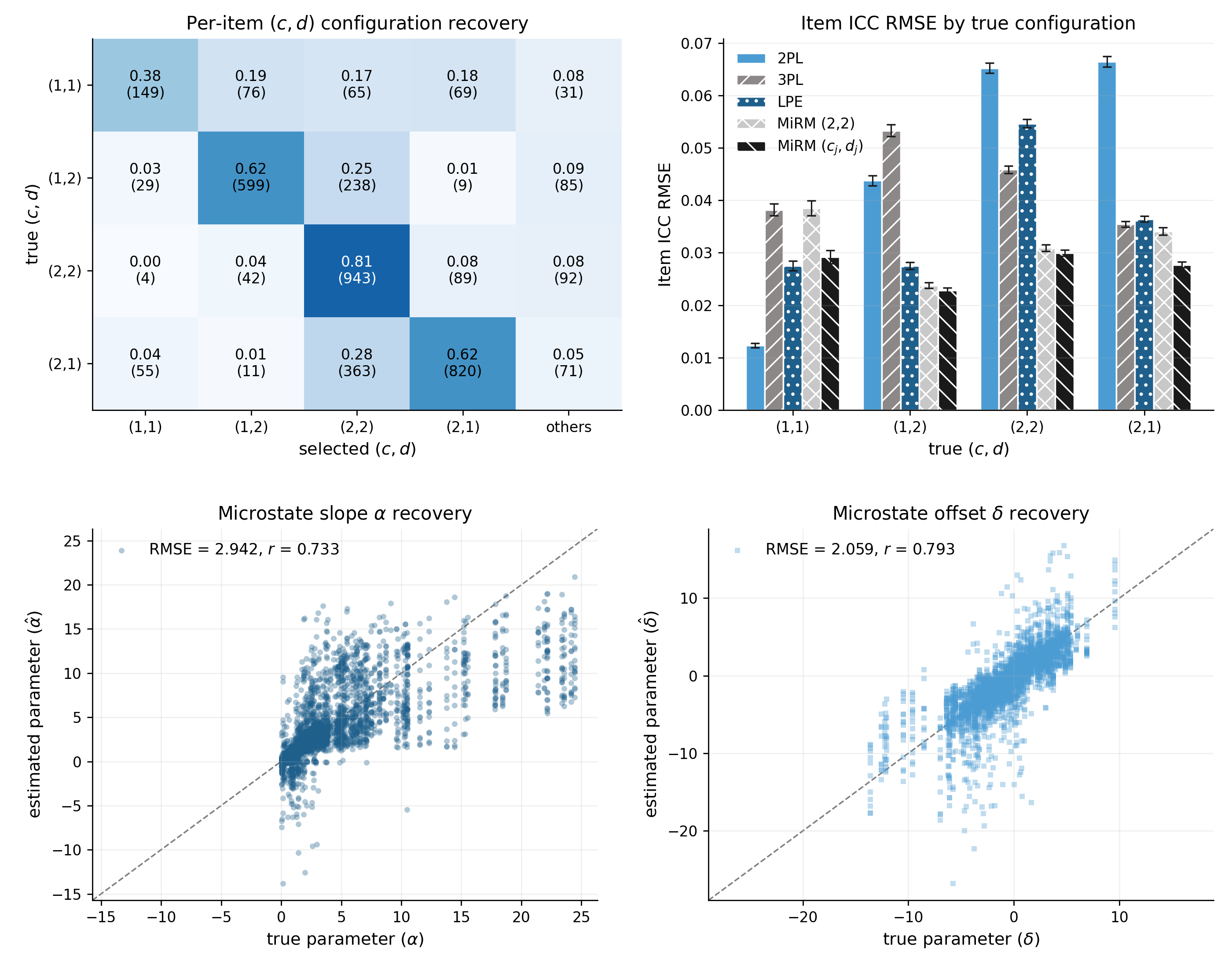}}
  {\caption{Recovery of item characteristics and microstate parameters under mixed MiRM configurations across 30 replications ($I=4{,}096, J=128$). Upper left: selected ($c$,$d$) by true configurations; each cell shows the row-wise proportion and the item count in parentheses. Upper right: item ICC RMSE grouped by true configuration; error bars show $\pm1$ standard error. Lower: estimated versus true microstate slopes $\alpha$ and offsets $\delta$, respectively, for items whose ($c$,$d$) configuration was correctly recovered; dashed lines indicate $y=x$}\label{fig:recovery}}
\end{figure}

The upper-left panel of Figure~\ref{fig:recovery} shows that the correct configuration was selected for 38\%, 62\%, 81\%, and 62\% of items generated from (1,1), (1,2), (2,2), and (2,1), respectively. Items generated from (1,1) were often assigned larger configurations, while (2,2) was selected for 28\% of the (2,1) items. The exact state counts were not always recovered, but selection errors tended to involve nearby configurations, and recovery was strongest for (2,2). This can be explained by Appendix~\ref{app:sub:micro}: a larger configuration always subsumes a smaller one. Thus, keeping items with a small true configuration at that configuration is a challenge for any selection algorithm. Among the four tested configurations, a configuration that does not subsume the true one was rarely selected: 4\%, 12\%, and 5\% for the (1,2), (2,2), and (2,1) items, respectively. On the other hand, it is also important to limit overly large configurations without an arbitrary penalty. Only 5\% to 9\% of the items were assigned configurations beyond the (2,2) grid. This shows that the algorithm does not keep increasing the configuration to fit the noise in the data.

This can also be confirmed in the ICC recovery. In the upper-right panel, the per-item MiRM had ICC RMSEs similar to or lower than those of the fixed \mirm{2,2} across the four true configurations. Unlike the fixed-configuration result in Table~\ref{tab:sim_precision}, selecting configurations reduced the error for (1,1) items, which have the same configuration as the 2PL. Configuration selection therefore does more than report state counts: by adjusting complexity to the item, it also improves ICC estimation.

The lower panels compare microstate parameters for items whose configurations were correctly selected. Permuting microstates within the same macrostate leaves the ICC unchanged, so we matched true and estimated states using the Hungarian algorithm \citep{kuhn1955hungarian}. The RMSE and correlation were 2.942 and 0.733 for the slopes $\alpha$, and 2.059 and 0.793 for the offsets $\delta$. This is one of the expected consequences of the non-identifiability of microstates (see Appendix~\ref{app:sub:micro}).

\subsection{Testing Monotonicity}\label{sub:nonmono_sim}
For the monotonicity study, we generated data for 4,096 students and 128 items using item parameters from per-item MiRM fits to the three PISA domains. Thirteen items, approximately 10\% of the bank, had a true local dip; the remaining items had nearly monotone ICCs. Dip items were drawn from fitted items whose ICCs fell by 0.05--0.30 over the ability range $[-4,4]$, with the trough in $[-2.5,2.5]$. Each bank contained dip items but satisfied PRM at the test level. 

To vary the difficulty of detecting these dips, we introduced several factors that may induce manifest nonmonotonicity. First, we generated responses from three latent population groups $g\in\{1,2,3\}$ with proportions of 60\%, 30\%, and 10\%. The groups differed in ability and in two additional traits:
\begin{align}
  \mu_g\sim\mathcal{N}(0,\sigma_\theta^2),\quad\theta_i=\mu_{g_i}+\mathcal{N}(0,1),\qquad\nu_{gm}\sim\mathcal{N}(0,\sigma_U^2),\quad u_{im}=\nu_{g_im}+\mathcal{N}(0,1),
\end{align}
for $m\in\{1,2\}$, and the ability $\theta_i$ was then standardized. With probability 0.25, item $j$ was affected by one of the two additional traits, $m(j)\in\{1,2\}$, chosen with equal probability, with a loading $\lambda_j\sim\mathcal{N}(0,\sigma_\lambda^2)$; the other items had $\lambda_j=0$. The response to item $j$ was generated from
\begin{align}
  \theta_{ij}=\theta_i+\lambda_ju_{im(j)}.
\end{align}
Second, we varied four conditions: the proportion of observed responses (20\%, 50\%, or 100\%), the separation between groups in ability ($\sigma_\theta\in\{0,0.75,1.5\}$), the separation between groups in the additional traits ($\sigma_U\in\{0,1,2\}$), and the strength of the additional traits' effects on items ($\sigma_\lambda\in\{0,0.5,1\}$). We repeated each combination 10 times, yielding $3^4\times10=810$ datasets, and fitted the 2PL, and \mirm{2,2} with and without PRM. An item was classified as having a dip if its fitted ICC fell at least 0.05 below an earlier maximum over $[-4,4]$. We then compared these classifications with the items' true dip status, and the ability estimates with the true abilities $\theta_i$ using Kendall's correlation $\tau$.

\begin{figure}[t]%
  \FIG{\includegraphics[width=\textwidth]{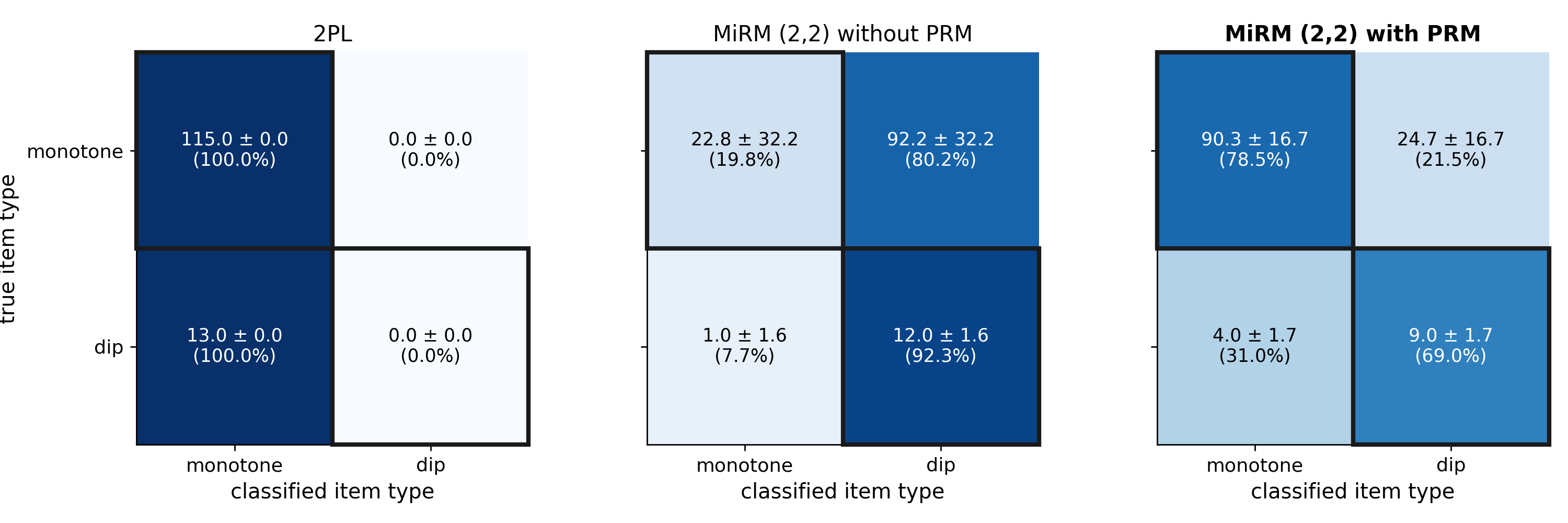}} 
  {\caption{Confusion matrices for detecting latent nonmonotonicity with 2PL (left), \mirm{2,2} fitted without PRM (center), and with PRM (right). Each cell shows the mean $\pm$ SD number of items per replicate (out of 115 true monotone and 13 true dip items) and the row-wise percentage in parentheses. Black borders highlight correct classifications on the diagonal cells}\label{fig:confusion}}
\end{figure}

Figure~\ref{fig:confusion} shows how the 2PL and \mirm{2,2} classified items with a latent dip and monotone items. Because the 2PL enforces monotonicity by construction, it classified all items as monotone (left), missing all 13 true dips. In contrast, \mirm{2,2} without PRM classified most items as having a dip (center). \mirm{2,2} with PRM produced a much clearer diagonal pattern (right).

\mirm{2,2} with PRM classified 21.5\% of the monotone items as having a dip, whereas the 2PL classified all monotone items correctly. It also found 69.0\% of the true dips, whereas \mirm{2,2} without PRM found 92.3\%. However, the 2PL classified all monotone items correctly because it classifies every item as monotone, and \mirm{2,2} without PRM found most true dips because it classified most items as having a dip. This can be measured by Youden's index \citep{youden1950index}, which is 0 for a classifier that does not distinguish the two classes and 1 for a perfect classifier. The index was 0 for the 2PL and 0.12 for \mirm{2,2} without PRM, whereas it was 0.47 with PRM.

Unlike many traditional models, MiRM can detect true dips if they exist. On the other hand, PRM makes the ability estimated by such a model preserve its order, so that the model reflects the property of tests. Hence, what we propose is not simply a dip classifier, and PRM alone does not fully establish that an individual item is nonmonotone. Detecting the dip is the natural consequence of reconstructing the true ICC and ability, but not the goal itself. From this point of view, the diagonal pattern in Figure~\ref{fig:confusion} can be read as a first step toward the true values, between traditional models that achieve a sound measure by a monotonicity assumption and flexible models that use ability only as a predictor of responses.

Kendall's $\tau$ shows this numerically. Without PRM, the ability scale was reversed ($\tau<0$) in 513 of the 810 datasets. With PRM, it was never reversed, and the mean $\tau$ was 0.817, close to that of the 2PL (0.823). Moreover, \mirm{2,2} with PRM recovered the order of abilities more accurately than the 2PL when at least 50\% of the responses were observed and the additional traits had no effect on items ($\sigma_\lambda=0$). In all 180 datasets under these conditions, its $\tau$ was higher than that of the 2PL; the mean $\tau$ was 0.894 for \mirm{2,2} and 0.887 for the 2PL with 100\% of the responses. This advantage disappeared when false dips were frequent: with 20\% of the responses, 42.8 of the 115 monotone items were classified as having a dip, and \mirm{2,2} had a higher $\tau$ than the 2PL in only 9 of the 270 datasets. Thus, when true dips exist, capturing them accurately may help to recover the order of abilities, whereas this gain is lost in conditions where false dips are frequent.

\section{Empirical Study}\label{sec:emp}
We examined item characteristics, model performance, nonmonotonicity, and ability paths in PISA 2015 Reading, Science, and Mathematics \citep{OECD2016PISAVolI}. PISA is a large public dataset: our data contain 2.5 to 9.1 million responses per domain. The three domains provide three separate analyses, and since many students took multiple domains, they can also be fitted as a single bank (Section~\ref{sub:ndim}). We chose the 2015 cycle because it was mainly computer-based and was the last cycle without adaptive testing. We used valid responses to every computer-scored item from the computer-based assessment, followed the official scoring rules, and dichotomized full-credit versus other valid responses. Missing responses (not reached, omitted, or invalid) were excluded. Details are given in Appendix~\ref{appendix:datasets}.

Alongside the 2PL and 3PL, the comparisons included RH and 3PLQ, which had performed well in an earlier PISA 2009 analysis \citep{robitzsch2022choice}. For the prediction comparison, we chose the number of spline basis functions of the monotone spline model \citep{ramsay1989binomial,ramsay1988monotone} so that the spline had one more free parameter per item than \mirm{5,5}, avoiding a disadvantage in parameter count for the spline. The LPE, which was included in the simulations, was omitted here because it showed no distinct performance advantage over the 3PL and RH.

\subsection{Item Characteristics}

\begin{figure}[t]%
  \FIG{\includegraphics[width=\textwidth]{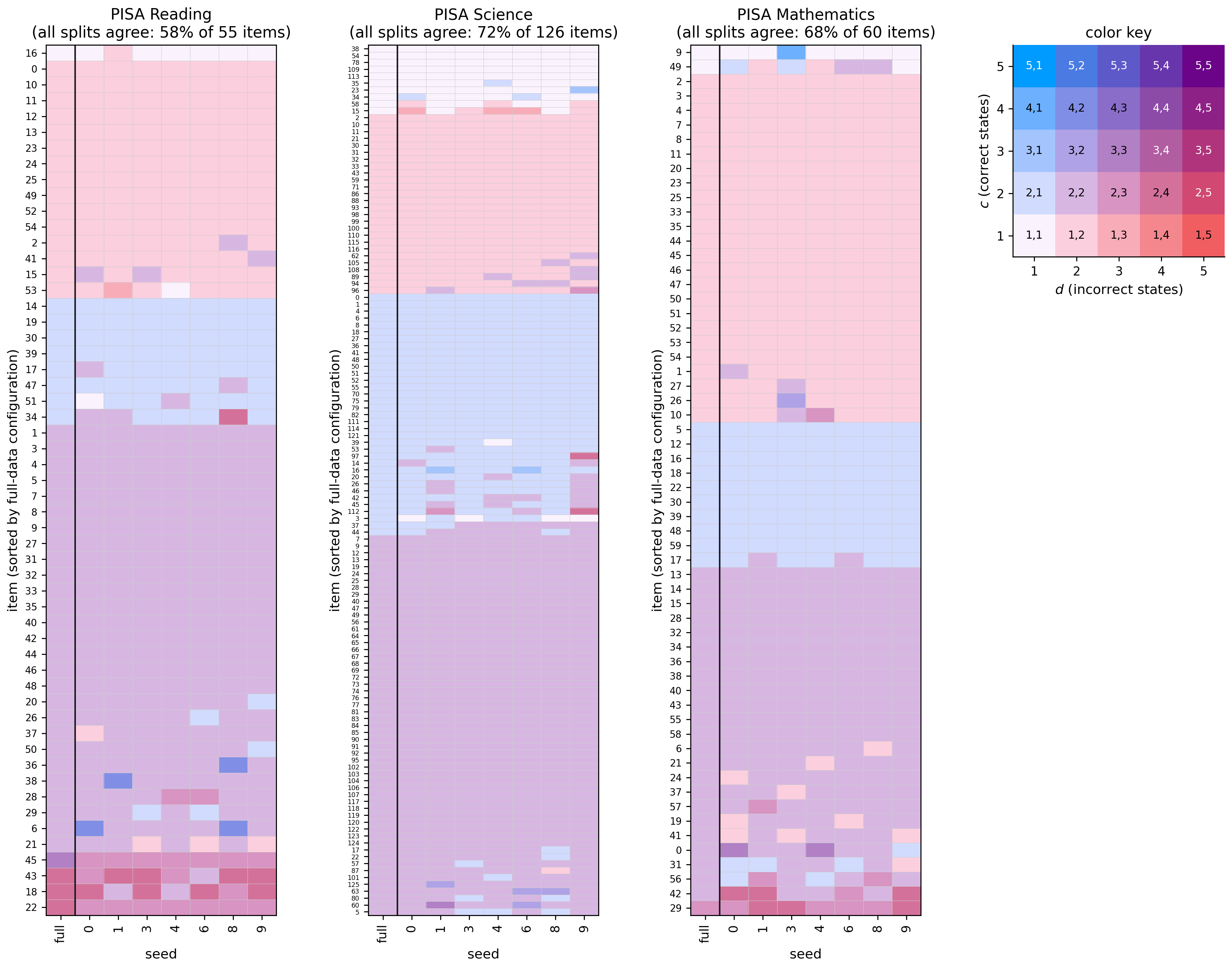}}
  {\caption{Microstate configurations selected for each item in the full data (first column) and in each random split (remaining columns; seeds from the digits of the Boltzmann constant). Rows are items sorted by their full-data configuration; colors follow the key on the right, in which similar tones denote similar configurations (bluish tones: more correct microstates $c$; reddish tones: more distractor microstates $d$; lighter tones: fewer microstates). Panel titles give the share of items whose configuration agrees across all splits}\label{fig:cd}}
\end{figure}

Figure~\ref{fig:cd} compares itemwise microstate configurations selected from the full data with those selected in seven random splits. Each split used 80\% of the students randomly sampled from the full data; the same splits are used for the held-out prediction in Section~\ref{sub:pfm}. The configuration agreed across the full data and every split for 58\% of the 55 Reading items, 72\% of the 126 Science items, and 68\% of the 60 Mathematics items. Even when selections did not agree exactly, they generally remained within similar groups of configurations. Thus, much of the variation in the selected configurations across items persisted across data splits.

\begin{figure}[t]%
  \FIG{\includegraphics[width=\textwidth]{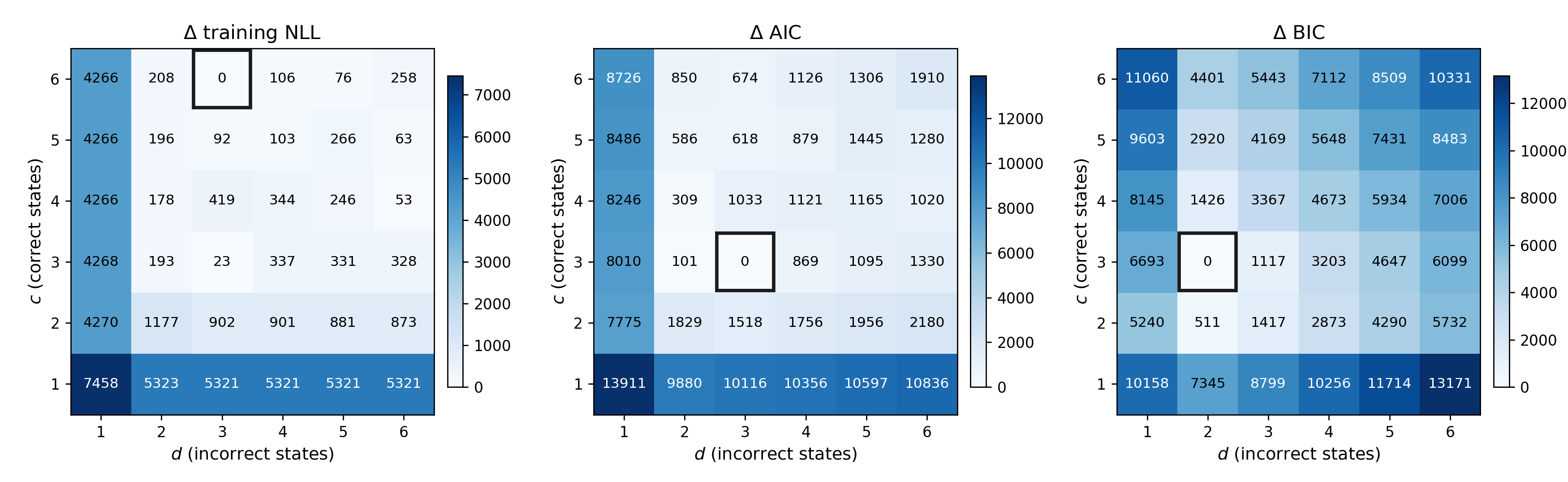}} 
  {\caption{Differences in training negative log-likelihood (NLL), AIC, and BIC from the best configuration over the global microstate configuration (c,d) grid for PISA 2015 Mathematics. Light colors denote better performance. The boxed cell marks the minimum of each panel}\label{fig:grid}}
\end{figure}

Figure~\ref{fig:grid} shows the fit of a common configuration for all items in Mathematics, taken as a representative domain, to examine how global state counts trade fit against complexity. The training negative log-likelihood was lowest at (6,3) rather than (6,6) due to early stopping and optimization noise in larger models (see Appendix~\ref{appendix:estimation}), while AIC \citep{akaike1974new} was lowest at (3,3) and BIC \citep{schwarz1978estimating} at (3,2). Across all three criteria, the largest improvement occurred when moving from $c=1$ or $d=1$ to (2,2), whereas adding more states to every item quickly increased the BIC penalty. This shows both why \mirm{2,2} serves as a practical common benchmark and why assigning configurations per item is better than using a large fixed $(c,d)$ for all items. Still, some items in Figure~\ref{fig:cd} consistently selected configurations larger than (2,2), which were not examined in Section~\ref{sub:param}.

\begin{figure}[t]%
  \FIG{\includegraphics[width=\textwidth]{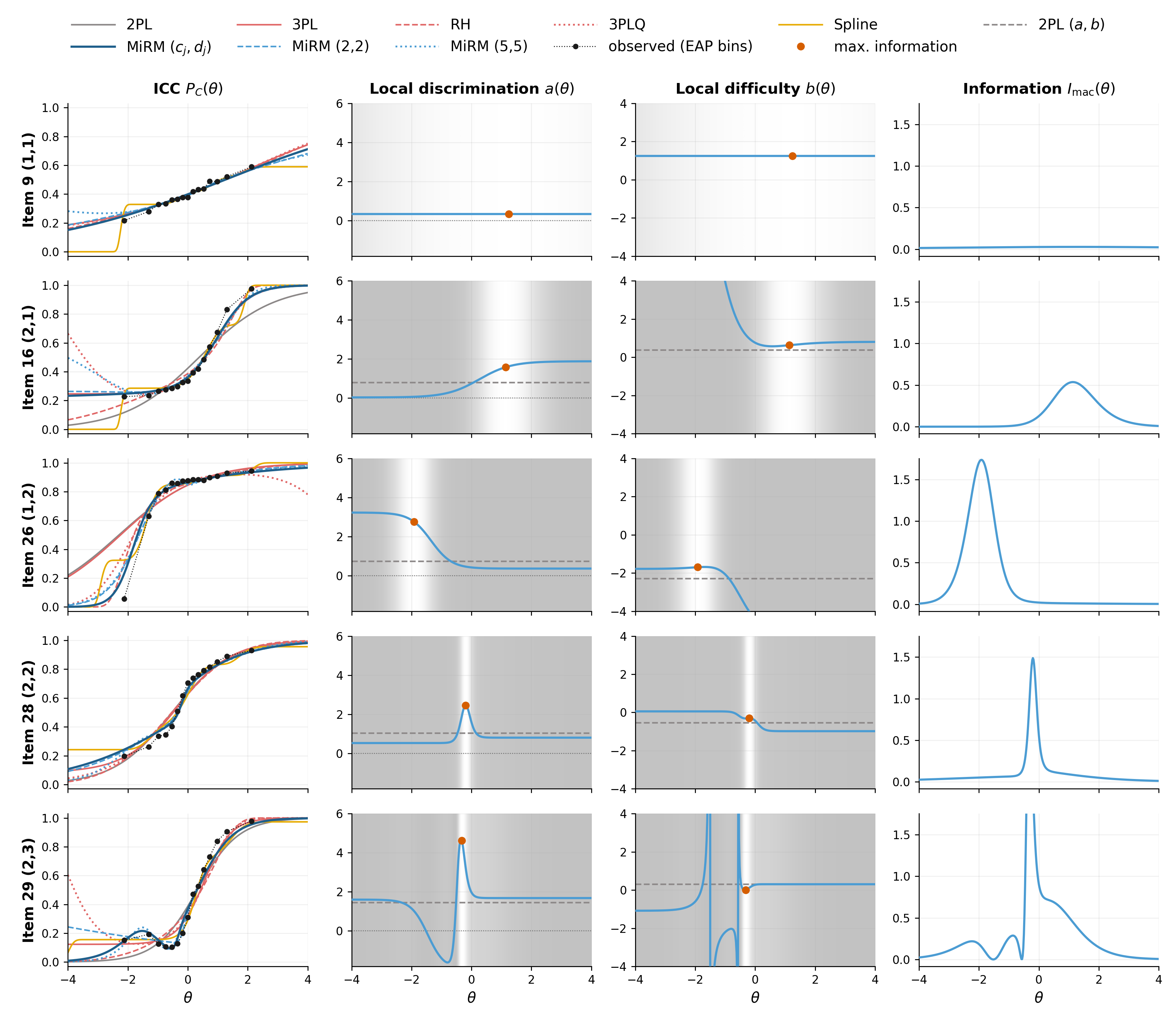}} 
  {\caption{Item characteristics of five PISA 2015 Mathematics items: ICC, discrimination, difficulty, and information (left to right). Parentheses in the row labels give the (c,d) selected by per-item MiRM. ICCs are shown for MiRM (blue; per-item: solid, (2,2): dashed, (5,5): dotted), 2PL (gray), other parametric models (red; 3PL: solid, RH: dashed, 3PLQ: dotted), and a monotone spline (yellow). Black points show the observed proportion correct in bins of per-item MiRM EAP ability. Darker gray backgrounds denote lower information, and orange points mark the maximum of information. Gray dashed lines show the 2PL discrimination and difficulty}\label{fig:character}} 
\end{figure}

Figure~\ref{fig:character} shows the item characteristic curves (ICCs), local discrimination $a(\theta)$, local difficulty $b(\theta)$, and item information $I(\theta)$ across ability $\theta$ for five PISA 2015 Mathematics items. The items were selected to illustrate the characteristics of different microstate configurations. The black points and connecting lines show observed proportions correct in ability bins based on posterior estimates from the per-item MiRM; they are a visual reference, not evidence of a better fit.

The ICCs also illustrate differences among the fitted models. For the 3PLQ (red dotted line), the probability correct rises sharply in the left tail of Item 16 and falls in the right tail of Item 26, as expected from its quadratic logit. The fitted spline (yellow) produces step-like ICCs that differ in shape from those of the other models, regardless of its fit. MiRM produces different ICC shapes according to the selected microstate configuration. Item 16 was assigned (2,1), a configuration that contains the guessing model 3PL as a submodel. Its MiRM ICC has a nonzero lower asymptote, and among the traditional models, the fitted 3PL follows this curve most closely. Item 26 was assigned (1,2) and has an asymmetric MiRM ICC. Here, the closest curve is given by RH (red dashed line), which can represent asymmetry; the 2PL and 3PL differ from it at low ability. Configurations with (2,2) or more states can represent more complex item characteristics. They capture not only the local dip in Item 29 but also the change in curvature in Item 28. \mirm{2,2} and \mirm{5,5} generally follow the per-item MiRM ICCs for items assigned smaller configurations.

Unlike the constant discrimination and difficulty of the 2PL, local discrimination $a(\theta)$ and difficulty $b(\theta)$ vary with ability in MiRM, with patterns that differ across configurations. Local difficulty can extend far beyond the usual range because its definition involves division by $a(\theta)$. It is therefore useful to focus on regions where item information $I(\theta)$ is sufficiently large; there, $b(\theta)$ is generally close to the 2PL difficulty in these examples. The values of $a(\theta)$ and $b(\theta)$ at the point of maximum information can also serve as summaries of an item. Unlike information, local discrimination retains its sign. Item 29 illustrates why this matters: $a(\theta)$ is negative over part of the ability range, marking the local decline in its ICC.

\subsection{Performance}\label{sub:pfm}

\begingroup
\renewcommand\tablebodyfont{\fontsize{7.5}{9}\selectfont}
\renewcommand\tablecolheadfont{\fontsize{7.5}{9}\selectfont}
\begin{table}[t]
  \tabcolsep=0pt%
  \TBL{\caption{Performance of each model on PISA 2015 datasets\label{tab:pfm_pisa}}}
  {\begin{fntable}
  \begin{tabular*}{\textwidth}{@{\extracolsep{\fill}}lcccccccccccc@{}}\toprule%
  \multirow[c]{2}{*}[-0.4ex]{\TCH{Model}} & \multicolumn{4}{c}{\TCH{Reading}} & \multicolumn{4}{c}{\TCH{Science}} & \multicolumn{4}{c}{\TCH{Math}} \\[-0.4ex]
  \cmidrule{2-5}\cmidrule{6-9}\cmidrule{10-13}\noalign{\vskip -0.4ex}
  &\TCH{\# params}&\TCH{$\Delta$AIC}&\TCH{$\Delta$BIC}&\TCH{$\Delta$ho-NLL}
  &\TCH{\# params}&\TCH{$\Delta$AIC}&\TCH{$\Delta$BIC}&\TCH{$\Delta$ho-NLL}
  &\TCH{\# params}&\TCH{$\Delta$AIC}&\TCH{$\Delta$BIC}&\TCH{$\Delta$ho-NLL}\\\midrule
  \TCH{2PL}   & 110 & (2.6M) & (2.6M) & (0.5000\textsubscript{$\pm$6})
              & 252 & (10.3M) & (10.3M) & (0.5487\textsubscript{$\pm$4})
              & 120 & (2.9M) & (2.9M) & (0.5157\textsubscript{$\pm$9}) \\
  \TCH{3PL}   & 165 & $-$4.7 & $-$4.1 & $-$14.9\textsubscript{$\pm$1.0}
              & 378 & $-$20.5 & $-$19.1 & $-$17.0\textsubscript{$\pm$0.5}
              & 180 & $-$6.1 & $-$5.5 & $-$20.1\textsubscript{$\pm$0.8} \\
  \TCH{RH}    & 165 & $-$3.5 & $-$3.0 & $-$10.1\textsubscript{$\pm$0.8}
              & 378 & $-$16.0 & $-$14.6 & $-$12.7\textsubscript{$\pm$0.6}
              & 180 & $-$7.5 & $-$6.9 & $-$21.3\textsubscript{$\pm$1.4} \\
  \TCH{3PLQ}  & 165 & $-$6.0 & $-$5.4 & $-$18.4\textsubscript{$\pm$0.8}
              & 378 & $-$24.6 & $-$23.3 & $-$19.8\textsubscript{$\pm$0.8}
              & 180 & $-$10.3 & $-$9.7 & $-$30.5\textsubscript{$\pm$1.2} \\
  \TCH{Spline}& 1045 & $-$6.3 & $+$3.2 & $-$23.4\textsubscript{$\pm$0.9}
              & 2394 & $-$27.6 & $-$4.0 & \textbf{$-$24.9}\textsubscript{$\pm$0.8}
              & 1140 & $-$10.7 & $-$0.3 & $-$35.6\textsubscript{$\pm$2.0} \\
  \cmidrule{2-5}\cmidrule{6-9}\cmidrule{10-13}\noalign{\vskip -0.4ex}
  \TCH{\mirm{2,2}}
              & 330 & \textbf{$-$7.7} & $-$5.5 & $-$24.5\textsubscript{$\pm$0.9}
              & 756 & \textbf{$-$29.7} & $-$24.2 & $-$24.5\textsubscript{$\pm$0.8}
              & 360 & $-$12.1 & $-$9.6 & $-$36.5\textsubscript{$\pm$1.4} \\
  \TCH{\mirm{5,5}}
              & 990 & \textbf{$-$7.7} & $+$1.2 & \textbf{$-$28.2}\textsubscript{$\pm$1.1}
              & 2268 & $-$28.1 & $-$5.9 & \textbf{$-$24.8}\textsubscript{$\pm$0.9}
              & 1080 & \textbf{$-$12.4} & $-$2.7 & \textbf{$-$39.9}\textsubscript{$\pm$1.0} \\
  \TCH{\mirm{c_j,d_j}}
              & 296 & $-$7.6 & \textbf{$-$5.7} & $-$24.4\textsubscript{$\pm$1.2}
              & 594 & $-$29.4 & \textbf{$-$25.6} & $-$24.2\textsubscript{$\pm$0.9}
              & 286 & $-$11.9 & \textbf{$-$10.2} & $-$35.5\textsubscript{$\pm$1.2} \\
  \botrule
  \end{tabular*}%
  \footnotetext[]{{Note:} $\Delta$AIC and $\Delta$BIC are differences from 2PL in units of $10^3$, and $\Delta$ho-NLL in units of $10^{-4}$; negative values favor the comparison model. The values of 2PL are given in parentheses (M $=10^6$). ho-NLL: held-out negative log-likelihood across 7 random splits (test split 20\%, 5-fold cross-validation); \# params: number of free parameters. Bold denotes the best value (not applied to \# params), including values tied with it within the error. Standard deviations of $\Delta$ho-NLL are those of the differences within each seed, and are in units of $10^{-4}$. A tie means that the difference from the best model within each seed is within one standard deviation of zero; all the other models are worse than the best model in every seed. Itemwise ($c$,$d$) configurations were selected separately for each seed.}
  \end{fntable}}
\end{table}
\endgroup

Table~\ref{tab:pfm_pisa} reports information criteria and held-out negative log-likelihood (ho-NLL) for the three domains. Among the traditional comparators, 3PLQ had the lowest BIC in every domain, while the flexible spline had the lowest ho-NLL. The ho-NLL of \mirm{2,2} was lower than that of 3PLQ by 6.1, 4.7, and 6.0 $\times10^{-4}$ in Reading, Science, and Mathematics, and lower than that of the spline except in Science. Compared with the interpretable three-parameter baselines (3PL and RH), the gains in both BIC and ho-NLL were much larger across all three domains. Thus, even a small common MiRM mostly improved prediction over the strongest traditional comparators.

The per-item MiRM used 5.4, 4.7, and 4.8 free parameters per item in Reading, Science, and Mathematics. These counts were about 10\%, 21\%, and 21\% below the 6 parameters per item of the fixed \mirm{2,2}, while ho-NLL was higher by only 0.1, 0.4, and 1.0 $\times10^{-4}$. The per-item MiRM had the lowest BIC of all compared models in each domain, about 300, 2,300, and 500 below 3PLQ, respectively. To examine where this advantage came from, we randomly reassigned the selected $(c_j,d_j)$ configurations among the 60 Mathematics items in 10 refits, keeping the total of 286 parameters unchanged. Among these replicates, BIC worsened by 3,400--6,100 (mean 4,800), falling back to the level of the 3PL. By contrast, refitting the original assignment from different random initializations changed BIC by less than 8. Together with the repeated selections in Figure~\ref{fig:cd}, these results indicate that the fit did not come simply from adding complexity: parameters were allocated differently across items. 

\mirm{5,5} had the lowest ho-NLL in every split in Reading and Mathematics. Science was the only domain where the best model differed across splits: the spline was best in four splits and \mirm{5,5} in three. Across the three domains, \mirm{5,5} used one fewer parameter per item than the spline and also had lower BIC. The microstate representation thus offered comparable flexibility with fewer parameters. Nevertheless, the BIC of \mirm{5,5} exceeded that of the per-item MiRM by about 6,900, 19,700, and 7,600 in Reading, Science, and Mathematics. The small predictive gains of (5,5) and the parameter savings from itemwise selection serve different purposes.

\subsection{Empirical ICC Local Dip}\label{sub:evidence}
In Section~\ref{sub:nonmono_sim}, the fitted ICCs alone flagged many false dips. Hence, a fitted dip alone does not establish that an item is nonmonotone. We therefore examined three items together with evidence that does not depend on the fitted model.

\begin{figure}[t]%
  \FIG{\includegraphics[width=\textwidth]{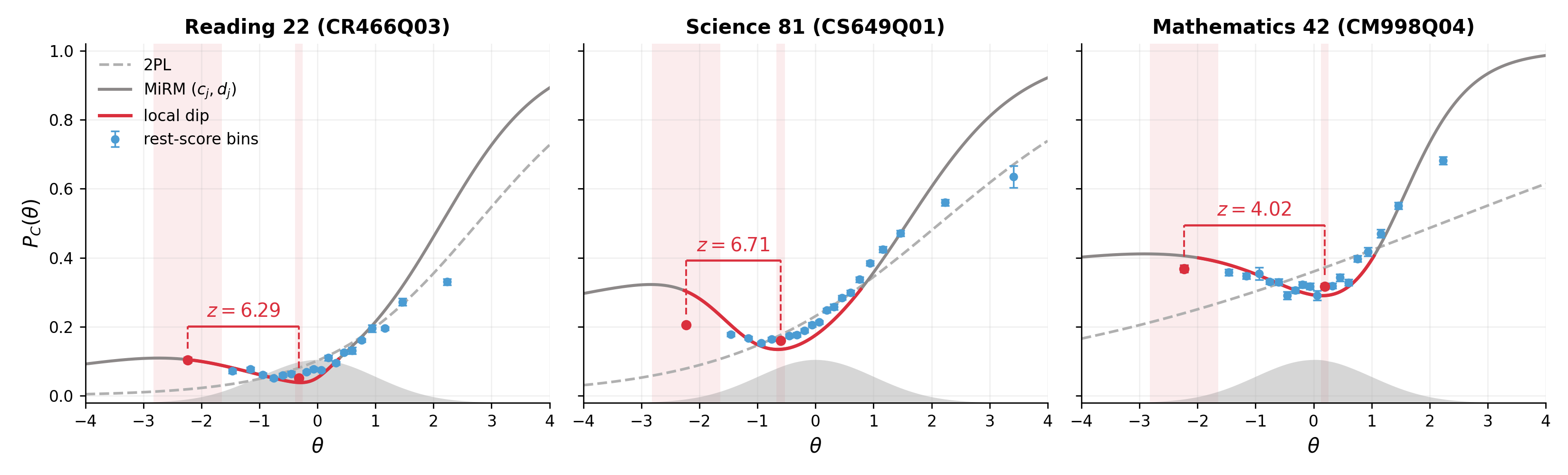}}
  {\caption{Nonmonotone ICCs of three PISA 2015 items: Reading 22 (CR466Q03, three binary judgments; left), Science 81 (CS649Q01, four-option multiple choice; center), and Mathematics 42 (CM998Q04, two binary judgments; right). Solid gray line: \mirm{c_j,d_j} ICC, with the dip in red; dashed line: 2PL ICC; blue points with error bars: observed proportion correct ($\pm 1$ SE) in rest-score bins, placed at standard normal quantiles; red points and vertical red bands: the peak-dip bins compared by the dip test (with the $z$-statistic of the difference in proportion correct shown above the two bins); bottom gray region: respondent ability density}\label{fig:nonmono}}
\end{figure}

Figure~\ref{fig:nonmono} shows one item from each domain for which the per-item MiRM fitted a local dip. For all three, a decline in the fitted ICC was accompanied by a decline in the observed proportion correct across rest-score bins. The $z$ value in each panel is the difference in observed proportions correct between the rest-score bins corresponding to the fitted ICC's pre-dip shoulder and minimum, divided by its standard error. The values were 6.29 for Reading 22, 6.71 for Science 81, and 4.02 for Mathematics 42. These items were chosen as cases in which the local decline can be examined alongside the response structure, rather than as a representative sample of all items with dips. Additional checks using alternative ability bins, complete 15-item blocks, and within-country stratified analyses are reported in Appendix~\ref{appendix:nonmono}.

Reading 22 is scored correct only when all three yes-or-no judgments are correct. The probability of getting all three right by random guessing is 0.125, almost identical to the overall proportion correct of 0.126. Yet the fitted ICC is near chance at low ability, falls below chance at intermediate ability, and rises again at high ability. An overall proportion near chance does not mean that everyone guessed. Above-chance performance at high ability and below-chance performance at intermediate ability can offset each other in the overall average; the ICC separates the two. In every data split, the per-item MiRM selected at least two correct microstates and more than two distractor microstates for Reading 22 (Figure~\ref{fig:cd}). This is consistent with an item that can be answered correctly through knowledge or chance, but can be answered incorrectly in several ways. The observed subitem responses show this pathway directly: while the correct pattern (Yes, No, No) accounts for 12.6\%, the most common pattern is (Yes, No, Yes) at 56.4\%, missing only the final judgment. Across rest-score deciles (proportion correct on the remaining items in the domain), the first two judgments increase monotonically, whereas the final judgment drops from 33\% in the first decile to 17\% in the fifth before rising to 38\% in the tenth. Intermediate students who resolve the first two judgments but favor the wrong answer on the last one therefore fall below the all-three guessing rate.

Science 81, \emph{Weather Balloon}, is a four-option multiple-choice item. The correct option (Option 2) states that the balloon's material cannot be stretched any farther, whereas 79\% of incorrect respondents selected the option that the pressure inside increases (Option 4). Across rest-score deciles, the selection rates of Options 1 and 3 decrease monotonically, whereas Option 4 peaks at 68\% in the sixth decile as Option 2 dips to 16\% in the third decile, showing that the local decline is driven by a single competing response rather than errors dispersed across the options. 

Mathematics 42 requires both of two yes-or-no judgments to be correct, giving a chance probability of 0.25. The left portion of its fitted ICC is higher than that of Reading 22, which requires three correct judgments; here too, the observed proportion correct and the ICC both decline at intermediate ability. In these three examples, nonmonotonicity is a decline over a limited ability range, not a reversal over the entire scale. Its location and height can be read alongside the item's scoring and option structure.

\subsection{Full-Bank Estimation With and Without PRM}\label{sub:ndim}
We showed that PRM guarantees orderability and directionality of ability in Section~\ref{sub:prm}, and mentioned the correspondence between MiRM and multidimensional NRM in Table~\ref{tab:rel}. The guarantees and the correspondence are closely related (Appendix~\ref{app:sub:instrument}), and these together provide clear intuition about how PRM aligns the global ability ordering, as shown in Section~\ref{sub:nonmono_sim}. As unidimensional IRT models place all respondents on a single scale, they seek the ability path that the test supports the most, even if the true ability space is multidimensional. Here, PRM prevents the path from being reversed in the ability space (Theorem~\ref{thm:instrument}\ref{thm:item:direction}). However, investigating this property empirically is challenging; the true ability space is not given in real data. The PISA dataset provides such an opportunity: since many students take forms across domains, we can regard the full-bank data combining the three domains as a single dataset and then fit the model. Furthermore, since the full bank contains the items of each domain, the ability scale of the full bank can be compared with the abilities obtained from each domain in the previous analyses. We therefore draw the ability scale of the full bank as a path in the three-dimensional space made of the three domain abilities, and examine how the path is drawn.

\begin{figure}[t]%
  \FIG{\includegraphics[width=\textwidth]{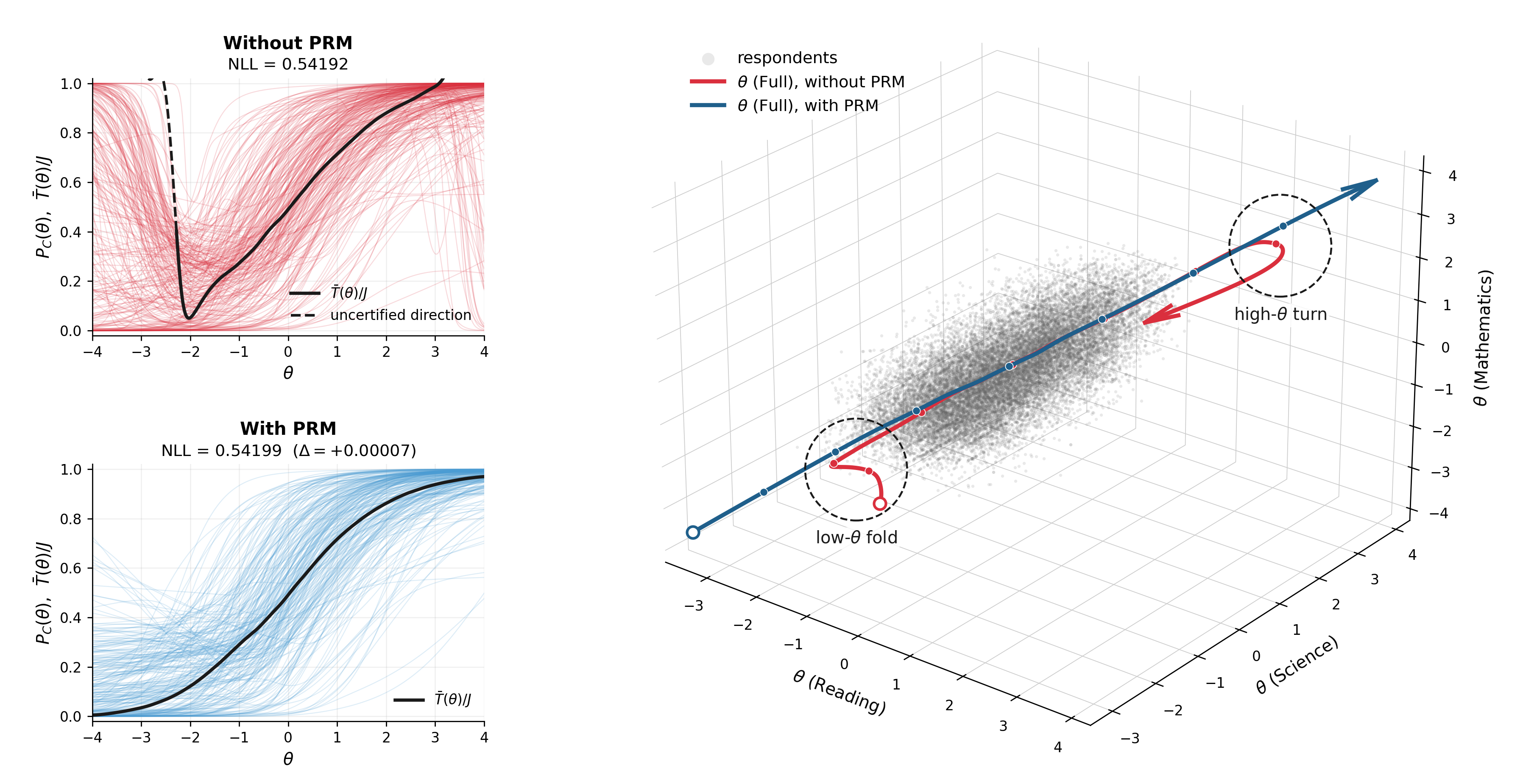}} 
  {\caption{Comparison of per-item MiRM fitted to the pooled PISA 2015 bank (Reading, Science, and Mathematics combined) without PRM (red) and with PRM (blue). Left: individual item characteristic curves $P_C(\theta)$ (colored curves) and the normalized expected generalized score $\bar{T}(\theta)/J$ (black curve; dashed where $a_{\mathcal{J}}(\theta)\leq 0$ for most response patterns, labeled `uncertified direction'). NLL denotes the training negative log-likelihood per observed response. Right: full-bank ability paths embedded in the three-dimensional space of domain-specific PRM ability coordinates ($\theta$ for Reading, Science, and Mathematics). Gray dots represent individual respondents' domain MAP ability estimates; markers along each path indicate equally spaced $\theta$ grid ticks, with arrowheads pointing toward increasing $\theta$}\label{fig:wo_prm}}
\end{figure}

We fitted the per-item MiRM to the three PISA domains together as a single 241-item bank, once with PRM and once without PRM (Figure~\ref{fig:wo_prm}). Without PRM, the constraint was removed from both the configuration selection and the final fit. To draw the full-bank ability $\theta$ in the three-dimensional domain space (right), we used the per-item MiRM fitted to each domain in the previous sections as the three domain coordinates. As the full-bank $\theta$ varies from $-4$ to $4$, its position on each domain axis is the domain ability that gives the same all-correct probability for that domain's items. Gray dots show the domain MAP ability estimates of the students who took all three domains.

The training negative log-likelihood per response was 0.54192 without PRM, 0.00007 below the 0.54199 obtained with PRM. A lower training loss is expected without the constraint; what matters here is the path accompanying this small gain. 

Without PRM, the ability path folds back toward parts of the domain space near both ends of the ability axis (right). Around $\theta\approx-2$, many ICCs also reverse direction (upper left), causing the expected generalized-score curve $\bar{T}(\theta)/J$ to spike and enter an uncertified region (dashed black curve). When different combinations of domain performance are placed on one dimension, a single path cannot branch; it can turn back to follow another direction and thereby represent some of those differences. This representation comes with a small gain in training likelihood, but within a folded region a larger $\theta$ no longer means better overall performance. By contrast, under PRM, the path moves in one direction, and $\bar{T}(\theta)/J$ remains bounded within $[0,1]$ (see Theorem~\ref{thm:instrument}\ref{thm:item:boundary}). PRM thus preserves the ordering interpretation of the full-bank ability scale without uniformly removing local dips from individual items. Figure~\ref{fig:wo_prm} suggests that differences in performance across domains contributed to the folding of the axis.

\section{Discussion}\label{sec:disc}

\subsection{On Microstates}
Traditional IRT models have worked well because they provide economical descriptions of corresponding psychometric item characteristics. Parameters for guessing-like behavior, slipping-like behavior, item complexity, or residual heteroscedasticity efficiently represent lower and upper asymptotes, and ICC asymmetry. However, item characteristics vary even among items within the same testlet, and introducing a separate parameterization for each departure does not explain why these characteristics are related. MiRM offers a different level of explanation through a single principle: an observed response is a coarse-grained outcome of competition among latent response states.

From this perspective, the results of this paper are not independent features added to a flexible item model. The model relations in Table~\ref{tab:rel} reinterpret conventional item parameters through the number, arrangement, or spectrum of microstates. Local nonmonotonicity arises when the mean slopes of the correct and distractor macrostates reverse their ordering (Theorem~\ref{thm:nonmono}). The giant-MiRM result (Theorem~\ref{thm:giant}) follows by combining item-level microstates across a test. PRM and the generalized total score then determine the direction and credit of the resulting giant item. Item characteristics, ICC geometry, test composition, and scoring are therefore different consequences of the same latent-state construction. In this sense, the microstate representation is mathematically coherent and structurally informative rather than merely a device for adding parameters.

This structural claim should nevertheless be distinguished from a cognitive one. The response distribution, its local discrimination and difficulty, and a reduced microstate representation are statistical objects. Calling a particular state a strategy, misconception, reason, or response pathway is an interpretation placed on those objects. However, this distinction is not peculiar to MiRM: a lower-asymptote parameter can be identified under the 3PL without establishing that the corresponding responses were actually produced by guessing, just as an upper asymptote does not by itself establish slipping. Likewise, the present results do not show that every fitted microstate corresponds one-to-one with a distinct cognitive process. Even if such a literal correspondence fails, the geometric, compositional, and information-theoretic results of MiRM remain intact. Our narrower claim is that microstates provide a common explanatory vocabulary in which previously separate item characteristics can be understood and empirically investigated.

Coarse-graining also places a fundamental limit on that investigation. The information decomposition in Eq.~\ref{eq:info} shows that observing only a macrostate loses information, in the amount of the expected within-macrostate variance of the microstate slopes. When several reasons for responding are reduced to the same category, some ability-relevant variation becomes unavailable. MiRM cannot recover information absent from the observations; it instead uses the structure that remains visible in the aggregate response curves. Not every microstate, however, must remain entirely latent. Option choices, response times, eye movements, intermediate work, or verbal reports may anchor particular states and restrict the representations compatible with the data. Open-ended assessment can be understood similarly as reducing coarse-graining by retaining part of the reasoning that selected-response scoring discards. With such external anchors, the information lost through coarse-graining may eventually become useful for assessing how clearly an item separates its underlying reasons for correct and incorrect responses.

A stronger cognitive interpretation requires validation against response processes established independently of binary outcomes. Concept inventories and diagnostic datasets such as the FCI \citep{hestenes1992force}, FMCE 
\citep{thornton1998assessing}, and Eedi \citep{king2024eedi} offer promising settings because their options are informed by prior accounts of misconceptions or solution processes. Microstates estimated from dichotomized responses could be compared with withheld option choices, process measurements, or expert classifications. Such analyses would directly address whether statistically inferred states predict independently observed response-generating processes. 
The information decomposition specifies how much ability-relevant information has been destroyed by dichotomization, although translating that quantity into a bound on the recovery of individual state labels would require additional assumptions. 

The behavior of the 3PLQ provides a more indirect illustration of the explanatory value of microstates. Its quadratic logit corresponds to a Gaussian spectrum of slopes in one macrostate, with the magnitude of the quadratic coefficient representing the variance of that spectrum (see Appendix~\ref{app:sub:3plq}). The model can consequently summarize diffuse pathway heterogeneity with a single parameter, helping to explain its strong BIC performance (Table~\ref{tab:pfm_pisa}). The same account explains its limitation: a Gaussian spectrum is unbounded, whereas a finite MiRM has extremal slopes. This appears as the extremal reversal of the 3PLQ's ICCs in Figure~\ref{fig:character}. Thus, the central flexibility and problematic tail behavior of the 3PLQ arise from the same structural assumption. Explaining both the strength and the weakness of a competing model in a common language supports the explanatory usefulness of the microstate representation without requiring every inferred state to be cognitively literal.

\subsection{Relations Between MiRM and PRM}
When introducing PRM in Section~\ref{sub:prm}, we noted that PRM provides several theoretical guarantees for modeling ability, whereas the fit of a flexible model without PRM is unreliable. Figures~\ref{fig:confusion} and \ref{fig:wo_prm} show this clearly; in particular, the upper-left panel of Figure~\ref{fig:wo_prm} is a typical example of the ICCs that MiRM produces when fitted without PRM. PRM is therefore essential for fitting MiRM. Conversely, although PRM was proposed as a property of a test independent of MiRM, it loses its practical meaning in most existing models because they always satisfy PRM. This shows why MiRM and PRM need to be introduced together. Moreover, the PRM conditions are available in closed form in MiRM, making their joint use practical. However, this does not mean that PRM is the single invariant answer for ability measurement. While PRM has the conceptual strength of scoring invariance, other test-level monotonicity conditions with different properties can also be proposed.

Moving the monotonicity condition from the item to the test has a broader implication. PRM as a test-level condition implies that item characteristics may change slightly depending on how the test is composed. Remark~\ref{rem:single} shows this well: a nonmonotonic item within a test is refitted as monotonic if the item itself is used as a test. This is not a defect of the condition, but its intended meaning. What we measure through a test is not the true ability itself. We only define from the response patterns the tendency that the test, as a measuring device, explains best, and expect that this tendency follows the true ability. Therefore, whether a test measures the true ability is a question of validity, and it cannot be answered inside the model. The model only shows the direction on which the items of a test agree. Whether this direction is the trait that the test was designed for must be checked outside the model. Without such an investigation, it is difficult to claim that an item's characteristics are strictly intrinsic to the item. One of the best practical approaches may be to regard the microstates or item characteristics that survive across different test sets and respondent groups as the item's own.

Meanwhile, since an ideal test will automatically generate the best unidimensional explanation, it may pass PRM as a diagnostic criterion even when PRM is not imposed as a constraint. Furthermore, if the ability path folds back or some ICCs reverse direction when fitted without PRM, as in Figure~\ref{fig:wo_prm}, this may be read as a sign that the test is intrinsically multidimensional.

\subsection{Implications for Practice}
MiRM's unification framework helps not only theory, but also practical item analysis by placing several departures on a single layer. Before MiRM, practitioners had to choose a specific model in advance to reflect an item's expected behavior---for instance, the 3PL for guessing or the RH model for asymmetry. However, items within the same test often exhibit different departures, leaving mismatches under any single traditional model, while fitting multiple competing models and comparing their parameters is cumbersome. In MiRM, the data make this choice instead of the practitioner. The model captures these departures simultaneously within one model family and summarizes them on a common scale. Although slight instability in configuration selection and heavier computation remain practical costs, we address these in Section~\ref{sub:limit}.

Putting item characteristics on a single layer also changes how we handle nonmonotonicity. We do not claim that nonmonotonic ICCs are practically healthy, nor do we suggest using nonmonotonic items or negative item credits. Our claim is narrower: itemwise nonmonotonicity should be detected rather than assumed away, and in rare cases a correct response can indeed serve as negative evidence for ability. By the results in Section~\ref{ssub:nonmono}, MiRM can serve as both a diagnostic tool and an operational model in a single step. If a practitioner prefers all items to be monotonic, MiRM can directly flag nonmonotonic items for revision or removal without a separate misfit analysis, and the stability of $\omega_j(\theta)$ across ability can guide practical item weights for sum scores. On the other hand, an educator or researcher who wishes to retain dip items for substantive analysis can use the fitted MiRM directly as the operational scoring model under PRM.

A local dip can also be informative in its own right. In educational assessment, a dip may capture U-shaped behavioral growth \citep{strauss1991u}, where intermediate learners temporarily perform worse than novices as they transition from superficial rules to deeper concepts, as suggested by Science 81 in Section~\ref{sub:evidence}. Our analysis was restricted to computer-scored items, many of which are selected-response items, where the chance to guess naturally provides a second correct pathway that satisfies Corollary~\ref{cor:dip} ($c\geq2$). Detecting a dip in an open-ended or constructed-response item would therefore be an even stronger substantive finding.

Beyond educational testing, nonmonotonic curves over an ordered trait are also natural in psychological measurement. In Kohlberg's Heinz dilemma \citep{kohlberg1981philosophy}, for example, the binary decision of whether Heinz should steal the drug can oscillate across stages of moral development even as the underlying reasoning matures monotonically. Many psychological items may share this structure when distinct reasons are compressed into a binary choice.

Finally, from the perspective of preventing unintended nonmonotonicity in item writing, Corollary~\ref{cor:dip} offers a counterintuitive lesson. Traditional item development has focused heavily on crafting plausible distractors, yet a local dip requires at least two correct pathways ($c\geq2$, such as intended mastery alongside guessing or a shortcut). To prevent a dip, ensuring that the correct answer is reachable through a single intended pathway ($c=1$) is therefore important, rather than focusing only on distractors. Conversely, Corollary~\ref{cor:peak} and Section~\ref{sec:gen} show that diversifying attractive distractors ($d\geq2$) --- long treated as an unquestioned virtue in item design --- may enable ICC peaks and flatten discrimination at high ability.

\subsection{Outside of MiRM}
MiRM is primarily a unidimensional parametric IRT model, yet it bridges parametric and nonparametric approaches through the Universal Approximation Theorem (Theorem~\ref{thm:uat}). \citet{wei2026beta} pursued a similar goal with the Beta-Bernstein IRT model, establishing an expressive space comparable to that of MiRM, though its direct readings into traditional parametric models were limited. By contrast, MiRM recovers the structural interpretations of classical models (Section~\ref{sec:gen}) and restores the 2PL parameters locally by microstate statistics (Eq.~\ref{eq:ab}). Consequently, fitting a MiRM to any nonparametric ICC can serve as an algebraic parameterization filter that equips the curve with interpretable derivatives and state-level readings. 

Moreover, because a finite MiRM with sufficiently many states can approximate any continuous ICC on a compact domain, it represents the theoretical ceiling of what a single ordered ability coordinate can explain. From this perspective, the small gains in performance across models in Table~\ref{tab:pfm_pisa} do not mean that MiRM contributes little; rather, they may indicate that the standard 2PL already captures the bulk of the unidimensional signal and that spline and MiRM approach the ceiling, with the per-item MiRM reaching this level parsimoniously. In fact, since the 3PLQ and spline are used mainly for fit rather than interpretation, it is fairer to compare MiRM with interpretable models such as the 2PL, 3PL, and RH, where the performance gain is much clearer. 

What truly stands outside MiRM are the alternative explanations for unidimensional IRT itself: latent dimensions (Multidimensional IRT; \cite{reckase200618}), latent populations \citep[{Mixture IRT;}][]{mislevy1990modeling}, group--item relations \citep[{DIF;}][]{holland1993differential}, or item--item relations (Network Psychometrics; \cite{epskamp2018network}). They account for features that lie outside a single universal, comparable ability $\theta$. However, just as with the departures within unidimensional IRT models, their representation spaces can still overlap \citep{kruis2016three}. They can even overlap with the unidimensional departures: \citet{liao2026approximating} showed that ICC asymmetry can arise when the data are generated from multidimensional latent traits, while MiRM is algebraically related to mixture models (Table~\ref{tab:rel}). The fundamental question is therefore which features can be truly reduced to a single ability and which cannot. Selecting among these external accounts before characterizing the full expressive capacity of an individual item may be conceptually premature. As we measure people on a single scale rather than teaching or treating them, what the test directly controls is the item, not the person; attributing extra causes of item--person or item--item interactions to each item should therefore precede introducing external traits. Establishing such a within-item reference family provides a basis against which additional structures can be added. The RH model \citep{molenaar2015heteroscedastic} exemplifies this approach: population-level heteroscedasticity can become an itemwise asymmetry parameter once the pattern arises systematically from the item. In the same sense, MiRM is important because it does as much of this work as possible inside the item.

\subsection{Limitations}\label{sub:limit}
Although attributing unidimensional explanations to the item as much as possible is a useful starting point, the alternative accounts above are neither meaningless nor absent in real data. Because MiRM is a flexible model, it can unintentionally absorb local dependence \citep{bradlow1999bayesian}, population-level features, or extra dimensions into its curve, as occasionally seen in \mirm{5,5}. In addition, where data are sparse, extreme-tail reversals similar to those of the 3PLQ can occasionally appear, such as in \mirm{5,5} on Item 16 in Figure~\ref{fig:character}, while the sharp boundaries in the local characteristics of Item 29 may come from insufficiently dense Gauss--Hermite quadrature nodes. Empirically, these issues could be mitigated by bounding the slopes $\alpha_s$, applying PRM to smaller item subsets, using early stopping when the loss plateaus, or increasing the number of quadrature nodes. However, these remedies are all heuristic, and a more fundamental approach is needed.

Because the algorithm that selects $(c,d)$ affects all of these issues, improving configuration selection is an important task for future work. The consistency results in Figure~\ref{fig:cd}, together with the performance and item characteristics in Section~\ref{sec:emp}, show that configurations are selected fairly consistently and that item-specific configurations exist in data. One may improve the selection consistency by selecting the most frequent configuration of each item across multiple splits. Yet we do not claim that our current procedure always finds the optimal configuration for every item. Specifically, in a preliminary simulation, items generated from (3,1) were recovered poorly and were mostly assigned (2,1). The algorithm is thus conservative for configurations beyond the (2,2) grid, and the state counts in Figure~\ref{fig:cd} may be underestimated. Part of this limit is fundamental, since ratios of exponential sums are not fully identified (Appendix~\ref{app:sub:micro}).

Connecting the theoretical quantities of MiRM to real-world features is equally important. Although $a(\theta)$ and $b(\theta)$ are theoretically clear and offer intuitive empirical readings, further validation is needed to establish what cognitive features they actually reflect. A natural candidate for $a(\theta)$ is how deterministically ability governs the response at that level, as seen in the RH derivation (Section~\ref{sub:rh}). If these local characteristics are understood more concretely, they could be directly useful in computerized adaptive testing (CAT) and adaptive learning.

Finally, an important direction for future work lies in the choice of prior. Fundamentally, MiRM makes no structural assumption on the prior distribution of ability. As noted earlier, the Gaussian prior $\theta\sim\mathcal{N}(0,1)$ is a safe and convenient baseline for comparing models, but there is no guarantee that true ability follows a normal distribution on that scale. In statistical mechanics, choosing the ability scale corresponds to setting the scale of inverse temperature. Just as temperature has a physical lower bound, setting upper or lower bounds on the ability domain may be substantively meaningful in psychometrics, suggesting that a normal distribution over all real numbers may not always be the most natural choice. However, if the prior is changed or learned from data in future work, some MiRM characteristics, such as $a(\theta)$ and $b(\theta)$, will also change with the prior. This implies that even the prior distribution of ability may be an exchangeable feature in psychometrics, just as itemwise departures, multidimensionality, population characteristics, and interactions are.


 
\begin{Backmatter}


\paragraph{Funding Statement}
This research received no specific grant from any funding agency, commercial or not-for-profit sectors.

\paragraph{Competing Interests}
The authors declare none.

\paragraph{Data Availability Statement}
The PISA 2015 dataset analyzed in this study is publicly available from the OECD (\url{https://www.oecd.org/pisa/data/2015database/}). The simulation and analysis code are publicly available at \url{https://github.com/hwater27/mirm}.

\paragraph{Ethical Standards}
The research meets all ethical guidelines, including adherence to the legal requirements of the study country.

\paragraph{Author Contributions}
Conceptualization: S.L. Methodology: S.L. Software: S.L. Validation: S.L. Formal analysis: S.L. Investigation: S.L; J.J. Data curation: S.L. Data visualisation: S.L. Supervision: J.J. Writing original draft: S.L. Resources: J.J. Writing review \& editing: S.L; J.J. All authors approved the final draft.


\printbibliography 

@article{bock1972estimating,
  title={Estimating item parameters and latent ability when responses are scored in two or more nominal categories},
  author={Bock, R Darrell},
  journal={Psychometrika},
  volume={37},
  number={1},
  pages={29--51},
  year={1972},
  publisher={Springer-Verlag}
}

@book{rasch1960studies,
  title={Probabilistic models for some intelligence and attainment tests},
  author={Rasch, Georg},
  year={1960},
  publisher={Danish Institute for Educational Research}
}

@incollection{birnbaum1968some,
  title={Some latent trait models and their use in inferring an examinee's ability},
  author={Birnbaum, Allan},
  editor={Lord, Frederic M and Novick, Melvin R},
  booktitle={Statistical theories of mental test scores},
  pages={397--479},
  year={1968},
  publisher={Addison-Wesley}
}

@article{barton1981upper,
  title={An upper asymptote for the three-parameter logistic item-response model},
  author={Barton, Mark A and Lord, Frederic M},
  journal={ETS Research Report Series},
  volume={1981},
  number={1},
  pages={i--8},
  year={1981},
  publisher={Wiley Online Library},
}

@article{samejima2000logistic,
  title={Logistic positive exponent family of models: Virtue of asymmetric item characteristic curves},
  author={Samejima, Fumiko},
  journal={Psychometrika},
  volume={65},
  number={3},
  pages={319--335},
  year={2000},
  publisher={Springer},
}

@article{molenaar2015heteroscedastic,
  title={Heteroscedastic latent trait models for dichotomous data},
  author={Molenaar, Dylan},
  journal={Psychometrika},
  volume={80},
  number={3},
  pages={625--644},
  year={2015},
  publisher={Cambridge University Press \& Assessment},
}

@article{mislevy1990modeling,
  title={Modeling item responses when different subjects employ different solution strategies},
  author={Mislevy, Robert J and Verhelst, Norman},
  journal={Psychometrika},
  volume={55},
  number={2},
  pages={195--215},
  year={1990},
  publisher={Cambridge University Press \& Assessment},
}

@article{smith2022motivations,
  title={Motivations for using the item response theory nominal response model to rank responses to multiple-choice items},
  author={Smith, Trevor I and Bendjilali, Nasrine},
  journal={Physical Review Physics Education Research},
  volume={18},
  number={1},
  pages={010133},
  year={2022},
  publisher={APS},
}

@article{kruis2016three,
  title={Three representations of the {I}sing model},
  author={Kruis, Joost and Maris, Gunter},
  journal={Scientific Reports},
  volume={6},
  number={1},
  pages={34175},
  year={2016},
  publisher={Nature Publishing Group UK London}, 
}

@article{junker2000latent,
  title={Latent and manifest monotonicity in item response models},
  author={Junker, Brian W and Sijtsma, Klaas},
  journal={Applied Psychological Measurement},
  volume={24},
  number={1},
  pages={65--81},
  year={2000},
  publisher={Sage Publications Sage CA: Thousand Oaks, CA},
}

@article{feuerstahler2024monotonicity,
  title={On the monotonicity of the residual heteroscedasticity item response model},
  author={Feuerstahler, Leah M and Ahn, J Rachael and Chen, Xing and Lorenzi, Daniel and Plourde, Jay},
  journal={Behaviormetrika},
  volume={51},
  number={1},
  pages={99--123},
  year={2024},
  publisher={Springer},
}

@article{bock1981marginal,
  title={Marginal maximum likelihood estimation of item parameters: Application of an {EM} algorithm},
  author={Bock, R Darrell and Aitkin, Murray},
  journal={Psychometrika},
  volume={46},
  number={4},
  pages={443--459},
  year={1981},
  publisher={Springer-Verlag},
}

@article{liu1989limited,
  title={On the limited memory {BFGS} method for large scale optimization},
  author={Liu, Dong C and Nocedal, Jorge},
  journal={Mathematical Programming},
  volume={45},
  number={1},
  pages={503--528},
  year={1989},
  publisher={Springer},
}

@article{bolt2018application,
  title={Application of asymmetric {IRT} modeling to discrete-option multiple-choice test items},
  author={Bolt, Daniel M and Lee, Sora and Wollack, James and Eckerly, Carol and Sowles, John},
  journal={Frontiers in Psychology},
  volume={9},
  pages={2175},
  year={2018},
  publisher={Frontiers Media SA}
}

@article{thissen1989multiple,
  title={Multiple-choice models: The distractors are also part of the item},
  author={Thissen, David and Steinberg, Lynne and Fitzpatrick, Anne R},
  journal={Journal of Educational Measurement},
  volume={26},
  number={2},
  pages={161--176},
  year={1989},
  publisher={Wiley Online Library}
}

@article{robitzsch2022choice, 
  title={On the choice of the item response model for scaling {PISA} data: Model selection based on information criteria and quantifying model uncertainty},
  author={Robitzsch, Alexander},
  journal={Entropy},
  volume={24},
  number={6},
  pages={760},
  year={2022},
  publisher={MDPI}
}

@article{feuerstahler2019metric,
  title={Metric transformations and the filtered monotonic polynomial item response model},
  author={Feuerstahler, Leah M},
  journal={Psychometrika},
  volume={84},
  number={1},
  pages={105--123},
  year={2019},
  publisher={Cambridge University Press \& Assessment}
}

@article{schwarz1978estimating,
  title={Estimating the dimension of a model},
  author={Schwarz, Gideon},
  journal={The Annals of Statistics},
  volume={6},
  number={2},
  pages={461--464},
  year={1978},
  publisher={JSTOR}
}

@article{bolt2022item,
  title={Item complexity: A neglected psychometric feature of test items?},
  author={Bolt, Daniel M and Liao, Xiangyi},
  journal={Psychometrika},
  volume={87},
  number={4},
  pages={1195--1213},
  year={2022},
  publisher={Cambridge University Press \& Assessment}
}

@article{liao2021item,
  title={Item characteristic curve asymmetry: A better way to accommodate slips and guesses than a four-parameter model?},
  author={Liao, Xiangyi and Bolt, Daniel M},
  journal={Journal of Educational and Behavioral Statistics},
  volume={46},
  number={6},
  pages={753--775},
  year={2021},
  publisher={SAGE Publications Sage CA: Los Angeles, CA}
}

@article{lee2018asymmetric,
  title={Asymmetric item characteristic curves and item complexity: Insights from simulation and real data analyses},
  author={Lee, Sora and Bolt, Daniel M},
  journal={Psychometrika},
  volume={83},
  number={2},
  pages={453--475},
  year={2018},
  publisher={Cambridge University Press \& Assessment}
}

@article{hestenes1992force,
  title={Force concept inventory},
  author={Hestenes, David and Wells, Malcolm and Swackhamer, Gregg},
  journal={The Physics Teacher},
  pages={141--158},
  volume={30},
  number={3},
  year={1992}
}

@book{landau2013statistical,
  title={Statistical physics},
  author={Landau, Lev Davidovich and Lifshitz, Evgenii Mikhailovich},
  volume={5},
  year={2013},
  publisher={Elsevier}
}

@article{weierstrass1885uber,
  title={{\"U}ber die analytische {D}arstellbarkeit sogenannter willk{\"u}rlicher {F}unctionen einer reellen {V}er{\"a}nderlichen},
  author={Weierstrass, Karl},
  journal={Sitzungsberichte der K{\"o}niglich Preu{\ss}ischen Akademie der Wissenschaften zu Berlin},
  volume={2},
  pages={633--639},
  year={1885}
}

@article{sijtsma1998methodology,
  title={Methodology review: Nonparametric {IRT} approaches to the analysis of dichotomous item scores},
  author={Sijtsma, Klaas},
  journal={Applied Psychological Measurement},
  volume={22},
  number={1},
  pages={3--31},
  year={1998},
  publisher={Sage Publications, Inc. 2455 Teller Road, Thousand Oaks, CA 91320-2218}
}

@article{thornton1998assessing,
  title={Assessing student learning of {N}ewton's laws: The force and motion conceptual evaluation and the evaluation of active learning laboratory and lecture curricula},
  author={Thornton, Ronald K and Sokoloff, David R},
  journal={American Journal of Physics},
  volume={66},
  number={4},
  pages={338--352},
  year={1998},
  publisher={American Association of Physics Teachers}
}

@incollection{reckase200618,
  title={Multidimensional item response theory},
  author={Reckase, Mark D},
  booktitle={Handbook of Statistics},
  volume={26},
  pages={607--642},
  year={2006},
  publisher={Elsevier}
}

@article{feuerstahler2026defining,
  title={Defining asymmetry in item response theory},
  author={Feuerstahler, Leah M and Verkuilen, Jay and Setti, Fabio and Johnson, Peter},
  journal={British Journal of Mathematical and Statistical Psychology},
  year={2026},
  publisher={Wiley Online Library}
}

@article{junker1993conditional,
  title={Conditional association, essential independence and monotone unidimensional item response models},
  author={Junker, Brian W},
  journal={The Annals of Statistics},
  volume={21},
  number={3},
  pages={1359--1378},
  year={1993},
  publisher={Institute of Mathematical Statistics}
}

@article{thissen1984response,
  title={A response model for multiple choice items},
  author={Thissen, David and Steinberg, Lynne},
  journal={Psychometrika},
  volume={49},
  number={4},
  pages={501--519},
  year={1984},
  publisher={Cambridge University Press \& Assessment}
}

@article{samejima1979new,
  title={A new family of models for the multiple-choice item},
  author={Samejima, Fumiko},
  journal={Office of Naval Research Report},
  year={1979},
  publisher={University of Tennessee}
}

@article{hemker1997stochastic,
  title={Stochastic ordering using the latent trait and the sum score in polytomous {IRT} models},
  author={Hemker, Bas T and Sijtsma, Klaas and Molenaar, Ivo W and Junker, Brian W},
  journal={Psychometrika},
  volume={62},
  number={3},
  pages={331--347},
  year={1997},
  publisher={Cambridge University Press \& Assessment}
}

@article{lee2018alternative,
  title={An alternative to the 3{PL}: Using asymmetric item characteristic curves to address guessing effects},
  author={Lee, Sora and Bolt, Daniel M},
  journal={Journal of Educational Measurement},
  volume={55},
  number={1},
  pages={90--111},
  year={2018},
  publisher={Wiley Online Library}
}

@article{shim2026asymmetry,
  title={On the asymmetry, complexity and predicted data patterns of nontraditional item response theory models},
  author={Shim, Hyejin and Bonifay, Wes},
  journal={British Journal of Mathematical and Statistical Psychology},
  year={2026},
  publisher={Wiley Online Library}
}

@article{strout1990new,
  title={A new item response theory modeling approach with applications to unidimensionality assessment and ability estimation},
  author={Stout, William F},
  journal={Psychometrika},
  volume={55},
  number={2},
  pages={293--325},
  year={1990},
  publisher={Cambridge University Press \& Assessment}
}

@article{muraki1992generalized,
  title={A generalized partial credit model: Application of an {EM} algorithm},
  author={Muraki, Eiji},
  journal={Applied Psychological Measurement},
  volume={16},
  number={2},
  pages={159--176},
  year={1992}
}

@book{holland1993differential,
  title={Differential item functioning},
  author={Holland, Paul W and Wainer, Howard},
  year={1993},
  publisher={Lawrence Erlbaum Associates}
}

@article{epskamp2018network,
  title={Network psychometrics},
  author={Epskamp, Sacha and Maris, Gunter and Waldorp, Lourens J and Borsboom, Denny},
  journal={The Wiley handbook of psychometric testing: A multidisciplinary reference on survey, scale and test development},
  volume={2},
  pages={953--986},
  year={2018},
  publisher={Wiley Online Library}
}

@article{holland1986conditional,
  title={Conditional association and unidimensionality in monotone latent variable models},
  author={Holland, Paul W and Rosenbaum, Paul R},
  journal={The Annals of Statistics},
  volume={14},
  number={4},
  pages={1523--1543},
  year={1986},
  publisher={JSTOR}
}

@article{de2012irtrees,
  title={{IRT}rees: Tree-based item response models of the {GLMM} family},
  author={De Boeck, Paul and Partchev, Ivailo},
  journal={Journal of Statistical Software},
  volume={48},
  pages={1--28},
  year={2012}
}

@article{bockenholt2012modeling,
  title={Modeling multiple response processes in judgment and choice.},
  author={B{\"o}ckenholt, Ulf},
  journal={Psychological Methods},
  volume={17},
  number={4},
  pages={665--678},
  year={2012},
  publisher={American Psychological Association}
}

@article{maris1995psychometric,
  title={Psychometric latent response models},
  author={Maris, Eric},
  journal={Psychometrika},
  volume={60},
  number={4},
  pages={523--547},
  year={1995},
  publisher={Cambridge University Press \& Assessment}
}

@article{suh2010nested,
  title={Nested logit models for multiple-choice item response data},
  author={Suh, Youngsuk and Bolt, Daniel M},
  journal={Psychometrika},
  volume={75},
  number={3},
  pages={454--473},
  year={2010},
  publisher={Cambridge University Press \& Assessment}
}

@article{jin2022exploring,
  title={Exploring the impact of random guessing in distractor analysis},
  author={Jin, Kuan-Yu and Siu, Wai-Lok and Huang, Xiaoting},
  journal={Journal of Educational Measurement},
  volume={59},
  number={1},
  pages={43--61},
  year={2022},
  publisher={Wiley Online Library}
}

@article{shigemasu1993new,
  title={A new item response model with parameters reflecting state of knowledge},
  author={Shigemasu, Kazuo and Ueno, Maomi},
  journal={Behaviormetrika},
  volume={20},
  number={2},
  pages={161--169},
  year={1993},
  publisher={Springer}
}

@article{zhang2022bayesian,
  title={Bayesian item response theory models with flexible generalized logit links},
  author={Zhang, Jiwei and Zhang, Ying-Ying and Tao, Jian and Chen, Ming-Hui},
  journal={Applied Psychological Measurement},
  volume={46},
  number={5},
  pages={382--405},
  year={2022},
  publisher={SAGE Publications Sage CA: Los Angeles, CA}
}

@article{hestenes1969multiplier,
  title={Multiplier and gradient methods},
  author={Hestenes, Magnus R},
  journal={Journal of Optimization Theory and Applications},
  volume={4},
  number={5},
  pages={303--320},
  year={1969},
  publisher={Springer}
}

@article{powell1969method,
  title={A method for nonlinear constraints in minimization problems},
  author={Powell, Michael JD},
  journal={Optimization},
  pages={283--298},
  year={1969},
  publisher={Academic Press}
}

@article{rockafellar1974augmented,
  title={Augmented {L}agrange multiplier functions and duality in nonconvex programming},
  author={Rockafellar, R Tyrrell},
  journal={SIAM Journal on Control},
  volume={12},
  number={2},
  pages={268--285},
  year={1974},
  publisher={SIAM}
}

@article{marcinkiewicz1939propriete,
  title={Sur une propri{\'e}t{\'e} de la loi de {G}auss},
  author={Marcinkiewicz, J{\'o}zef},
  journal={Mathematische Zeitschrift},
  volume={44},
  number={1},
  pages={612--618},
  year={1939},
  publisher={Springer}
}

@article{roberts2000general,
  title={A general item response theory model for unfolding unidimensional polytomous responses},
  author={Roberts, James S and Donoghue, John R and Laughlin, James E},
  journal={Applied Psychological Measurement},
  volume={24},
  number={1},
  pages={3--32},
  year={2000},
  publisher={Sage Publications Sage CA: Thousand Oaks, CA}
}

@article{vondavier2004partially,
  title={Partially observed mixtures of {IRT} models: An extension of the generalized partial-credit model},
  author={von Davier, Matthias and Yamamoto, Kentaro},
  journal={Applied Psychological Measurement},
  volume={28},
  number={6},
  pages={389--406},
  year={2004},
  publisher={Sage Publications Sage CA: Thousand Oaks, CA}
}

@book{OECD2016PISAVolI,
  author    = {{OECD}},
  title     = {{PISA} 2015 Results (Volume {I}): Excellence and Equity in Education},
  series    = {PISA},
  publisher = {OECD Publishing},
  address   = {Paris},
  year      = {2016}
}

@inproceedings{wei2026beta,
  title={The {B}eta-{B}ernstein Bridge: Closed-Form {B}ayesian Measurement Beyond Monotone Item Response Functions},
  author={Wei, Yumou},
  booktitle={Proceedings of the Artificial Intelligence in Measurement and Education Conference (AIME-Con): Full Papers},
  pages={639--648},
  year={2026}
}

@misc{king2024eedi,
  title={Eedi-mining misconceptions in mathematics},
  author={King, Jules and Burleigh, L and Woodhead, Simon and Kon, Panagiota and Baffour, Perpetual and Crossley, Scott and Reade, Walter and Demkin, Maggie},
  year={2024},
  note={Kaggle Competition},
  url={https://kaggle.com/competitions/eedi-mining-misconceptions-in-mathematics}
}

@article{masters1982rasch,
  title={A {R}asch model for partial credit scoring},
  author={Masters, Geoff N},
  journal={Psychometrika},
  volume={47},
  number={2},
  pages={149--174},
  year={1982},
  publisher={Cambridge University Press \& Assessment}
}

@article{ramsay1989binomial,
  title={Binomial regression with monotone splines: A psychometric application},
  author={Ramsay, James O and Abrahamowicz, Michal},
  journal={Journal of the American Statistical Association},
  volume={84},
  number={408},
  pages={906--915},
  year={1989},
  publisher={Taylor \& Francis}
}

@article{akaike1974new,
  title={A new look at the statistical model identification},
  author={Akaike, Hirotugu},
  journal={IEEE Transactions on Automatic Control},
  volume={19},
  number={6},
  pages={716--723},
  year={1974},
  publisher={Ieee}
}

@article{culpepper2016revisiting,
  title={Revisiting the 4-parameter item response model: Bayesian estimation and application},
  author={Culpepper, Steven Andrew},
  journal={Psychometrika},
  volume={81},
  number={4},
  pages={1142--1163},
  year={2016},
  publisher={Cambridge University Press \& Assessment}
}

@article{meng2020marginalized,
  title={Marginalized maximum a posteriori estimation for the four-parameter logistic model under a mixture modelling framework},
  author={Meng, Xiangbin and Xu, Gongjun and Zhang, Jiwei and Tao, Jian},
  journal={British Journal of Mathematical and Statistical Psychology},
  volume={73},
  pages={51--82},
  year={2020},
  publisher={Wiley Online Library}
}

@article{rost1990rasch,
  title={Rasch models in latent classes: An integration of two approaches to item analysis},
  author={Rost, J{\"u}rgen},
  journal={Applied Psychological Measurement},
  volume={14},
  number={3},
  pages={271--282},
  year={1990},
  publisher={Sage Publications, Sage CA: Thousand Oaks, CA}
}

@article{bradlow1999bayesian,
  title={A {B}ayesian random effects model for testlets},
  author={Bradlow, Eric T and Wainer, Howard and Wang, Xiaohui},
  journal={Psychometrika},
  volume={64},
  number={2},
  pages={153--168},
  year={1999},
  publisher={Cambridge University Press \& Assessment}
}

@article{kuhn1955hungarian,
  title={The {H}ungarian method for the assignment problem},
  author={Kuhn, Harold W},
  journal={Naval Research Logistics Quarterly},
  volume={2},
  number={1-2},
  pages={83--97},
  year={1955},
  publisher={Wiley Online Library}
}

@article{strauss1991u,
  title={U-shaped Behavioral Growth: Implications for Curriculum Development},
  author={Strauss, S},
  journal={The International Encyclopedia of Curriculum},
  year={1991},
  publisher={Pergamon}
}

@article{ramsay1988monotone,
  title={Monotone regression splines in action},
  author={Ramsay, James O},
  journal={Statistical Science},
  volume={3},
  number={4},
  pages={425--441},
  year={1988},
  publisher={JSTOR}
}

@article{liao2026approximating,
  title={Approximating multidimensionality with asymmetric unidimensional {IRT} models},
  author={Liao, Xiangyi and He, Yinqiu and Bolt, Daniel and Soland, James and Conner, Carlin and Solari, Emily J},
  journal={British Journal of Mathematical and Statistical Psychology},
  year={2026},
  publisher={Wiley Online Library}
}

@book{kohlberg1981philosophy,
  title={The philosophy of moral development: Moral stages and the idea of justice},
  author={Kohlberg, Lawrence},
  year={1981},
  publisher={Harper \& Row}
}

@article{rosenbaum1988items,
  title={Item bundles},
  author={Rosenbaum, Paul R},
  journal={Psychometrika},
  volume={53},
  number={3},
  pages={349--359},
  year={1988},
  publisher={Cambridge University Press \& Assessment}
}

@article{huynh1994equivalence,
  title={On equivalence between a partial credit item and a set of independent {R}asch binary items},
  author={Huynh, Huynh},
  journal={Psychometrika},
  volume={59},
  number={1},
  pages={111--119},
  year={1994},
  publisher={Cambridge University Press \& Assessment}
}

@article{huynh1996decomposition,
  title={Decomposition of a {R}asch partial credit item into independent binary and indecomposable trinary items},
  author={Huynh, Huynh},
  journal={Psychometrika},
  volume={61},
  number={1},
  pages={31--39},
  year={1996},
  publisher={Springer}
}

@article{wilson1995rasch,
  title={Rasch models for item bundles},
  author={Wilson, Mark and Adams, Raymond J},
  journal={Psychometrika},
  volume={60},
  number={2},
  pages={181--198},
  year={1995},
  publisher={Cambridge University Press \& Assessment}
}

@article{youden1950index,
  title={Index for rating diagnostic tests},
  author={Youden, William J},
  journal={Cancer},
  volume={3},
  number={1},
  pages={32--35},
  year={1950},
  publisher={New York}
}

\begin{appendix}
\markboth{\textit{Preprint}}{Lee \& Jo}
\counterwithin{figure}{section}
\counterwithin{table}{section}
\setcounter{figure}{0}
\setcounter{table}{0}

\section{Proofs}\label{appendix:proof}

\subsection{Identification of Microstates}\label{app:sub:micro}
Because only the response probabilities $P_k(\theta)=Z_k(\theta)/Z(\theta)$ are observed, four types of invariance leave the ICC unchanged:
\begin{enumerate}
  \item \emph{Global shift}: adding $\alpha_0$ to all $\alpha_s$ and $\delta_0$ to all $\delta_s$ multiplies every $Z_k$ by $e^{-\alpha_0\theta+\delta_0}$, which cancels in $Z_k/Z$.
  \item \emph{Within-macrostate permutation}: relabeling microstates within the same $\mathcal{S}_k$ leaves their sum $Z_k$ unchanged.
  \item \emph{Splitting tied slopes}: a state with weight $e^{-\alpha_s\theta+\delta_s}$ can be split into multiple states of the same slope $\alpha_s$ whose weights sum to $e^{\delta_s}$ (e.g., $e^{-\theta+\log2}=2e^{-\theta}=e^{-\theta}+e^{-\theta}$).
  \item \emph{Common subprocess}: multiplying every $Z_k(\theta)$ by a common positive exponential sum $Z_0(\theta)$ cancels in $Z_k/Z$ while increasing the number of terms (e.g., multiplying a $(1,1)$ model by $1+e^{-\theta}$ yields an equivalent $(2,2)$ model).
\end{enumerate}
Reindexing $(k,m)\rightarrow s$ in the order $\alpha_1\leq\alpha_2\leq\cdots\leq\alpha_S$ with $\alpha_1=\delta_1=0$ fixes (1) and (2) up to tied slopes, and combining terms with identical slopes within each macrostate removes (3). We addressed these issues practically in Sections~\ref{sub:cfg} and~\ref{sub:param}. For the remaining invariance (4), Theorem~\ref{thm:micro} shows that the representation without a common subprocess is unique, and that it has the fewest states among the representations connected by a positive common subprocess.

\begin{theorem}[uniqueness and restricted minimality of reduced representations]\label{thm:micro}
  Call a binary MiRM representation $(Z_C,Z_D)$ \emph{reduced} if terms with the same slope are combined within each macrostate and $Z_C$ and $Z_D$ share no non-monomial exponential-sum factor.
  \begin{enumerate}[label=(\alph*)]
    \item Two reduced representations of the same ICC on a nonempty open interval are identical up to a common monomial factor $\gamma_0 e^{-\alpha_0\theta}$ ($\gamma_0>0$, a global shift).\label{thm:item:a}
    \item For any reduced representation with $(c,d)$ states and any positive exponential sum $Z_0$ with $h$ distinct slopes, $(Z_0Z_C,Z_0Z_D)$ has at least $(c+h-1,d+h-1)$ states. Thus, within the family connected by a positive common subprocess $Z_0$, the reduced representation uniquely minimizes the state counts up to a global shift.\label{thm:item:b}
  \end{enumerate}
\end{theorem}

\begin{proof}
  The idea is to view $Z_C/Z_D$ as a fraction of polynomials in $e^{-\theta}$: once common factors are canceled, a reduced fraction is unique up to a monomial factor, and multiplying by any positive factor $Z_0$ can only increase the number of terms because positive coefficients never cancel.

  \noindent\emph{\ref{thm:item:a}} Because the slopes $\alpha_s$ are real numbers that need not be integer multiples of a single unit (e.g., $1$ and $\sqrt{2}$), let $\Gamma\subset\mathbb{R}$ be the additive subgroup generated by all slopes in the two representations being compared. Since only finitely many slopes appear, $\Gamma$ is a finitely generated torsion-free abelian group, so $\Gamma\approxeq\mathbb{Z}^q$ for some basis $\gamma_1,\ldots,\gamma_q$. Writing each slope as $\alpha=\sum_{l=1}^qn_l\gamma_l$ ($n_l\in\mathbb{Z}$) maps $e^{-\alpha\theta}$ to the Laurent monomial $z_1^{n_1}\cdots z_q^{n_q}$ with $z_l=e^{-\gamma_l\theta}$. Because exponential functions with distinct real slopes are linearly independent on any open interval, identities between exponential sums hold if and only if the corresponding Laurent polynomials are identical in the unique factorization domain (UFD)
  \begin{align}
    \mathbb{R}[z_1^{\pm1},\ldots,z_q^{\pm1}],
  \end{align}
  where a reduced pair $(Z_C,Z_D)$ is coprime ($\gcd(Z_C,Z_D)=1$).

  If two reduced pairs $(Z_C,Z_D)$ and $(\tilde{Z}_C,\tilde{Z}_D)$ yield the same ICC on an open interval, analytic continuation gives $Z_C\tilde{Z}_D=\tilde{Z}_CZ_D$ everywhere. Since $\gcd(Z_C,Z_D)=1$, Euclid's lemma implies that $Z_C$ divides $\tilde{Z}_C$ and $Z_D$ divides $\tilde{Z}_D$ in the Laurent ring, so $\tilde{Z}_C=Z_0Z_C$ and $\tilde{Z}_D=Z_0Z_D$ for some Laurent polynomial $Z_0$. Because $(\tilde{Z}_C,\tilde{Z}_D)$ is also coprime, $Z_0$ must be a unit of the ring. The only units with positive coefficients are single monomials $\gamma_0 z_1^{n_1}\cdots z_q^{n_q}$ ($\gamma_0>0$), which correspond back to $Z_0(\theta)=\gamma_0 e^{-\alpha_0\theta}$, a global shift of all $\alpha_s$ by $\alpha_0$ and all $\delta_s$ by $\log\gamma_0$.

  \noindent\emph{\ref{thm:item:b}} Let $\mathcal{A}_k=\{\alpha_1<\cdots<\alpha_m\}$ and $\mathcal{A}_0=\{\tilde{\alpha}_1<\cdots<\tilde{\alpha}_h\}$ be the sets of distinct slopes in $Z_k$ and $Z_0$. In the product $Z_0Z_k$, every term has slope $\alpha+\tilde{\alpha}$ with $\alpha\in\mathcal{A}_k$ and $\tilde{\alpha}\in\mathcal{A}_0$. Because all coefficients in $Z_k$ and $Z_0$ are positive, terms with the same sum of slopes only add and never cancel to zero, so the set of distinct slopes in $Z_0Z_k$ is $\mathcal{A}_k+\mathcal{A}_0$. The strictly ordered chain
  \begin{align}
    \alpha_1+\tilde{\alpha}_1<\cdots<\alpha_m+\tilde{\alpha}_1<\alpha_m+\tilde{\alpha}_2<\cdots<\alpha_m+\tilde{\alpha}_h
  \end{align}
  already contains $m+h-1$ distinct slopes, so $|\mathcal{A}_k+\mathcal{A}_0|\geq m+h-1$. Setting $m=c$ for $Z_C$ and $m=d$ for $Z_D$ gives at least $(c+h-1,d+h-1)$ states.
\end{proof}

Theorem~\ref{thm:micro} shows that a reduced representation is unique up to a global shift and has fewer states than any representation obtained by multiplying it by a positive common subprocess $Z_0$. Moreover, across all equivalent representations of an ICC, the local characteristics $a(\theta)$, $b(\theta)$, and $I_{\rm mac}(\theta)$ remain invariant, while option-level responses can further constrain the microstates.

\subsection{Test-Level Results}\label{app:sub:test}

\begin{proof}[Proof of Theorem~\ref{thm:giant}]
  For response $k_j$ to item $j$, $P_j(X_j=k_j|\theta)=Z_{j,k_j}/Z_j$. Local independence therefore gives $P_{\mathcal{J}}(\mathbf{X}=\mathbf{k}|\theta)=\prod_jZ_{j,k_j}/\prod_jZ_j$. Since $Z_{j,k_j}=\sum_{s_j\in\mathcal{S}_{j,k_j}}e^{-\alpha_{s_j}\theta+\delta_{s_j}}$, expanding the product yields 
  \begin{align}
    \prod_jZ_{j,k_j}=\sum_{(s_1,\ldots,s_J)}e^{-\sum_j\alpha_{s_j}\theta+\sum_j\delta_{s_j}},
  \end{align}
  where the sum ranges over all tuples of microstates belonging to the selected response in each item. Treating a tuple as a microstate of the giant item and $\sum_j\alpha_{s_j}$ and $\sum_j\delta_{s_j}$ as its slope and offset, respectively, gives a MiRM numerator. The denominator $\prod_jZ_j$ is a sum of the same form over all tuples. Distinct patterns have disjoint sets of tuples, establishing the macrostate partition.

  If patterns assigned the same value by the credit $w_{\mathcal{J}}$ are grouped together, the partition function of the new macrostate is the sum of $Z_{\mathbf{k}}^\star$ over the grouped patterns, which remains an exponential sum with positive coefficients. The denominator is unchanged, so the distribution of the coarsened response $w_{\mathcal{J}}(\mathbf{X})$ is a MiRM.
\end{proof}

\subsubsection{Credit Identities}\label{app:sub:gpcm}
Under the 2PL, $a_j$ is constant, so $\omega_j=Ja_j/\sum_\ell a_\ell$ is independent of $\theta$ and 
\begin{align}
  \hat{T}=J\frac{\sum_ja_jw_{j,k_j}}{\sum_ja_j},
\end{align}
which is the weighted sum score. Under the Rasch model, $a_j=a$ is common, so $\omega_j=J/J=1$ and $\hat{T}=\sum_jw_{j,k_j}$, reducing to the sum score. Under the 3PL, differentiating $P_{jC}$ in $a_j(\theta)=P_{jC}'/(P_{jC}P_{jD})$ yields
\begin{align}
  a_j(\theta)=a_j\frac{P_{jC}-p_{c,j}}{P_{jC}(1-p_{c,j})},
\end{align}
which depends on $\theta$ and matches the locally best weight derived by \citet{birnbaum1968some}.

Under the partial credit model (PCM) and generalized partial credit model (GPCM), each item $j$ with maximum credit $m_j$ is \mirm{1,\ldots,1} with integer credits $w_{jk}\in\{0,1,\ldots,m_j\}$ and category slopes $\alpha_{jk}=-a_jw_{jk}$. Taking the zero-credit and full-credit patterns as the two anchors $\mathbf{D}$ and $\mathbf{C}$ with maximum total credit $J=\sum_jm_j$ in Eq.~\ref{eq:credit} gives $\bar{\alpha}_\mathbf{D}^\star-\bar{\alpha}_\mathbf{k}^\star=\sum_ja_jw_{j,k_j}$ and $\bar{\alpha}_\mathbf{D}^\star-\bar{\alpha}_\mathbf{C}^\star=\sum_ja_jm_j$, and hence
\begin{align}
  \hat{T}=\Bigl(\sum_jm_j\Bigr)\frac{\sum_ja_jw_{j,k_j}}{\sum_ja_jm_j},
\end{align}
which reduces to $\sum_jw_{j,k_j}$ under the PCM ($a_j=a$). Moreover, even for a single GPCM item ($J=m_1$), the same formula recovers the item's own partial credit $w_{1,k_1}$, showing that within-item partial credits and test-level total scores are governed by the same expression.

\subsubsection{Relation Between PRM and SOM}\label{app:sub:som}
Let $X_+=\sum_jw_{j,X_j}$. Under SOM, $P(X_+\geq x|\theta)$ is nondecreasing in $\theta$ for every $x$. Taking $x=J$ gives that $P_{\mathcal{J}}(\mathbf{C}|\theta)$ is nondecreasing, and taking $x=1$ gives that $P_{\mathcal{J}}(\mathbf{D}|\theta)=1-P(X_+\geq1|\theta)$ is nonincreasing. Positivity and differentiability of finite-state MiRM probabilities then yield $a_{\mathcal{J}C}\geq0$ and $a_{\mathcal{J}D}\geq0$, respectively. Hence SOM implies PRM.

\subsection{Guarantees of PRM and Dimensionality}\label{app:sub:instrument}
PRM and the generalized credit provide four guarantees for ability measurement. For an interval $\Theta_I\subseteq\Theta_T$, write $\mathbf{k}\succeq_{\Theta_I}\mathbf{k}'$ if the likelihood ratio $P_{\mathcal{J}}(\mathbf{X}=\mathbf{k}|\theta)/P_{\mathcal{J}}(\mathbf{X}=\mathbf{k}'|\theta)$ is nondecreasing on $\Theta_I$, meaning that pattern $\mathbf{k}$ provides stronger evidence for higher ability than $\mathbf{k}'$ across $\Theta_I$. In addition, if the true trait $\boldsymbol{\theta}\in\mathbf{\Theta}$ is multidimensional with vector slopes $\boldsymbol{\alpha}_s$, the scalar ability $\theta$ estimated from the test traces a piecewise-differentiable path $\boldsymbol{\theta}:\Theta\to\mathbf{\Theta}$; along this path, define the vector discriminations
\begin{align}
  \mathbf{a}_j=\bar{\boldsymbol{\alpha}}_{jD}-\bar{\boldsymbol{\alpha}}_{jC},\qquad\mathbf{a}_{\mathcal{J}C}=\sum_jP_{jD}\mathbf{a}_j,\qquad\mathbf{a}_{\mathcal{J}D}=\sum_jP_{jC}\mathbf{a}_j.
\end{align}

\begin{theorem}[instrumental guarantees]\label{thm:instrument}
  For a finite-state dichotomous MiRM test with fixed item parameters on a connected ability domain $\Theta$, suppose that $P_{\mathcal{J}}(\mathbf{C}|\cdot)$ is nonconstant. Then the following statements hold.
  \begin{enumerate}[label=(\roman*)]
    \item \emph{Identifiability.} Under PRM, $P_{\mathcal{J}}(\mathbf{C}|\theta)$ is strictly increasing; hence $\theta\mapsto P_{\mathcal{J}}(\cdot|\theta)$ is injective.\label{thm:item:identifiability}
    \item \emph{Direction.} PRM orients every smooth segment of $\boldsymbol{\theta}(\theta)$ by $\mathbf{a}_{\mathcal{J}C}^{\mathsf{T}}\boldsymbol{\theta}'\geq0$ and $\mathbf{a}_{\mathcal{J}D}^{\mathsf{T}}\boldsymbol{\theta}'\geq0$, and globally orders the entire path by $\theta_1<\theta_2\Longrightarrow P_{\mathcal{J}}(\mathbf{C}|\boldsymbol{\theta}(\theta_1))<P_{\mathcal{J}}(\mathbf{C}|\boldsymbol{\theta}(\theta_2))$.\label{thm:item:direction}
    \item \emph{Ordering.} For every interval $\Theta_I\subseteq\Theta_T$, $\mathbf{k}\succeq_{\Theta_I}\mathbf{k}'\Longleftrightarrow T(\mathbf{k};\theta)\geq T(\mathbf{k}';\theta)$ for all $\theta\in\Theta_I$.\label{thm:item:ordering}
    \item \emph{Boundary.} For every $\theta\in\Theta_T$, the PRM inequalities are equivalent to $0\leq\bar T(\theta)\leq J$.\label{thm:item:boundary}
  \end{enumerate}
\end{theorem}

\begin{proof}[Proof of Theorem~\ref{thm:instrument}]
  For a dichotomous item, $P_{jC}'=P_{jC}P_{jD}a_j$ gives $\frac{d}{d\theta}\log P_{jC}=P_{jD}a_j$ and $-\frac{d}{d\theta}\log P_{jD}=P_{jC}a_j$. By local independence, $P_{\mathcal{J}}(\mathbf{C}|\theta)=\prod_jP_{jC}(\theta)$, so
  \begin{align}
    \frac{d}{d\theta}\log P_{\mathcal{J}}(\mathbf{C}|\theta)=\sum_jP_{jD}(\theta)a_j(\theta)=a_{\mathcal{J}C}(\theta).
  \end{align}

  \noindent\textbf{(i) Identifiability.} Under PRM, $a_{\mathcal{J}C}(\theta)\geq0$, so $P_{\mathcal{J}}(\mathbf{C}|\theta)$ is nondecreasing. If $P_{\mathcal{J}}(\mathbf{C}|\theta_1)=P_{\mathcal{J}}(\mathbf{C}|\theta_2)$ for some $\theta_1<\theta_2$, then
  \begin{align}
    0=\log\frac{P_{\mathcal{J}}(\mathbf{C}|\theta_2)}{P_{\mathcal{J}}(\mathbf{C}|\theta_1)}=\int_{\theta_1}^{\theta_2}a_{\mathcal{J}C}(t)dt.
  \end{align}
  Because $a_{\mathcal{J}C}$ is continuous and nonnegative, it must be zero on the entire interval $[\theta_1,\theta_2]$. Since every finite-state MiRM probability is a ratio of finite exponential sums and hence real analytic, vanishing on $[\theta_1,\theta_2]$ implies that $a_{\mathcal{J}C}$ vanishes across the entire connected domain $\Theta$, making $P_{\mathcal{J}}(\mathbf{C}|\cdot)$ constant---contrary to assumption. Thus $P_{\mathcal{J}}(\mathbf{C}|\theta)$ is strictly increasing, which makes the full distribution map $\theta\mapsto P_{\mathcal{J}}(\cdot|\theta)$ one-to-one.

  \noindent\textbf{(ii) Direction.} In the multidimensional space $\mathbf{\Theta}$, let $\bar{\boldsymbol{\alpha}}_{jk}=-\nabla_{\boldsymbol{\theta}}\log Z_{jk}=\sum_{s\in\mathcal{S}_{jk}}\rho_{s|jk}\boldsymbol{\alpha}_s$ be the mean slope vector of category $k$. Then $\nabla_{\boldsymbol{\theta}}\log P_{jC}=P_{jD}(\bar{\boldsymbol{\alpha}}_{jD}-\bar{\boldsymbol{\alpha}}_{jC})=P_{jD}\mathbf{a}_j$, and summing over $j$ gives $\nabla_{\boldsymbol{\theta}}\log P_{\mathcal{J}}(\mathbf{C}|\boldsymbol{\theta})=\mathbf{a}_{\mathcal{J}C}$ and $\nabla_{\boldsymbol{\theta}}[-\log P_{\mathcal{J}}(\mathbf{D}|\boldsymbol{\theta})]=\mathbf{a}_{\mathcal{J}D}$. Applying the chain rule along any differentiable segment of $\boldsymbol{\theta}(\theta)$ yields
  \begin{align}
    a_{\mathcal{J}C}(\theta)=\frac{d}{d\theta}\log P_{\mathcal{J}}(\mathbf{C}|\boldsymbol{\theta}(\theta))=\mathbf{a}_{\mathcal{J}C}^{\mathsf{T}}\boldsymbol{\theta}'(\theta),\qquad a_{\mathcal{J}D}(\theta)=-\frac{d}{d\theta}\log P_{\mathcal{J}}(\mathbf{D}|\boldsymbol{\theta}(\theta))=\mathbf{a}_{\mathcal{J}D}^{\mathsf{T}}\boldsymbol{\theta}'(\theta),
  \end{align}
  so PRM is equivalent to $\mathbf{a}_{\mathcal{J}C}^{\mathsf{T}}\boldsymbol{\theta}'\geq0$ and $\mathbf{a}_{\mathcal{J}D}^{\mathsf{T}}\boldsymbol{\theta}'\geq0$ on each smooth segment. Global ordering across the whole path follows directly from \ref{thm:item:identifiability} because $P_{\mathcal{J}}(\mathbf{C}|\boldsymbol{\theta}(\theta))=P_{\mathcal{J}}(\mathbf{C}|\theta)$.

  \noindent\textbf{(iii) Ordering.} Since $w_{jC}=1$ and $w_{jD}=0$, we can write $P_j(X_j=k_j|\theta)=\left[P_{jC}(\theta)\right]^{w_{j,k_j}}\left[P_{jD}(\theta)\right]^{1-w_{j,k_j}}$, so the ratio for two patterns $\mathbf{k}$ and $\mathbf{k}'$ satisfies
  \begin{align}
    \frac{d}{d\theta}\log\frac{P_{\mathcal{J}}(\mathbf{X}=\mathbf{k}|\theta)}{P_{\mathcal{J}}(\mathbf{X}=\mathbf{k}'|\theta)}&=\sum_j(w_{j,k_j}-w_{j,k_j'})\frac{d}{d\theta}\log\frac{P_{jC}(\theta)}{P_{jD}(\theta)}\\&=\frac{a_{\mathcal{J}}(\theta)}{J}[T(\mathbf{k};\theta)-T(\mathbf{k}';\theta)].
  \end{align}
  Because $a_{\mathcal{J}}(\theta)>0$ on $\Theta_I\subseteq\Theta_T$, the derivative of the log-likelihood ratio is nonnegative throughout $\Theta_I$ if and only if $T(\mathbf{k};\theta)\geq T(\mathbf{k}';\theta)$ for all $\theta\in\Theta_I$.

  \noindent\textbf{(iv) Boundary.} Since $\mathbb{E}_\theta[w_{j,X_j}]=P_{jC}(\theta)$, taking the expectation of $T(\mathbf{X};\theta)$ gives
  \begin{align}
    \bar{T}(\theta)=\frac{J}{a_{\mathcal{J}}(\theta)}\sum_ja_j(\theta)P_{jC}(\theta)=J\frac{a_{\mathcal{J}D}(\theta)}{a_{\mathcal{J}}(\theta)},
  \end{align}
  and $J-\bar{T}(\theta)=Ja_{\mathcal{J}C}(\theta)/a_{\mathcal{J}}(\theta)$ from $a_{\mathcal{J}}=a_{\mathcal{J}C}+a_{\mathcal{J}D}$. Since $a_{\mathcal{J}}>0$ on $\Theta_T$, we have $a_{\mathcal{J}D}\geq0\Longleftrightarrow\bar{T}\geq0$ and $a_{\mathcal{J}C}\geq0\Longleftrightarrow\bar{T}\leq J$.
\end{proof}

These four properties ensure that the scalar ability $\theta$ defined by the test is identifiable, moves in a consistent direction, and orders response patterns according to their generalized credit. Conversely, under local independence, a failure of PRM suggests that the responses cannot be summarized along a single directed ability axis, which may imply multidimensionality (as seen in Figure~\ref{fig:wo_prm}). Viewing this multidimensional space $\mathbf{\Theta}\subseteq\mathbb{R}^{\dim\mathbf{\Theta}}$ through a multidimensional NRM (MNRM) over microstates $s\in\mathcal{S}$ with vector slopes $\boldsymbol{\alpha}_s\in\mathbb{R}^{\dim\mathbf{\Theta}}$ and weights $e^{-\boldsymbol{\alpha}_s^{\mathsf{T}}\boldsymbol{\theta}+\delta_s}$ further clarifies how a scalar MiRM reads out multidimensional traits (Table~\ref{tab:rel}): a latent path $\boldsymbol{\theta}(\theta)$ can preserve the individual microstate occupancies $\rho_s(\boldsymbol{\theta})$ along piecewise-linear segments, or preserve the observed macrostate probabilities $P_k(\boldsymbol{\theta})=\sum_{s\in\mathcal{S}_k}\rho_s(\boldsymbol{\theta})$ after coarsening.

Specifically, suppose that $\theta$ traces piecewise-linear segments $\boldsymbol{\theta}_h(\theta)=\mathbf{u}_h\theta+\mathbf{v}_h$ for $\theta\in\Theta_h$ ($\Theta=\bigcup_h\Theta_h$). Along segment $h$, the exponent of state $s$ becomes
\begin{align}
  -\boldsymbol{\alpha}_s^{\mathsf{T}}\boldsymbol{\theta}_h(\theta)+\delta_s=-\alpha_{s,h}\theta+\delta_{s,h},
\end{align}
where $\alpha_{s,h}=\boldsymbol{\alpha}_s^{\mathsf{T}}\mathbf{u}_h$ and $\delta_{s,h}=\delta_s-\boldsymbol{\alpha}_s^{\mathsf{T}}\mathbf{v}_h$. Projecting onto each direction $\mathbf{u}_h$ thus orders the multidimensional states by their scalar slopes $\alpha_{s,h}$; if these effective slopes and offsets agree across segments up to constants common to all states, coarsening $\mathcal{S}=\bigsqcup_k\mathcal{S}_k$ yields the same fixed scalar MiRM on every segment. 

More generally, because the observed probabilities depend on the microstates only through each $Z_k$, a path can also bend without changing the scalar MiRM. In the saturated space of dimension $\dim\mathbf{\Theta}=S-1$, where each coordinate is the log-weight contrast of one microstate against a reference state, a scalar MiRM is a straight line. Any two microstates $s,s'\in\mathcal{S}_k$ with different slopes have equal weights at a single ability $\theta_{ss'}$. If the path exchanges the two states at $\theta_{ss'}$, so that $s$ follows the weight of $s'$ for $\theta>\theta_{ss'}$ and vice versa, the path remains continuous, every $Z_k(\theta)$ is unchanged, and only the log-weights of $s$ and $s'$ change direction, exchanging their slopes $\alpha_s$ and $\alpha_{s'}$. Exchanging one microstate in each category with each of the others in turn produces at most $\sum_{k=1}^K(M_k-1)=S-K$ bends whose turns are linearly independent. The same fixed scalar MiRM is therefore the exact readout of a continuous piecewise-linear path spanning at most $S-K+1$ dimensions, which for a binary item is $c+d-1$, the full contrast space. Each additional microstate thus accommodates one bend toward a new dimension: the number of microstates bounds the dimensionality of the latent path rather than identifying it, while PRM secures a single global direction along any such path.

\subsection{Universal Approximation}
\begin{proof}[Proof of Theorem~\ref{thm:uat}]
  On the compact interval $\Theta_0=[\theta_{\mathbf{D}},\theta_{\mathbf{C}}]$, let $A_f(\theta)=\log[f(\theta)/(1-f(\theta))]\in C(\Theta_0)$ be the target logit. By the Weierstrass theorem \citep{weierstrass1885uber}, $A_f$ can be uniformly approximated on $\Theta_0$ by polynomials in $\theta$. Meanwhile, each monomial $\theta^r$ is uniformly approximated on $\Theta_0$ as $h\to0$ by
  \begin{align}
    h^{-r}(e^{-h\theta}-1)^r\longrightarrow(-\theta)^r,
  \end{align}
  whose binomial expansion is a finite exponential sum. Hence $A_f$ can be uniformly approximated on $\Theta_0$ by a signed exponential sum $E(\theta)=\sum_{m=1}^M\gamma_me^{-\alpha_m\theta}$ with real coefficients $\gamma_m\in\mathbb{R}$.

  Writing $\gamma_m^+=\max(\gamma_m,0)$ and $\gamma_m^-=\max(-\gamma_m,0)$, define $U(\theta)=1+\sum_{m=1}^M\gamma_m^+e^{-\alpha_m\theta}$ and $V(\theta)=1+\sum_{m=1}^M\gamma_m^-e^{-\alpha_m\theta}$, so that $E=U-V$ and both $U$ and $V$ have only positive coefficients. For $n\in\mathbb{N}$, define
  \begin{align}
    Z_C(\theta)=\left(1+\frac{U(\theta)}{n}\right)^n,\qquad Z_D(\theta)=\left(1+\frac{V(\theta)}{n}\right)^n.
  \end{align}
  Expanding the powers and collecting like slopes shows that $Z_C$ and $Z_D$ are finite positive exponential sums of the form $\sum_s e^{-\alpha_s\theta+\delta_s}$, so $A_n(\theta)=\log[Z_C(\theta)/Z_D(\theta)]$ is the logit of a finite-state binary MiRM. Because $U$ and $V$ are bounded on $\Theta_0$, $n\log(1+x/n)\to x$ uniformly, giving $A_n=n\log(1+U/n)-n\log(1+V/n)\to U-V=E$ uniformly on $\Theta_0$. Choosing $E$ close to $A_f$ and $n$ sufficiently large makes $A_n$ uniformly close to $A_f$; since the logistic map $A\mapsto 1/(1+e^{-A})$ is $1/4$-Lipschitz, the resulting MiRM ICC $P_C(\theta)=1/(1+e^{-A_n(\theta)})$ is uniformly within $\epsilon_0$ of $f$ on $\Theta_0$.
\end{proof}

\section{Notation}\label{appendix:notation}
See Table~\ref{tab:notation} for the item model, and Table~\ref{tab:notation_test} for the test model.

\begingroup
\renewcommand\tablebodyfont{\fontsize{7}{9}\selectfont}
\renewcommand\tablecolheadfont{\fontsize{7}{9}\selectfont}
\begin{table*}[t]
    \tabcolsep=0pt
    \TBL{\caption{Notation for the item model\label{tab:notation}}}
    {\begin{fntable}
    \begin{tabular*}{\textwidth}{@{\extracolsep{\fill}}llll@{}}
    \toprule
    \TCH{Level} & \TCH{Symbol} & \TCH{Definition} & \TCH{Identity / reading}\\
    \midrule
    \multicolumn{4}{@{}l}{\emph{Person level (ability $\boldsymbol{\theta}\in\mathbf{\Theta}\subseteq\mathbb{R}^{\dim\mathbf{\Theta}}$; test $\mathcal{J}:\boldsymbol{\theta}\rightarrow\theta\in\mathbb{R}$)}}\\
    & $\theta_i$ & ordered ability of person $i$ by a test $\mathcal{J}$ & $\theta_i\sim\mathcal{N}(0,1)$ as a working prior; index $i$ further suppressed\\
    & $\Theta$ & $\mathcal{J}(\mathbf{\Theta})\subseteq\mathbb{R}$ & range of $\theta$ (ability domain)\\
    \multicolumn{4}{@{}l}{\emph{Microstate level (microstate $s\in\mathcal{S}_k$, $M_k=|\mathcal{S}_k|$)}}\\
    & $\alpha_s,\ \delta_s$ & slope, offset of pathway $s$ & energy level, log-degeneracy\\
    & $\epsilon_s$ & $\alpha_s\theta-\delta_s$ & energy of $s$ at ability $\theta$\\
    & $\rho_s,\ \rho_{s|k}$ & $e^{-\epsilon_s}/Z,\ e^{-\epsilon_s}/Z_k$ & occupancy, within-macrostate occupancy\\
    & $\beta_s$ & $\delta_s-\log\rho_{s|k}=\alpha_s\theta+\Psi_k$ & offset plus surprisal\\
    \multicolumn{4}{@{}l}{\emph{Macrostate level (macrostate $k\in\mathcal{K}=\{1,\ldots,K\}$, $\mathcal{S}=\bigsqcup_k\mathcal{S}_k$, $S=|\mathcal{S}|$)}}\\
    & $Z_k,\ \Psi_k$ & $\sum_{s\in\mathcal{S}_k}e^{-\epsilon_s},\ \log Z_k$ & restricted partition function, its logarithm\\
    & $P_k$ & $Z_k/Z$ & response probability\\
    & $\bar\alpha_k$ & $-\Psi_k'=\sum_{s\in\mathcal S_k}\rho_{s|k}\alpha_s$ & mean slope (mean energy); $\partial_\theta\log P_k=\bar\alpha-\bar\alpha_k$\\
    & $\sigma^2_{\alpha,k}$ & $\Psi_k''=\sum_{s\in\mathcal S_k}\rho_{s|k}(\alpha_s-\bar\alpha_k)^2$ & slope fluctuation\\
    & $H_k$ & $-\sum_{s\in\mathcal S_k}\rho_{s|k}\log\rho_{s|k}$ & latent pathway entropy\\
    & $\bar\beta_k$ & $\bar\delta_k+H_k=\bar\alpha_k\theta+\Psi_k$ & intercept of the tangent $\Psi_k(\theta)=-\bar\alpha_k\theta+\bar\beta_k$\\
    \multicolumn{4}{@{}l}{\emph{Item level (item index $j\in\mathcal{J}=\{1,\ldots,J\}$ suppressed until this level)}}\\
    & $\mathcal{P}(\theta;\boldsymbol{\psi})$ & $\{P_k(\theta)\}_{k\in\mathcal{K}}$ & item response model; $\boldsymbol{\psi}=\{\alpha_s,\delta_s\}$\\
    & $Z,\ \Psi$ & $\sum_kZ_k,\ \log Z$ & partition function, its logarithm\\
    & $\bar\alpha,\ \sigma^2_\alpha$ & $-\Psi',\ \Psi''$ & $\bar\alpha=\sum_kP_k\bar\alpha_k$\\
    & $I_{\rm mic},\ I_{\rm mac}$ & $\sigma^2_\alpha,\ \sum_kP_k(\bar\alpha_k-\bar\alpha)^2$ & microstate/response-carried Fisher information\\
    & $I_{\rm loss}$ & $I_{\rm mic}-I_{\rm mac}=\sum_kP_k\sigma^2_{\alpha,k}$ & information loss by coarse-graining\\
    \multicolumn{4}{@{}l}{\emph{Dichotomous item: \mirm{c,d} ($k\in\{C,D\}$, $c=M_C$, $d=M_D$)}}\\
    & $A$ & $\log(P_C/P_D)=\Psi_C-\Psi_D$ & item logit\\
    & $a$ & $A'=\bar\alpha_D-\bar\alpha_C$ & local discrimination; $P_C'=P_CP_Da$; $I_{\rm mac}=P_CP_Da^2$\\
    & $a'$ & $\sigma^2_{\alpha,C}-\sigma^2_{\alpha,D}$ & local acceleration; constant $2q_2$ for 3PLQ\\
    & $b$ & $(\bar\beta_D-\bar\beta_C)/(\bar\alpha_D-\bar\alpha_C)$ & local difficulty, defined where $a\neq0$; $A=a(\theta-b)$\\
    & $p_c,\ p_d$ & $P_C(-\infty),\ P_D(+\infty)$ & lower/upper asymptotes\\
    \multicolumn{4}{@{}l}{\emph{Scored polytomous item: \mirm{\mathbf{M};\mathbf{w}} ($C=1$, $D_{k-1}=k$ for $k\geq2$)}}\\
    & $\mathbf{M}$ & $(M_1,\ldots,M_K)$ & microstate configuration; nominal $(c,d_1,\ldots,d_{K-1})$\\
    & $w,\ \mathbf{w}$ & $w:\mathcal{K}\to\mathcal{R}\subseteq\mathbb{R}$, $(w_1,\ldots,w_K)$ & credit rule and credit vector\\
    & $Y$ & $w(X)$ & credit of a response\\
    & $\bar w(\theta)$ & $\sum_kw_kP_k$ & credit expectation curve\\
    \botrule
    \end{tabular*}
    \footnotetext[]{{Note:} Primes denote $d/d\theta$; the correct macrostate is expected to carry the microstate with the lowest slope $\alpha_1$.}
\end{fntable}}
\end{table*}

\begin{table*}[t]
  \tabcolsep=0pt
  \TBL{\caption{Notation for the test model\label{tab:notation_test}}}
  {\begin{fntable}
  \begin{tabular*}{\textwidth}{@{\extracolsep{\fill}}llll@{}}
  \toprule
  \TCH{Level} & \TCH{Symbol} & \TCH{Definition} & \TCH{Identity / reading}\\
  \midrule
  \multicolumn{4}{@{}l}{\emph{Test level (test $\mathcal{J}$, data $\mathcal{D}$; pattern $\mathbf{k}=(k_1,\ldots,k_J)\in\mathcal{K}_{\mathcal J}\equiv\prod_j\mathcal{K}_j$)}}\\
  & $P_{\mathcal J}(\mathbf{X}=\mathbf{k}|\theta)$ & $\prod_jP_j(X_j=k_j|\theta)$ & local independence\\
  & $\mathcal{P}^\star(\theta;\boldsymbol{\psi})$ & MiRM on $\mathcal{K}_{\mathcal J}$ with $Z^\star_{\mathbf{k}}$ & giant model; a test is one scored item\\
  & $Z^\star_{\mathbf{k}},\ Z^\star$ & $\prod_jZ_{j,k_j},\ \prod_jZ_j$ & giant partition functions; $P_{\mathcal J}(\mathbf{X}=\mathbf{k}|\theta)=Z^\star_{\mathbf{k}}/Z^\star$\\
  & $\bar\alpha^\star,\ \bar\alpha^\star_{\mathbf{k}}$ & $\sum_j\bar\alpha_j,\ \sum_j\bar\alpha_{j,k_j}$ & giant mean slopes; $\partial_\theta\log P_{\mathcal J}(\mathbf{X}=\mathbf{k}|\theta)=\bar\alpha^\star-\bar\alpha^\star_{\mathbf{k}}$\\
  & $\mathbf{C},\ \mathbf{D}$ & $(C,\ldots,C),\ (D,\ldots,D)$ & all-correct and all-distractor response patterns\\
  & $a_{\mathcal{J}C},\ a_{\mathcal{J}D}$ & $\partial_\theta\log P_{\mathcal J}(\mathbf{C}|\theta),\ -\partial_\theta\log P_{\mathcal J}(\mathbf{D}|\theta)$ & equals $\sum_jP_{jD}a_j,\ \sum_jP_{jC}a_j$; $a_{\mathcal J}=a_{\mathcal{J}C}+a_{\mathcal{J}D}$\\
  & $\operatorname{PRM}_{\mathcal J}(\theta)$ & $a_{\mathcal{J}C}(\theta),a_{\mathcal{J}D}(\theta)\geq0$ & perfect-response monotonicity\\
  & $I_{\mathcal J}$ & $\sum_jP_{jC}P_{jD}a_j^2$ & test information; $=\mathrm{Var}_\theta(\bar\alpha^\star_{\mathbf X})$\\
  \multicolumn{4}{@{}l}{\emph{Instrument level (a test satisfying $\operatorname{PRM}_{\mathcal J}(\theta)$ for all $\theta$)}}\\
  & $\hat\theta(\mathbf{k})$ & posterior mean of $\theta$ given $\mathbf{k}$ & ability estimate (EAP under $\mathcal{N}(0,1)$)\\
  & $\theta_{\mathbf D},\ \theta_{\mathbf C}$ & $\hat\theta(\mathbf D),\ \hat\theta(\mathbf C)$ & anchor abilities; under PRM, $\theta_{\mathbf D}\leq E_\pi[\theta]\leq\theta_{\mathbf C}$\\
  & $J$ & $\sum_j\max_kw_{jk}$ & maximum credit for binary items ($w_{jC}=1, w_{jD}=0$)\\
  & $\Theta_T$ & $\{\theta\in\Theta:a_{\mathcal J}(\theta)>0\}$ & positive score-direction domain\\
  & $\omega_j(\theta)$ & $Ja_j(\theta)/a_{\mathcal J}(\theta)$ & local item weight on $\Theta_T$; $\sum_j\omega_j=J$\\
  & $T(\mathbf{k};\theta)$ & $\sum_j\omega_jw_{j,k_j}=J(\bar\alpha^\star_{\mathbf D}-\bar\alpha^\star_{\mathbf{k}})/(\bar\alpha^\star_{\mathbf D}-\bar\alpha^\star_{\mathbf C})$ & generalized credit; $\partial_\theta\log P_{\mathcal J}=a_{\mathcal J}(T-\bar T)/J$\\
  & $\hat T(\mathbf{k})$ & $T(\mathbf{k};\hat\theta(\mathbf{k}))$ & generalized total score\\
  & $\bar T(\theta)$ & $E_\theta[T]=Ja_{\mathcal{J}D}/a_{\mathcal J}$ & expected generalized score; on $\Theta_T$, PRM $\iff0\leq\bar T\leq J$\\
  \multicolumn{4}{@{}l}{\emph{Estimation ($\boldsymbol{\psi}\in\Omega_{\mathbf M}$; PRM-feasible set $\Omega_{\mathbf M;\mathbf w}$, data $\mathcal D$)}}\\
  & $\Theta_{\rm P},\ Q_{\rm P}$ & PRM grid $\{\theta_q\}$ and its size & $Q_{\rm P}=2Q-1$ for $Q$ Gauss--Hermite nodes\\
  & $\tilde a_{\mathcal{J}C},\ \tilde a_{\mathcal{J}D}$ & normalized margins for $a_{\mathcal{J}C},\ a_{\mathcal{J}D}$ & lie in $[-1,1]$ and have the same signs as $a_{\mathcal{J}C},\ a_{\mathcal{J}D}$\\
  & $g_q^k,\ \lambda_q^k,\ \rho$ & $-\tilde a_{\mathcal{J}k}(\theta_q)\leq0$, multiplier, penalty & parameters for PHR augmented Lagrangian $\operatorname{PRML}$\\
  \botrule
  \end{tabular*}
\end{fntable}}
\end{table*}
\endgroup

\section{Additional Relations}\label{appendix:relations}

\subsection{Latent Response Model Family}
\citet{maris1995psychometric} defined a latent response model using a vector of latent responses $\mathbf{X}=(X_1,\ldots,X_H)$ and a deterministic condensation rule $Y=C(\mathbf{X})$. Suppose that the latent responses are conditionally independent, with $P(X_h=1\mid\theta)=e^{\eta_h(\theta)}/[1+e^{\eta_h(\theta)}]$, where each $\eta_h(\theta)$ is linear in $\theta$. A complete pattern $\mathbf{k}\in\{0,1\}^H$ then has probability
\begin{align}
P(\mathbf{X}=\mathbf{k}\mid\theta)
=\frac{\exp[\sum_{h=1}^H k_h\eta_h(\theta)]}
{\sum_{\mathbf{k}'\in\{0,1\}^H}\exp[\sum_{h=1}^H k'_h\eta_h(\theta)]}.
\end{align}
Each pattern is thus a microstate, and the patterns satisfying $C(\mathbf{k})=k$ constitute macrostate $k$. The conjunctive Rasch model assigns only the all-success pattern to the correct macrostate, giving \mirm{1,2^H-1}; the disjunctive Rasch model similarly gives \mirm{2^H-1,1}. With identical conjunctive components, the former reduces to the integer-exponent LPE model.

Finite logistic IRTrees with linear node logits \citep{de2012irtrees,bockenholt2012modeling} follow the same construction. Completing each terminal path with the outcomes of nodes not reached gives a binary latent-response pattern; the terminal category supplies its condensation rule. Thus, these IRTrees are exact constrained MiRMs.

The 2-parameter nested logit model \citep{suh2010nested} instead combines a logistic correct-response decision with an NRM over $K-1$ distractors conditional on an incorrect response. Writing their linear logits as $A(\theta)$ and $\Psi_{D_k}(\theta)$, respectively, gives the MiRM weights
\begin{align}
Z_C(\theta)=\sum_{\ell=1}^{K-1}e^{A(\theta)+\Psi_{D_\ell}(\theta)},
\qquad
Z_{D_k}(\theta)=e^{\Psi_{D_k}(\theta)}.
\end{align}
It is therefore \mirm{K-1,1,\ldots,1}, with the correct-category microstates constrained to have logits $A+\Psi_{D_\ell}$. More generally, a \mirm{c,d} at the first stage gives \mirm{c(K-1),d,\ldots,d}, which also covers the 3- and 4-parameter nested logit models.
 
\subsection{Quadratic-Logit Model}\label{app:sub:3plq}
The quadratic-logit model (3PLQ; see \cite{feuerstahler2019metric,robitzsch2022choice}) is
\begin{align}
  P_C(\theta)=\frac{1}{1+e^{-(q_0+q_1\theta+q_2\theta^2)}}.
\end{align}
If the correct slopes form a continuous spectrum $\alpha\sim\mathcal{N}(\mu,\sigma^2)$, then
\begin{align}
  Z_C=\int e^{-\alpha\theta}d\mathcal{N}(\mu,\sigma^2)(\alpha)=e^{-\mu\theta+\sigma^2\theta^2/2}.
\end{align}
If the distractor has a single slope $\alpha_D$, then 
\begin{align}
  A=\log Z_C-\log Z_D=\frac{\sigma^2}{2}\theta^2+(\alpha_D-\mu)\theta+\operatorname{const},
\end{align}
a quadratic logit with $q_2=\sigma^2/2>0$. For $q_2<0$, place the spectrum on the distractor side. Conversely, for $\log Z_C$ to be a quadratic polynomial, the cumulant-generating function of the spectrum must be quadratic. By Marcinkiewicz's theorem \citep{marcinkiewicz1939propriete}, a distribution with a polynomial cumulant-generating function has degree at most two and is Gaussian. The cumulant-generating function of a finite spectrum approaches a linear function at both extremes, so a finite MiRM cannot reproduce the 3PLQ exactly. This is why the 3PLQ lies outside finite MiRM but is contained as a limit.

\subsection{Generalized Graded Unfolding Model}
In the generalized graded unfolding model (GGUM; \cite{roberts2000general}), the probability of category $k\in\{1,\ldots,K\}$ is proportional to 
\begin{align}
  Z_k(\theta)=e^{a[(k-1)(\theta-b)-\sum_{\ell=1}^k\tau_\ell]}+e^{a[(2K-k)(\theta-b)-\sum_{\ell=1}^k\tau_\ell]},
\end{align}
a sum of two exponentials, where $b$ is the item location and $\tau_\ell$ are threshold parameters, and the denominator is $\sum_{k=1}^KZ_k(\theta)$. Thus each category has two microstates, giving \mirm{2,\ldots,2}. This is a representative example of Corollary~\ref{cor:peak}: the paired low- and high-slope terms produce an extremal reversal, creating a peak at $\theta=b$. Cognitively, this reads as two mechanisms on either side of the ideal point that pull respondents away from endorsing the item.

\subsection{Mixture IRT Models}\label{app:sub:mixture}
Consider a mixture 2PL model \citep{mislevy1990modeling,vondavier2004partially}. Let group $g\in\{1,\ldots,G\}$ have mixing weight $\pi_g>0$, with $\sum_g\pi_g=1$, and define $z_g(\theta)=e^{-a_g(\theta-b_g)}$. The group-specific 2PL probabilities are
\begin{align}
  P_{C,g}(\theta)=\frac{1}{1+z_g(\theta)},\qquad P_{D,g}(\theta)=\frac{z_g(\theta)}{1+z_g(\theta)}.
\end{align}
The marginal correct-response probability is therefore
\begin{align}
  P_C(\theta)=\sum_{g=1}^G\frac{\pi_g}{1+z_g}=\frac{\sum_{g=1}^G\pi_g\prod_{h\neq g}(1+z_h)}{\prod_{g=1}^G(1+z_g)}.
\end{align}
Expanding the products over subsets $H\subseteq\{1,\ldots,G\}$ gives
\begin{align}
  \sum_{g=1}^G\pi_g\prod_{h\neq g}(1+z_h)=\sum_{H\subsetneq\{1,\ldots,G\}}\left(\sum_{g\notin H}\pi_g\right)\prod_{h\in H}z_h.
\end{align}
Likewise,
\begin{align}
  P_D(\theta)=\sum_{g=1}^G\frac{\pi_gz_g}{1+z_g}=\frac{\sum_{H\neq\varnothing}\left(\sum_{g\in H}\pi_g\right)\prod_{h\in H}z_h}{\prod_{g=1}^G(1+z_g)}.
\end{align}
For each subset $H$,
\begin{align}
  \prod_{h\in H}z_h=\exp\left[-\sum_{h\in H}a_h\theta+\sum_{h\in H}a_hb_h\right].
\end{align}
Thus every proper subset $H\subsetneq\{1,\ldots,G\}$ defines a correct microstate with
\begin{align}
  \alpha_{C,H}=\sum_{h\in H}a_h,\qquad\delta_{C,H}=\sum_{h\in H}a_hb_h+\log\left(\sum_{g\notin H}\pi_g\right),
\end{align}
whereas every nonempty subset $H\neq\varnothing$ defines a distractor microstate with
\begin{align}
  \alpha_{D,H}=\sum_{h\in H}a_h,\qquad\delta_{D,H}=\sum_{h\in H}a_hb_h+\log\left(\sum_{g\in H}\pi_g\right).
\end{align}
There are $2^G-1$ proper subsets and $2^G-1$ nonempty subsets. Hence a mixture of $G$ 2PL models is exactly a constrained \mirm{2^G-1,2^G-1}. If some subset sums of the slopes coincide, the corresponding exponential terms combine and the minimal number of microstates is smaller. The mixture 2PL is therefore a nested family within MiRM, with latent group membership represented by group-aware microstates.

Furthermore, for a fixed-weight mixture of positively oriented 2PL components ($a_g>0$) where $P(G=g\mid\theta)=\pi_g$ is independent of $\theta$, the marginal ICC satisfies
\begin{align}
P_C'(\theta)=\sum_{g=1}^G\pi_ga_gP_{C,g}(\theta)P_{D,g}(\theta)>0,
\end{align}
so a fixed-weight mixture of monotone 2PLs cannot produce a dip. More generally, if $w_g(\theta)=P(G=g\mid\theta)$ varies with ability, then
\begin{align}
P_C'(\theta)=\sum_{g=1}^Gw_g(\theta)a_gP_{C,g}(\theta)P_{D,g}(\theta)+\sum_{g=1}^Gw_g'(\theta)P_{C,g}(\theta).
\end{align}
Thus, under positively oriented group-specific ICCs, a mixture-induced dip requires the group-composition term to outweigh the positive within-group derivatives. Moreover, because a group whose membership depends this strongly on $\theta$ can itself be read as a component of ability, population heterogeneity may offer an alternative representation of a dip, but does not by itself refute the dip.

\section{Estimation Procedure}\label{appendix:estimation}
Let $\mathcal{D}$ be the set of observed respondent--item pairs $(i,j)$ with binary response $y_{ij}\in\{0,1\}$. Under the working prior $\theta\sim\mathcal{N}(0,1)$, the marginal negative log-likelihood is approximated by $Q=41$ Gauss--Hermite quadrature nodes $\theta_q=\sqrt{2}x_q$ and normalized weights $w_q=\tilde{w}_q/\sqrt{\pi}$ ($\sum_{q=1}^Qw_q=1$):
\begin{align}
  \operatorname{NLL}(\boldsymbol{\psi})=-\sum_{i=1}^I\log\!\left[\sum_{q=1}^Qw_q\exp\!\left\{\sum_{j:(i,j)\in\mathcal{D}}\log P_j(y_{ij}\mid\theta_q;\boldsymbol{\psi})\right\}\right].
\end{align}
The inner sum over items is evaluated by sparse matrix multiplication over whole-student chunks, so the exact marginal objective and its gradients are used at every step. For MiRM, raw parameters $\alpha_{j,s}$ and $\delta_{j,s}$ are initialized from $\mathcal{N}(0,0.1^2)$. At evaluation, the active state with the lowest slope (and largest offset if tied) is set as the reference state $\alpha_1=\delta_1=0$. All models are optimized in PyTorch using L-BFGS (learning rate 1, up to 20 iterations per epoch, history size 100, strong Wolfe line search) for up to 500 epochs, with early stopping when the per-response NLL change falls below $10^{-9}$ for 5 consecutive epochs. When PRM is imposed, the model from the feasible epoch with the lowest NLL is returned.

In the baseline models, positive parameters ($a$, $\xi$, and the I-spline coefficients on $[-4,4]$) are parameterized through the softplus function and the 3PL lower asymptote through the logistic function. For held-out evaluation (Table~\ref{tab:pfm_pisa}), each random split partitions respondents into 80\% training and 20\% test respondents, and all item parameters and per-item $(c_j,d_j)$ configurations are estimated using only the training respondents. On the test respondents, each respondent's observed items are randomly partitioned into 5 folds, and the response probability for each fold is predicted by integrating $P_j(y_{ij}\mid\theta_q)$ over the ability posterior computed from the remaining 4 folds.

\subsection{Details for PRML}\label{app:sub:prml}
For the total loss $\operatorname{NLL}+\operatorname{PRML}$, the PRM loss is
\begin{align}
  \operatorname{PRML}(\boldsymbol{\psi},\boldsymbol{\lambda})=\frac{N_{\rm obs}\ln2}{Q_{\rm P}}\sum_{k\in\{C,D\}}\sum_{q=1}^{Q_{\rm P}}\frac{[\lambda_q^k+\rho g_q^k(\boldsymbol{\psi})]_+^2-(\lambda_q^k)^2}{2\rho}.
\end{align}
The factor $N_{\rm obs}\ln2$ is a convention that places the normalized mean violation and likelihood loss on comparable numerical scales, and $\rho=1$ is the default heuristic on that scale. 

For each side $k\in\{C,D\}$, the PRM evaluation grid $\Theta_{\rm P}$ consists of the $Q$ fixed Gauss--Hermite nodes $\theta_1<\cdots<\theta_Q$ together with one movable cut in each of the $Q-1$ open intervals $(\theta_q,\theta_{q+1})$, giving $Q_{\rm P}=2Q-1$ points per side. The cuts start at the interval midpoints. At the end of each epoch, every cut is moved to the point in its interval that minimizes the normalized margin $\tilde{a}_{\mathcal{J}k}(\theta)$ among $32$ equally spaced interior scan points, all within-item microstate line crossings $\theta=(\delta_s-\delta_{s'})/(\alpha_s-\alpha_{s'})$ falling inside the interval, and the cut's current position. The multipliers are then updated by $\lambda_q^k\leftarrow[\lambda_q^k+\rho g_q^k(\boldsymbol{\psi})]_+$, and the L-BFGS curvature history is reset whenever a cut moves or $\boldsymbol{\lambda}$ or $\rho$ changes.

If grid violations remain after the prescribed epochs, training continues until they disappear; if violations fail to decrease for several epochs, $\rho$ is increased stepwise. Final PRM verification uses an adaptive interval certificate for the margins $a_{\mathcal{J}C}(\theta)$ and $a_{\mathcal{J}D}(\theta)$. Since $|a_{\mathcal{J}k}'(\theta)|\leq\frac{1}{4}\sum_j\alpha_{j,S_j}^2$, where $\alpha_{j,S_j}$ is the largest slope of item $j$, each margin is nonnegative throughout an interval of width $h$ if its values at both endpoints are at least $\frac{h}{8}\sum_j\alpha_{j,S_j}^2$. Only intervals failing this condition are bisected and classified as certified, violated, or uncertified at the maximum depth. When scoring is used, $a_{\mathcal{J}}(\theta)>0$ is checked separately.

\subsection{Per-Item MiRM Selection}\label{app:sub:peritem}

Let $c_j$ and $d_j$ denote the numbers of correct and distractor microstates for item $j$. Selection begins at $(c_j,d_j)=(1,1)$ for every item. At each iteration, we jointly fit the four configurations indexed by a grid
\begin{align}
  \mathcal{G}=\{(0,0),(1,0),(0,1),(1,1)\}
\end{align}
where $g=(u,v)$ denotes configuration $(c_j+u,d_j+v)$ for item $j$. The four configurations share the parameters of their common microstates. Rather than fitting four separate models, we minimize the mean marginal negative log-likelihood of the four configurations over these shared parameters. When PRM is used, its constraints are imposed on the full item bank corresponding to each configuration.

Let $\mathcal{D}$ be the set of observed respondent--item pairs, and let $(\theta_q,w_q)$ be the quadrature nodes and weights used to integrate over ability. Define respondent $i$'s conditional log-likelihood under configuration $g$ as
\begin{align}
  \ell_i^{\,g}(\theta_q)=\sum_{j:(i,j)\in\mathcal{D}}\log p_{j,g}(y_{ij}\mid\theta_q),
\end{align}
where $p_{j,g}$ is the response probability for item $j$ under that configuration. The fitting objective is
\begin{align}
  \mathcal{L}=-\frac{1}{4}\sum_{g\in\mathcal{G}}\sum_i\log\!\left[\sum_qw_q\exp\{\ell_i^{\,g}(\theta_q)\}\right].
\end{align}

After fitting, we construct one ability posterior for each respondent that is shared across the four configurations. The default implementation geometrically pools their conditional likelihoods:
\begin{align}
  \bar{\pi}_{iq}=\frac{w_q\exp\!\left\{\frac{1}{4}\sum_{g\in\mathcal{G}}\ell_i^{\,g}(\theta_q)\right\}}{\sum_{q'}w_{q'}\exp\!\left\{\frac{1}{4}\sum_{g\in\mathcal{G}}\ell_i^{\,g}(\theta_{q'})\right\}}.
\end{align}
Holding this posterior fixed, we compute the expected negative log-likelihood for each item and configuration,
\begin{align}
  L_{j,g}=-\sum_{i:(i,j)\in\mathcal{D}}\sum_q\bar{\pi}_{iq}\log p_{j,g}(y_{ij}\mid\theta_q).
\end{align}
Write $L_{j,uv}$ for the loss of configuration $(c_j+u,d_j+v)$, and define
\begin{align}
  \operatorname{UL}_j=L_{j,10}-L_{j,00},\qquad\operatorname{LU}_j=L_{j,01}-L_{j,00},\qquad\operatorname{UU}_j=L_{j,00}-L_{j,11}.
\end{align}
We add a correct microstate when
\begin{align}
  \operatorname{UL}_j<0,\qquad\operatorname{LU}_j+\operatorname{UU}_j>0.
\end{align}
The first inequality requires the addition to improve fit at $d_j$; the second requires it to improve fit at $d_j+1$, since $\operatorname{LU}_j+\operatorname{UU}_j=L_{j,01}-L_{j,11}$. Thus, the addition must help in both neighboring distractor configurations. Symmetrically, we add a distractor microstate when
\begin{align}
  \operatorname{LU}_j<0,\qquad\operatorname{UL}_j+\operatorname{UU}_j>0.
\end{align}
Both counts can increase in the same iteration when both addition rules hold.

A separate joint-move rule applies when neither single-addition rule holds, each single addition increases loss, and adding both types together yields a sufficiently large gain. Specifically, a joint move from $(c_j,d_j)$ to $(c_j+1,d_j+1)$ requires
\begin{align}
  0<\min(\operatorname{UL}_j,\operatorname{LU}_j)<\operatorname{UU}_j
\end{align}
and the failure of both single-addition rules. This joint move is therefore distinct from an iteration in which the two single-addition rules both hold.

To distinguish item-specific gains from an improvement common to the item bank, we apply the same decision rules twice: first to $(\operatorname{UL}_j,\operatorname{LU}_j,\operatorname{UU}_j)$, and then to
\begin{align}
  \bigl(\operatorname{UL}_j-\overline{\operatorname{UL}},\ \operatorname{LU}_j-\overline{\operatorname{LU}},\ \operatorname{UU}_j-\overline{\operatorname{UU}}\bigr).
\end{align}
The means are computed across all items, including those that cannot grow further. A move is accepted only if it is selected by both sets of contrasts. State-count limits are enforced so that all four configurations in the next iteration remain within the specified bounds. Newly exposed microstates are initialized with small weights.

We repeat fitting and selection until no item changes configuration in two consecutive iterations. We then retain only the selected microstates $(c_j,d_j)$, initialize their parameters from the shared-parameter fit, and refit the resulting item-specific MiRM.

\section{Configuration Statistics of Used Datasets}\label{appendix:datasets}
Table~\ref{tab:config_pisa} summarizes the three PISA 2015 domains used in Section~\ref{sec:emp}. From the official cognitive response file, we extracted all computer-based assessment (CBA) automatically scored items and dichotomized valid responses according to the official scoring rules: full credit was coded as 1, and partial credit or no credit as 0. Unadministered items from the rotated-booklet design and missing codes (not reached, omitted, or invalid) were excluded, retaining all students and items with at least one valid response.

\begingroup
\renewcommand\tablebodyfont{\fontsize{7}{9}\selectfont}
\renewcommand\tablecolheadfont{\fontsize{7}{9}\selectfont}
\begin{table}[t]
  \tabcolsep=0pt%
  \TBL{\caption{Configuration of PISA 2015 datasets\label{tab:config_pisa}}}
  {\begin{fntable}
  \begin{tabular*}{\textwidth}{@{\extracolsep{\fill}}lccc@{}}\toprule%
  &\TCH{PISA 2015 Reading}&\TCH{PISA 2015 Science}&\TCH{PISA 2015 Mathematics}\\\midrule  
  \TCH{Number of responses ($|\mathcal{D}|$)} & 2,483,550 & 9,106,213 & 2,730,800 \\
  \TCH{Number of students ($I$)} & 187,954 & 445,955 & 188,046 \\
  \TCH{Number of items ($J$)} & 55 & 126 & 60 \\
  \TCH{Number of choices ($K$)} & 2 (collapsed) & 2 (collapsed) & 2 (collapsed) \\
  \TCH{Density ($|\mathcal{D}|/(IJ)$)} & 0.240 & 0.162 & 0.242 \\
  \TCH{Response accuracy} & 0.574 & 0.506 & 0.476 \\
  \TCH{Mean (min/max) \# students per item} & 45,155 (13,066 / 57,251) & 72,272 (41,116 / 97,214) & 45,513 (10,078 / 57,497) \\
  \TCH{Mean (min/max) \# items per student} & 13.2 (1 / 20) & 20.4 (1 / 45) & 14.5 (1 / 18) \\
  \botrule
  \end{tabular*}%
  \footnotetext[]{{Note:} Response accuracy is the proportion of full-credit responses among all valid responses. Science was the major domain in PISA 2015.}
  \end{fntable}}
\end{table}
\endgroup

\section{Additional Verification for Nonmonotonicity}\label{appendix:nonmono}
 
\subsection{Full-Bank PRM and Ability Estimate Variants}
The dips in Figure~\ref{fig:nonmono} were shown using the fitted MiRM ICCs and rest-score bins. Here, we check whether they also appear under other ability estimates. Theoretically, conditioning on the total score $X_+=\sum_\ell X_\ell$ can produce manifest (estimated) nonmonotonicity even when all latent (true) ICCs are monotonic, whereas conditioning on the rest score $R_{-j}=X_+-X_j$ avoids this because $P(X_j=1\mid R_{-j}=r)$ is always nondecreasing under unidimensionality, local independence, and itemwise monotonicity \citep{junker2000latent}. Following this result, we used rest-score bins in Figure~\ref{fig:nonmono}.

Under full-bank PRM, however, the rest score also sharpens a genuine dip. Because PRM aligns the ability direction across the whole bank, a large number of remaining items provides positive evidence for ability; if item $j$ reverses locally ($a_j(\theta)<0$), removing it makes the positive direction of the remaining items even stronger:
\begin{align}
a_{\mathcal{J}\setminus{j},C}(\theta)=\sum_{\ell\neq j}P_{\ell D}(\theta)a_\ell(\theta)=a_{\mathcal{J}C}(\theta)-P_{jD}(\theta)a_j(\theta)> a_{\mathcal{J}C}(\theta)\geq0.
\end{align}
This implies both that imposing PRM on small individual forms is not always preferable, and that as the bank grows, spurious dips are suppressed while genuine dips survive. Conversely, because the total score $X_+=R_{-j}+X_j$ includes item $j$ itself, answering item $j$ correctly raises $X_+$ by one point and partially smooths out a true dip. Even so, Figure~\ref{fig:dip_bins} shows that the local declines in all three items remain visible when respondents are binned by rank-normalized total scores (black circles) or 2PL EAP estimates (dark blue triangles), while aligning most closely with the fitted curves under the per-item MiRM EAP estimates (light blue squares).

\begin{figure}[t]%
  \FIG{\includegraphics[width=\textwidth]{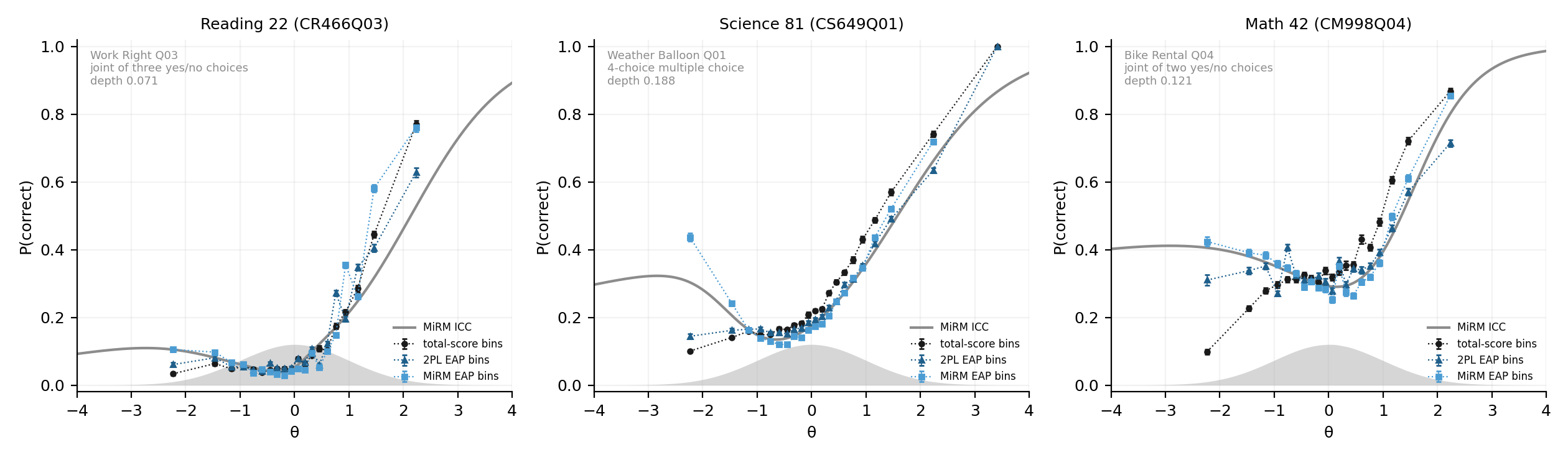}}
  {\caption{Nonmonotone ICCs of the three PISA 2015 items in Figure~\ref{fig:nonmono}, with observed proportions correct ($\pm 1$ SE) binned by rank-normalized total scores (black circles), 2PL EAP ability estimates (dark blue triangles), and per-item MiRM EAP ability estimates (light blue squares)}\label{fig:dip_bins}}
\end{figure}

\subsection{Addressing Alternative Explanations for Nonmonotonicity}
Here we examine alternative explanations for the dips in Figure~\ref{fig:nonmono}. Note that our aim is not to reject every alternative account. We view nonmonotonicity not as always strictly intrinsic to an item, but as interchangeable with representations of multidimensionality, asymmetry, or group characteristics. However, saying that nonmonotonicity can be re-expressed through such accounts is different from saying that it lacks reality and must be absorbed into them. Our claim is the former, and we examine below whether the dips survive under several alternative explanations.

\begin{figure}[t]%
  \FIG{\includegraphics[width=\textwidth]{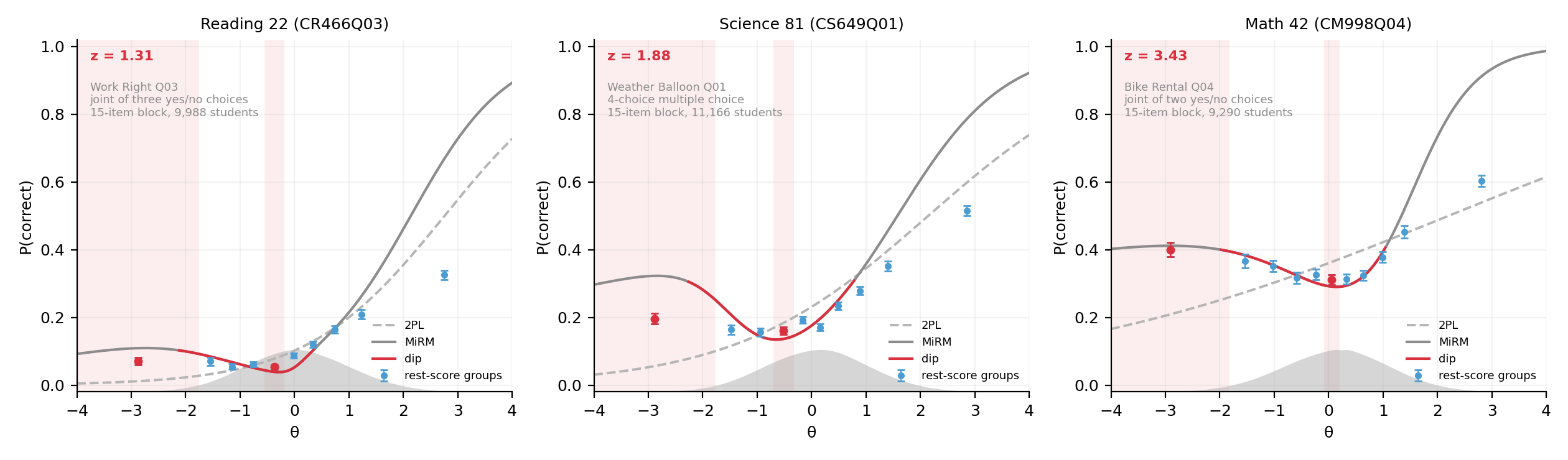}}
  {\caption{Nonmonotone ICCs and rest-score group proportions ($\pm 1$ SE) of the three PISA 2015 showcase items within complete $15$-item blocks in which every respondent answered all $15$ items}\label{fig:dip_block}} 
\end{figure}

Several factors can be pointed out as sources of nonmonotonicity, especially data sparsity and group differences. First, because PISA administers items through a rotated-booklet design, respondents answer different subsets of items, and simulation results in Section~\ref{sub:nonmono_sim} showed that data sparsity can increase false-positive dip detections. To examine whether the empirical dips in Figure~\ref{fig:nonmono} are artifacts of missingness or incomparable rest-score compositions, we extracted for each showcase item the largest complete respondent-by-item submatrix of 15 items in which every respondent answered all 15 items (9,988 respondents for Reading 22, 11,166 for Science 81, and 9,290 for Mathematics 42). Within each complete block, the rest score $R_{-j}\in\{0,\ldots,14\}$ is computed over the exact same $14$ items for every respondent; consecutive rest-score values were merged from the bottom until each group contained at least 10\% of the block sample and placed on the ability axis via the expected rest score $\mathbb{E}[R_{-j}\mid\theta]$. As shown in Figure~\ref{fig:dip_block}, restricting to a single 15-item block substantially reduces sample size and statistical power, yet the downward direction of the observed proportions correct is preserved in all three items ($z=1.31$, 1.88, and 3.43 for Reading 22, Science 81, and Mathematics 42, respectively) and remains statistically significant on its own for Mathematics 42. For Reading 22 and Science 81, this block-level check is complemented by the higher-powered stratified analysis below.

\begin{figure}[t]%
  \FIG{\includegraphics[width=\textwidth]{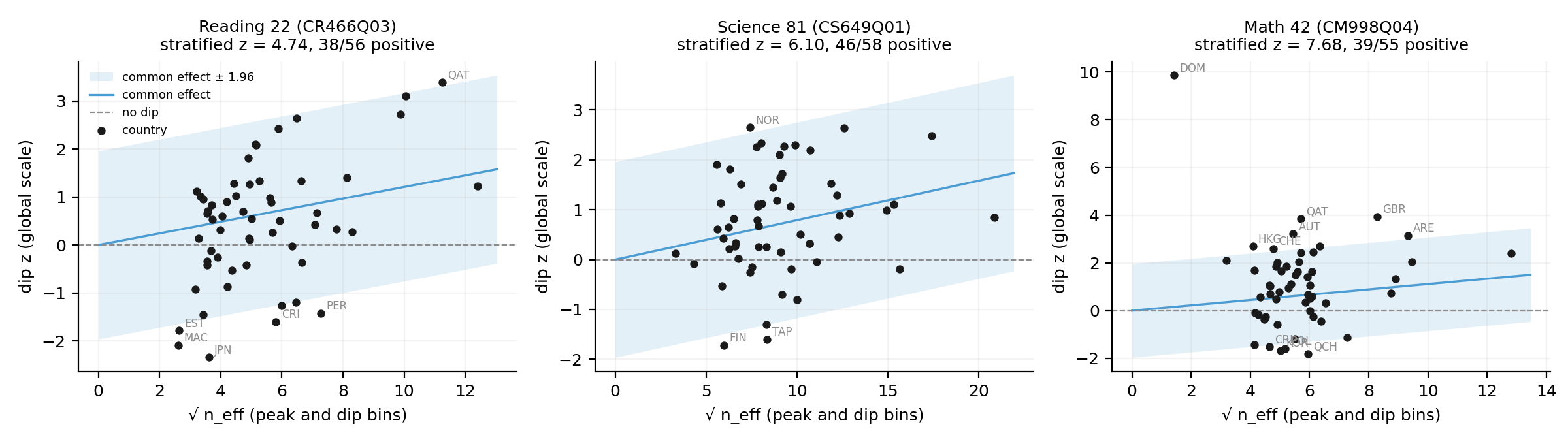}}
  {\caption{Within-country dip $z$-scores plotted against the square root of the effective sample size $\sqrt{n_{\rm eff}}$ in the compared peak and dip bins for each showcase item. Blue lines and shaded bands show the common-effect slope through the origin and its 95\% interval}\label{fig:dip_dif}}
\end{figure}

A second alternative explanation is population heterogeneity across countries: as shown in Appendix~\ref{app:sub:mixture}, pooling subpopulations that differ in ability distribution and item functioning (DIF) could in principle produce a marginal dip even when within-group ICCs are monotonic. To test whether the observed dips are an artifact of cross-national aggregation, we evaluated the peak-minus-dip contrast separately within each participating country or economy with at least 300 respondents on the item (56, 58, and 55 countries for Reading 22, Science 81, and Mathematics 42, respectively) on the common ability scale. Figure~\ref{fig:dip_dif} plots each country's dip $z$-score against the square root of its effective sample size $\sqrt{n_{\rm eff}}$ in the compared bins. The within-country dip contrast is positive in the majority of countries---38 of 56 countries for Reading 22 (stratified $z=4.74$), 46 of 58 for Science 81 (stratified $z=6.10$), and 39 of 55 for Mathematics 42 (stratified $z=7.68$)---and increases systematically with $\sqrt{n_{\rm eff}}$ along the common-effect line. Because the local decline persists after conditioning on country, cross-national composition can be ruled out as the sole source of the nonmonotonicity.

\end{appendix}

\end{Backmatter}


\end{document}